\PassOptionsToPackage{usenames,dvipsnames}{xcolor}
\documentclass{article}

\usepackage{arxiv}

\usepackage[utf8]{inputenc}
\usepackage[T1]{fontenc}
\usepackage{times}
\usepackage[hidelinks]{hyperref}
\usepackage{url}
\usepackage{booktabs}
\usepackage{amsfonts}
\usepackage{amsmath}
\usepackage{amssymb}
\usepackage{bm}
\usepackage{siunitx}
\DeclareSIUnit{\volpercent}{vol.\%}
\DeclareSIUnit{\cal}{cal}
\DeclareSIUnit{\angstrom}{\text{Å}}
\DeclareSIUnit{\elementarycharge}{\text{\ensuremath{e}}}
\DeclareMathOperator{\tr}{tr}
\usepackage{nicefrac}
\usepackage{microtype}
\usepackage{lipsum}
\usepackage{graphicx}
\usepackage[]{placeins}
\usepackage{tabularx}
\usepackage[numbers]{natbib}

\newcommand{\acite}[1]{\citeauthor{#1}\,\cite{#1}}

\usepackage{soul}

\usepackage{float}

\usepackage{tikz}
\usetikzlibrary{external}
	\tikzsetexternalprefix{pdfs/}

\usepackage{xcolor}
\usepackage[
	margin=10pt,
	format=plain,
	font=small,
	labelfont=bf,
	labelsep=colon,
	skip=0pt,
]{caption}
\usepackage[super]{nth}
\usepackage{enumitem}
\usepackage{extarrows}

\usepackage{ifthen}

\usepackage{mathtools}

\newcommand{\ZeroTensor}{\mathbf{0}}

\newcommand{\CrackLengthLEFM}{a}
\newcommand{\CrackDimension}{a}

\newcommand{\SampleWidthLEFM}{b}

\newcommand{\SampleLengthLEFM}{c}

\newcommand{\Stiffness}{C}

\newcommand{\NonAffineDisp}{D}
\newcommand{\DissipEn}{d}
\newcommand{\LoadLengthLEFM}{d}

\newcommand{\InfinitesInt}{\,\mathrm{d}}
\newcommand{\RateDefTens}{\mathbf{d}}

\newcommand{\Divergence}{\mathrm{div}}

\newcommand{\GreeLagStra}{E}
\newcommand{\GreeLagStraTen}{\mathbf{E}}

\newcommand{\YoungsModulus}{E}

\newcommand{\KineticEnergy}{E^{\mathrm{k}}}

\newcommand{\TotalEnergy}{E^{\mathrm{t}}}

\newcommand{\ForceVector}{\mathbf{f}}

\newcommand{\GeometryFactorSih}{F}

\newcommand{\DefGradTensor}{\mathbf{F}}

\newcommand{\RDF}{g}

\newcommand{\SampleThicknessLEFM}{h}

\newcommand{\AtomOne}{i}
\newcommand{\AtomTwo}{j}

\newcommand{\Jac}{J}

\newcommand{\StressIntensFactor}{K}

\newcommand{\StressIntensFactorSih}{k}
\newcommand{\AnchorSpringStiffness}{k^{\mathrm{AP}}}

\newcommand{\BoltzmannConst}{k_{\mathrm{B}}}

\newcommand{\LengthCurrent}{l}

\newcommand{\LengthInitial}{L}

\newcommand{\Mass}{m}

\newcommand{\Moment}{M}

\newcommand{\NormalVectorCurrent}{\mathbf{n}}

\newcommand{\NumberAtoms}{n_{\mathrm{A}}}

\newcommand{\NumberBonds}{n_{\mathrm{B}}}

\newcommand{\NumberGaussPoints}{n_{\mathrm{gp}}}
\newcommand{\IterationsPerLoadStep}{\Delta n_{\mathrm{FEMD}}}
\newcommand{\MDTimeStepsPerIteration}{\Delta n_{\mathrm{MD}}}

\newcommand{\PressureTensor}{\mathbf{p}}
\newcommand{\DipoleMoment}{p}
\newcommand{\DipoleMomentVector}{\mathbf{p}}

\newcommand{\ForceSih}{P}

\newcommand{\Charge}{q}
\newcommand{\InterparticleDistance}{r}

\newcommand{\PositionVectorCur}{\mathbf{r}}

\newcommand{\RadiusPlasticZone}{r_{\mathrm{p}}}

\newcommand{\RadiusKDetField}{R}

\newcommand{\LongitShearStressLEFM}{s}

\newcommand{\StrainEnergyDensityFactor}{S}

\newcommand{\InnerSpan}{S_{\mathrm{i}}}
\newcommand{\OuterSpan}{S_{\mathrm{o}}}
\newcommand{\SecPioKirStress}{S}
\newcommand{\SecPioKirStressTensor}{\mathbf{S}}
\newcommand{\Time}{t}
\newcommand{\Traction}{t}
\newcommand{\TractionCurrent}{t}
\newcommand{\TractionVector}{\mathbf{t}}
\newcommand{\TractionVectorCurrent}{\mathbf{t}}
\newcommand{\PeriodDuration}{\mathfrak{T}}
\newcommand{\Temperature}{T}

\newcommand{\Displacement}{u}

\newcommand{\DisplacementVectorCurrent}{\mathbf{u}}

\newcommand{\VelocityCurrent}{v}

\newcommand{\VelocityVectorCurrent}{\mathbf{v}}

\newcommand{\VelocityVector}{\mathbf{v}}
\newcommand{\Volume}{V}

\newcommand{\CrackSampleRatio}{\alpha}

\newcommand{\WeightingFE}{\alpha}

\newcommand{\ModWeightingMD}{\bar{\beta}}

\newcommand{\CrackOpeningDisp}{\delta}

\newcommand{\SmallStrain}{\varepsilon}

\newcommand{\VolStrain}{\GreeLagStra^{\mathrm{hyd}}}
\newcommand{\ShearStrain}{\GreeLagStra^{\mathrm{vM}}}

\newcommand{\EngineeringShearStrain}{\gamma}

\newcommand{\ShearModulus}{\mu}

\newcommand{\PoissonsRatio}{\nu}

\newcommand{\MassDensity}{\rho}

\newcommand{\CauchyStress}{\sigma}

\newcommand{\VolStress}{\CauchyStress^{\mathrm{hyd}}}

\newcommand{\Angle}{\theta}

\newcommand{\SmallStrainTensor}{\boldsymbol{\varepsilon}}

\newcommand{\DPDFricCoeff}{\gamma}

\newcommand{\BodyBoundary}{\partial\Omega}

\newcommand{\CauchyStressTensor}{\boldsymbol{\sigma}}

\newcommand{\InPlaneShearStressLEFM}{\tau}
\newcommand{\KirchhoffStressTensor}{\boldsymbol{\tau}}
\newcommand{\TemperatureRate}{\dot{\Temperature}}

\newcommand{\YieldStress}{\sigma^{\mathrm{y}}}

\newcommand{\Eqref}[1]{Eq.~\eqref{#1}}
\newcommand{\Figref}[1]{Figure~\ref{#1}}

\newcommand{\Tabref}[1]{Table~\ref{#1}}

\newdimen\imageheight

\newcommand{\includetikz}[2]{
	\begin{minipage}{#2\textwidth}
    \centering
	\includegraphics[]{#1.pdf}
	\end{minipage}
}

\def\dotbar{\mathpalette\fooaux}
\def\fooaux#1#2{
	\mkern1mu
	\setbox0=\hbox{\mathsurround=0pt$#1\bar{\mkern-2mu #2 \mkern1mu}$}
	\setbox1=\hbox to \wd0{\hss$#1\cdot$\hss}
	\vbox{\offinterlineskip\copy1\vskip-.3\ht1\box0}
	\mkern-2mu
}

\title{The Capriccio method as a versatile tool for quantifying the fracture properties of glassy materials under complex loading conditions with chemical specificity}
\date{}

\newboolean{showauthors}
\setboolean{showauthors}{true}

\ifthenelse{\boolean{showauthors}}{

\author{ 
\href{https://orcid.org/0000-0002-5151-8075}{\includegraphics[scale=0.06]{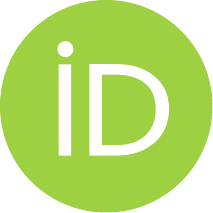}\hspace{1mm}Felix Weber} \\
Friedrich-Alexander-Universit\"at Erlangen-N\"urnberg (FAU)\\
Institute of Applied Mechanics \&\\
Competence Unit for Scientific Computing (CSC)\\
Egerlandstrasse 5, 91058 Erlangen\\
Germany \\
\texttt{felix.w.weber@fau.de} \\
\And
\href{https://orcid.org/0000-0003-2975-0625}{\includegraphics[scale=0.06]{orcid.pdf}\hspace{1mm}Maxime Vassaux} \\
 Université de Rennes, CNRS\\
  Institut de Physique de Rennes (IPR) - UMR 6251\\
  F-35000 Rennes\\
  France\\
\texttt{maxime.vassaux@univ-rennes.fr} \\
\And
\href{https://orcid.org/0009-0009-4524-2621}{\includegraphics[scale=0.06]{orcid.pdf}\hspace{1mm}Lukas Laubert} \\
Friedrich-Alexander-Universit\"at Erlangen-N\"urnberg (FAU)\\
Institute of Applied Mechanics\\
Egerlandstrasse 5, 91058 Erlangen\\
Germany \\
\texttt{lukas.laubert@fau.de} \\
\And
\href{https://orcid.org/0000-0001-8577-5048}{\includegraphics[scale=0.06]{orcid.pdf}\hspace{1mm}Sebastian Pfaller} \\
Friedrich-Alexander-Universit\"at Erlangen-N\"urnberg (FAU)\\
Institute of Applied Mechanics\\
Egerlandstrasse 5, 91058 Erlangen\\
Germany \\
\texttt{sebastian.pfaller@fau.de} \\
}

}{

}

\renewcommand{\shorttitle}{}

\begin{document}
\maketitle

\ifthenelse{\boolean{showauthors}}{
\pagebreak
}{
}

\begin{abstract}

Molecular dynamics (MD) simulations are widely used to provide insights into fracture mechanisms while maintaining chemical specificity. However, particle-based techniques such as MD are limited in terms of accessible length scales and applicable boundary conditions, which restricts the investigation of fracture phenomena in typical engineering settings. In an attempt to overcome these limitations, we apply the partitioned-domain Capriccio method to couple atomistic MD samples representing silica glass with the finite element (FE) method. With this approach, we perform mode I (rectangular panel under tension, three-, and four-point bending), mode II as well as mode III (rectangular panel under in-plane or out-of-plane shear) simulations. Thereby, we investigate multiple criteria to identify the onset of crack propagation based on the virial stress, the number of pair interactions, the kinetic energy/temperature, the crack velocity, and the crack opening displacement. The approach presented provides quantitatively plausible results for the critical stress intensity factors $\StressIntensFactor_{\mathrm{Ic}}$, $\StressIntensFactor_{\mathrm{IIc}}$, and $\StressIntensFactor_{\mathrm{IIIc}}$. This contribution shows that the Capriccio method is a flexible means of performing fracture simulations that take into account boundary conditions typical of experimental test setups with atomistic specificity near the crack tip. While also pointing out the current limitations of the Capriccio method, we demonstrate its potential to integrate atomistic insights into FE models with significantly larger overall dimensions.

\end{abstract}

\keywords{
Fracture mechanics 
\and Multiscale modeling
\and Molecular dynamics
\and Finite element method
\and Glassy materials
}

\renewcommand*\contentsname{Contents (just for draft)}
\setcounter{tocdepth}{4}
\setcounter{secnumdepth}{3}

\clearpage\newpage

\section*{Nomenclature}

\subsection*{Abbreviations}

\begin{tabularx}{\textwidth}{p{1.5cm} l}
3PB & Three-point bending \\
4PB & Four-point bending \\
DPD & Dissipative particle dynamics \\
FE & Finite element \\
FEM & Finite element method \\
LEFM & Linear elastic fracture mechanics \\
MD & Molecular dynamics \\  
O & Oxygen \\
Si & Silicon \\
\end{tabularx}

\subsection*{Variables}

\begin{tabularx}{\textwidth}{p{1.5cm} l}
$\CrackLengthLEFM$ & Crack length \\
$\SampleWidthLEFM$ & Sample width \\
$\SampleLengthLEFM$ & Half sample length \\
$\Stiffness_{11}$ & Longitudinal modulus\\
$\YoungsModulus$ & Young's modulus \\
$\YoungsModulus'$ & Plane strain stiffness \\
$\KineticEnergy$ & Kinetic energy \\
$\GreeLagStraTen$ & Green-Lagrange strain tensor \\ 
$\ForceVector$ & Force vector \\
$\DefGradTensor$ & Deformation gradient tensor \\
$\SampleThicknessLEFM$ & Sample thickness \\
$\BoltzmannConst$ & Boltzmann constant \\
$\StressIntensFactor$ & Stress intensity factor \\
$\LengthCurrent$ & Current length \\
$\LengthInitial$ & Initial length \\
$\Moment$ & Moment \\
$\NumberAtoms$ & Total number of atoms \\
$\NumberBonds$ & Total number of bonds \\
$\Charge$ & Point charge \\
$\ForceSih$ & Boundary force \\
$\PositionVectorCur$ & Position vector \\
$\StrainEnergyDensityFactor$ & Strain energy density factor \\
$\InnerSpan$ & Inner span \\
$\OuterSpan$ & Outer span \\
$\Time$ & Time \\
$\TractionVector$ & Surface traction vector \\
$\Temperature$ & Temperature \\
$\DisplacementVectorCurrent$ & Displacement vector \\ 
$\VelocityVector$ & Velocity vector \\
$\Volume$ & Volume \\[10pt]

$\CrackSampleRatio$ & Ratio of crack length to sample width\\
$\CrackOpeningDisp$ & Crack opening displacement \\
$\SmallStrainTensor$ & Engineering strain tensor \\
$\PoissonsRatio$ & Poisson's ratio \\
$\MassDensity$ & Mass density \\
$\CauchyStressTensor$ & Cauchy stress tensor\\
\end{tabularx}

\clearpage\newpage

\section{Introduction}
\label{sec:Introduction}

All-atom molecular dynamics (MD) simulations are widely used in materials science. They provide essential insights into material properties while maintaining chemical specificity. Various mechanical properties can be readily investigated using MD, which includes the modeling of fracture.
A common approach is to prescribe the analytical displacement field resulting from linear elastic fracture mechanics (LEFM) at the boundary of pre-notched cylindrical samples, allowing for relaxation in the interior of the system, see e.g. \cite{Andric2018,Jones2018,Rimsza2019,Lakshmipathy2022}.  This so-called $\StressIntensFactor$-test procedure allows for  predictions of the classical fracture mechanical quantities such as the critical $J$-integral or the critical stress intensity factor
that agree well with experimental values.
However, if arbitrary boundary conditions arising from engineering applications are to be imposed, the crack tip fields to be used in the $\StressIntensFactor$-test simulations can become complex \cite{Andric2018}.
Furthermore, the LEFM solution must be adapted as the crack propagates, which requires sophisticated updating schemes \cite{Wilson2019,Andric2018}. 
In general, the boundary conditions applicable in pure MD setups limit the fracture mechanical loading conditions that can be considered. In fact, most studies using MD simulations to model fracture deal with mode I or mode II fracture or a combination of both \cite{Tabarraei2015,Wei2018,Stepanova2018,Lu2005,Zhou2008}. However, there is only a limited number of investigations under mode III conditions available in the literature, with demanding force-based boundary conditions subjected to subgroups of atoms \cite{Shimizu2007,Uhnakova2011}. 
Although mode I is commonly regarded as the most relevant fracture mode \cite{Varadarajan2009}, in many applications the crack propagates in a direction characterized by a combination of different modes. Consequently, other fracture modes and combinations thereof must also be taken into account \cite{Song2016}. 
Moreover, the simulation of macroscopic engineering settings is unfeasible due to the tremendous amount of degrees of freedom to be considered \cite{Frenkel2001}.

Multiscale approaches coupling continuum-based and particle-based techniques can provide an effective remedy for the above-mentioned challenges.
The introduction of  macroscopic boundary conditions makes it possible to reproduce  experimental test setups while capturing the microscopic motion of atoms in a single simulation. Among the variety of multiscale methods (see the classification proposed by \acite{Tadmor2011}), concurrent coupling techniques with partitioned domains are particularly relevant to achieving these objectives. Here, the boundary conditions can be  specified to a surrounding macroscale model, which facilitates the consideration of a variety of failure modes without a requirement to prescribe the LEFM displacement fields. Concurrent partitioned-domain coupling techniques include, but are not limited to, the Arlequin method \cite{BenDhia1998,Bauman2008}, the Finite Element-Atomistic (FEAt) coupling scheme \cite{Gumbsch1995}, the Bridging Domain method \cite{Belytschko2003,Xiao2004,Guidault2007,Xu2008,Xu2010} and the quasi-continuum method \cite{Tadmor1996,Shenoy1999,Eidel2009a}. Nevertheless, most of these methods were limited to highly simplified MD regions (same resolution for MD and FE mesh), crystalline materials, or
investigations at zero temperature.

In this work, we perform mode I, mode II, and mode III fracture simulations using the Capriccio method 
\cite{Pfaller2013,Pfaller2019,Pfaller2021,Zhao2021a,Jain2022,Weber2023,Laubert2024,Laubert2024a}, which employs elements of the Arlequin method and couples a particle-based model to a continuum model. The former is solved using MD, while the latter is discretized and solved using the finite element method (FEM). The Capriccio method was specifically designed for coupled FE-MD simulations of amorphous materials. We take amorphous silicon dioxide (silica) prepared at various quenching rates as a model material. Indeed,  the fracture of glasses is governed by processes at the atomic level \cite{Wu2025}, in contrast to, e.g., polymers, which develop craze regions that span hundreds of nanometers \cite{Kausch1978}. However, the general applicability of the Capriccio method for modeling fracture-related phenomena of thermoplastic polymers in the immediate vicinity of the crack was recently demonstrated by \citeauthor{Zhao2024a}~\cite{Zhao2023,Zhao2024a}. 
The objective of the present study is to investigate the capabilities of the Capriccio method as a tool for accurately determining the critical stress intensity factor under test conditions that cannot be achieved through pure MD simulations. We  apply mode I fracture to (i)~a single edge crack in a rectangular panel and in (ii)~three-point bending (3PB) and (iii)~four-point bending (4PB) test setups. Afterwards, we perform iv)~mode II and v)~mode III simulations with a single edge crack in a rectangular panel under in-plane shear and out-of-plane shear, respectively. To precisely determine the critical stress intensity factor, we investigate several criteria for identifying the onset of crack propagation
based on the virial stress, the number of pair interactions, the kinetic energy/temperature, the crack velocity, and the crack opening displacement. Therefore, we study 
not only the influence of the 
test conditions, but also of
the protocol used to determine the crack propagation. In this way, we also aim to increase the objectivity of the computational prediction of the critical stress intensity factor.
This contribution is structured as follows: First, we present the methodology used and the identification of parameters essential for the Capriccio systems (Section~\ref{sec:Methods}). We then provide the results of the fracture simulations (Section~\ref{sec:Results}), followed by a discussion of the latter (Section~\ref{sec:Discussion}). 
Finally, the key conclusions 
are summarized (Section~\ref{sec:Conclusions}).

\section{Methods}
\label{sec:Methods}

In this section, we outline the molecular model used (interatomic potential) and the preparation of the atomistic samples (Section~\ref{ssec:Molecular dynamics}), the coupling to an FE domain using the Capriccio  method (Section~\ref{ssec:Capriccio method}), and a general introduction to the setups of the mode I, mode II,
and mode III simulations, together with the respective calculation of the stress intensity factors (Section~\ref{ssec:Fracture setups and associated stress intensity factors}). Finally, Section~\ref{ssec:Definition of the quantities of interest} presents the quantities that are evaluated to identify the start of crack propagation.
Further details on the MD interaction potentials  (Supplementary Section~S1), the
material characterization
based on pure MD simulations (Supplementary Section~S2), and the setup of the Capriccio coupling (Supplementary Section~S3) can be found in the supplementary material.

\subsection{Molecular dynamics}
\label{ssec:Molecular dynamics}

\subsubsection{Atomistic interactions}
\label{sssec:Atomistic interactions}

In this study, we apply an atomistic MD representation of silica glass introduced by \citeauthor{Sundararaman2018}~\cite{Sundararaman2018,Sundararaman2019}, the so-called SHIK potential, which describes the interatomic interactions through a summation of a Coulomb and a Buckingham term. This model uses the Wolf truncation method \cite{Wolf1999} to calculate the long-range Coulomb interactions, whose accuracy under non-periodic boundary conditions has been demonstrated by \acite{Gdoutos2010}. Additionally, a short-range repulsive term is applied to prevent overlapping atoms. In \cite{Zhang2020}, \citeauthor{Zhang2020} find good agreement of the linear-elastic material parameters obtained by using the SHIK potential with experimental values. They also compare the results to those derived with other MD potentials that use the same functional formulation but different parametrizations, i.e., the BKS \cite{Beest1990}, GS \cite{Guillot2007}, Teter \cite{Cormack2002} and HO \cite{Habasaki1992} potentials, demonstrating that the SHIK potential is well suited for mechanical studies.  The analytical functions and the specific parameter values of the SHIK potential are given in Supplementary Section~S1.

\subsubsection{Specimen preparation}
\label{sssec:Specimen preparation}

In this study, we examine each load case by averaging over five independent replicas, indicating the standard deviation by  shaded areas around the resulting curves. Each of these samples is first generated by a random, stoichiometric packing of atoms in a simulation box with a mass density of $\MassDensity = \SI{2.22}{\gram\per\cubic\centi\metre}$, ensuring a minimum interatomic distance of \SI{1.7}{\angstrom}. The synthesis under periodic boundary conditions  is performed according to \cite{Zhang2020}, using an MD time step size $\Delta\Time_{\mathrm{MD}}$ of \SI{1.6}{\femto\second}. First, the atomic velocities are initialized to achieve a temperature of the melt of $\Temperature = \SI{3600}{\kelvin}$, prescribing a uniform distribution and a zero linear momentum of the velocities. After an initial equilibration phase of \SI{1.6}{\pico\second} in the microcanonical ensemble, in which the atoms are allowed to move a maximum of \SI{0.1}{\angstrom} per time step, \SI{320}{\pico\second} in the canonical ensemble and a further \SI{320}{\pico\second} in the isothermal-isobaric ensemble at zero external pressure, the systems are quenched to the target temperature of $\Temperature = \SI{300}{\kelvin}$ with various quenching rates in the range from $\TemperatureRate = \SI{0.26}{\kelvin\per\pico\second}$ to \SI{3.2e4}{\kelvin\per\pico\second}. 
We then equilibrate the systems for an additional time of \SI{320}{\pico\second} at zero external pressure. In all MD simulations under periodic boundary conditions, we apply a Nos\'{e}-Hoover thermostat \cite{Evans1985} and, in the simulations in which pressure is also controlled, a Nos\'{e}-Hoover barostat \cite{Martyna1994}. The resulting mass densities $\MassDensity$ increase with the quenching rate $\TemperatureRate$, which is reasonable in the range of $\TemperatureRate$ investigated here \cite{Vollmayr1996}, varying from $\MassDensity = \SI{2.224(0.009)}{\gram\per\cubic\centi\metre}$ for $\TemperatureRate = \SI{0.26}{\kelvin\per\pico\second}$ to $\MassDensity = \SI{2.277(0.004)}{\gram\per\cubic\centi\metre}$ for $\TemperatureRate = \SI{3.2e4}{\kelvin\per\pico\second}$, cf.\ Supplementary Section~S2.

\subsection{The Capriccio method}
\label{ssec:Capriccio method}

\subsubsection{Coupled particle-continuum systems}
\label{sssec:Setup of the coupled particle-continuum systems}

To apply non-affine deformations to the MD systems, as required in the fracture mechanical test setups to be investigated here, we use the Capriccio framework, see \Figref{fig:Silica_FEMD_SSP_modeI_Setup_SHIK}. The Capriccio method, first published in \cite{Pfaller2013} and further developed and investigated in subsequent contributions \cite{Pfaller2019,Pfaller2021,Zhao2021a,Jain2022,Weber2023,Laubert2024,Laubert2024a}, is a  technique that concurrently couples a continuum discretized by FEM and a particle-based inclusion controlled by MD. In the following, we present the details of the system setups used in this paper together with some insights into the fundamentals of the method. For a comprehensive overview of the foundations of the Capriccio method, the reader is referred to \cite{Pfaller2021}. In the inner part of the particle domain, the MD time integration is based on classical Newtonian particle dynamics (ND). For the coupling of the MD domain to the continuum, stochastic boundary conditions \cite{Rahimi2011} are applied. Information about forces and displacements is transferred between the two domains using virtual evaluation points, called anchor points. These anchor points are connected to a subset of atoms within the bridging domain by harmonic springs of equilibrium length zero, which act as a potential that penalizes a spatial mismatch between the FE and MD domains in the bridging domain. 

The Capriccio method uses a staggered solution scheme, conducting alternating FE and MD calculations: In (i)~each FE calculation, the MD domain is kept spatially fixed and the updated nodal displacements, anchor point positions, and Lagrange multipliers used to solve the underlying optimization problem are determined. In (ii)~each MD calculation,
the FE domain remains spatially fixed. Here, in addition to the new atomic positions, the forces acting on the anchor points via the harmonic springs are calculated, with their time-averaged values serving as additional Neumann boundary conditions in the next FE calculation.
In the linear elastic continuum domain, we solve the boundary value problem \cite{Holzapfel2000}
\begin{equation}
\begin{aligned}
    \label{eq:BVP}
    \Divergence \CauchyStressTensor &= \ZeroTensor \quad \mathrm{with} & & \\
    \DisplacementVectorCurrent & = \bar{\DisplacementVectorCurrent}   & \mathrm{on}   \quad & \BodyBoundary_{u} \quad \mathrm{and} \\
    \TractionVectorCurrent & = \boldsymbol{\sigma} \cdot \NormalVectorCurrent = \bar{\TractionVectorCurrent}    & \mathrm{on} \quad  & \BodyBoundary_{\sigma}
\end{aligned}
\end{equation}
with the (Cauchy) stress tensor $\CauchyStressTensor$, the displacement field $\DisplacementVectorCurrent$ with $\bar{\DisplacementVectorCurrent}$ prescribed on the Dirichlet boundary $\BodyBoundary_{u}$, the surface traction $\TractionVectorCurrent$ with $\bar{\TractionVectorCurrent}$ prescribed on the Neumann boundary $\BodyBoundary_{\sigma}$, and the normal vector $\NormalVectorCurrent$. We apply Neumann boundary conditions by prescribing the surface tractions as so-called dead loads, which are defined as deformation-independent, i.e., $\bar{\TractionCurrent}_{ti} = \bar{\Traction}_{0i}$ ($i = x,y,z$).
The surface tractions are employed  incrementally in load steps of size $\Delta\bar{\TractionCurrent}_{i}$. 

The load step size $\Delta\bar{\TractionCurrent}_{i}$ and the stiffness  of the  anchor point springs $\AnchorSpringStiffness$ must be carefully calibrated,  as they influence the deformation of the FE-MD systems and thus the fracture properties, see Supplementary Section~S4.4. 
\citeauthor{Laubert2024}~\cite{Laubert2024,Laubert2024a} found
that the staggered FE-MD solution procedure used by the Capriccio simulations introduces a resistance of the bridging domain with regard to motion in space, where \citeauthor{Laubert2024}
termed this effect ``motion resistance''.
Further details on this phenomenon and the calibration procedure performed in the present study are given in Supplementary Section~S3.3. In particular, we determine first estimates for the load step size $\Delta\bar{\TractionCurrent}_{i}$ and the anchor point spring stiffness $\AnchorSpringStiffness$ based on the elastic system response in uniaxial tensile tests on setups in which the  Capriccio coupling is only applied in the loading direction.
In the present study, we use  $\AnchorSpringStiffness = \SI{1.0}{\eV\per\square\angstrom}$ and  $\Delta\bar{\TractionCurrent}_{i} = \SI{1.6}{\mega\pascal}$.   

Different loading rates $\dotbar{\TractionCurrent}_{i}$ and load step sizes $\Delta\bar{\TractionCurrent}_{i}$ are applied  by scaling the number of MD time steps per load step $\MDTimeStepsPerIteration$ via 
\nopagebreak\begin{align}
\label{eq:LoadingRateFExx}
\dot{\bar{\TractionCurrent}}_{i} = \frac{\Delta\bar{\TractionCurrent}_{i}}{\IterationsPerLoadStep\, \MDTimeStepsPerIteration\,\Delta\Time_{\mathrm{MD}}}.
\end{align}
Here, we employ a number of FE-MD iterations per load step $\IterationsPerLoadStep$ of  1, an MD time step size $\Delta\Time_{\mathrm{MD}}$ of \SI{1.6}{\femto\second}, and, unless otherwise stated,  $\MDTimeStepsPerIteration = 50$.  
To control the temperature in the particle domain, dissipative particle dynamics (DPD) \cite{Hoogerbrugge1992,Groot1997} is applied in addition to the SHIK interactions at the particle domain's boundary.
We use a DPD friction coefficient of $\DPDFricCoeff = \SI{0.1}{\eV \pico\second\per\square\angstrom}$ and prescribe a temperature of \SI{300}{\kelvin}.
Further details on the parametrization of the Capriccio systems can be found in Supplementary Section~S3.1.

\subsubsection{Constitutive model}
\label{sssec:Constitutive law}

We use Hooke's law to model the constitutive behavior of the FE domain, i.e.
\nopagebreak\begin{align}
	\label{eq:LinElast3d}
	\CauchyStress_{ij} = \dfrac{\YoungsModulus}{1 + \PoissonsRatio} \left[\varepsilon_{ij} + \dfrac{\PoissonsRatio}{1 - 2 \PoissonsRatio} \varepsilon_{kk} \delta_{ij}\right],
\end{align}
see e.g. \cite{Ponson2023}. Here,  $\CauchyStress_{ij}$ and $\SmallStrain_{ij}$ denote  the components of the stress and strain tensors $\CauchyStressTensor$ and $\SmallStrainTensor$ and  $\delta_{ij}$ represents the Kronecker delta. This simple material description is sufficient to model the system behavior up to failure in the present simulations, cf.\ Supplementary Section~S2.  Young's modulus  $\YoungsModulus$ and Poisson's ratio $\PoissonsRatio$ are determined from pure MD systems subjected to uniaxial tension in Supplementary Section~S2. 
\subsection{Fracture setups and associated stress intensity factors}
\label{ssec:Fracture setups and associated stress intensity factors}

\subsubsection{General remarks on the studied samples}
\label{sssec:General remarks on the studied samples}

\begin{figure}[ht!]
	\centering
	\includetikz{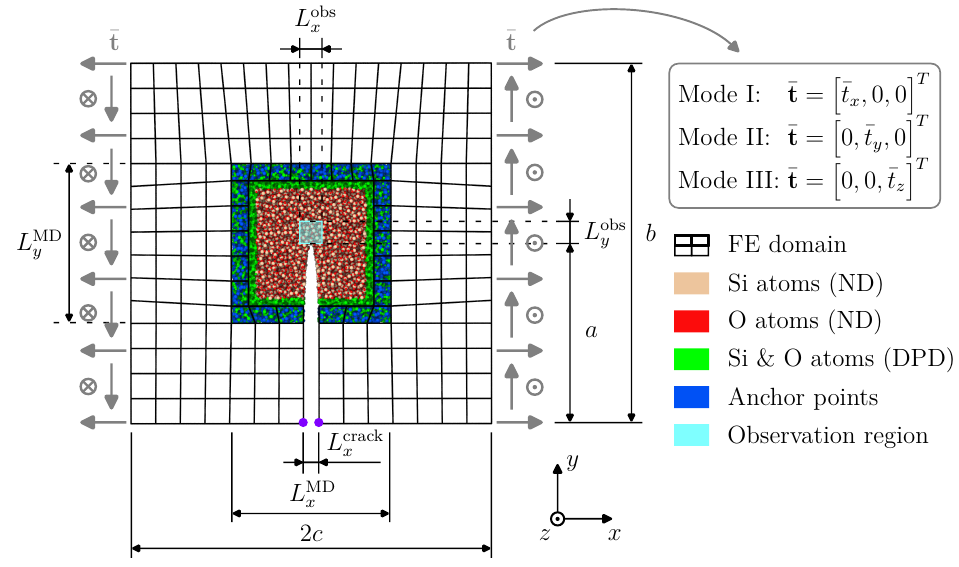}{1.0}
	\caption[General setup of the fracture simulations.]{General setup of the fracture simulations: 
	The samples  are pre-notched with notch length $\CrackLengthLEFM$ as well as notch width $\LengthInitial_{x}^{\mathrm{crack}}$. While loading the specimens with surface tractions $\bar{\TractionVectorCurrent}$, atomistic quantities are evaluated in an observation region of dimensions $\LengthInitial_{x}^{\mathrm{obs}} = \LengthInitial_{y}^{\mathrm{obs}}$ in front of the crack tip, whereas the crack opening displacement $\CrackOpeningDisp$ is measured as the change in the distance between the two purple points. The entire samples measure $\SampleWidthLEFM$ in width, $2\SampleLengthLEFM$ in length and $\SampleThicknessLEFM$ in thickness.} 
	\label{fig:Silica_FEMD_SSP_modeI_Setup_SHIK}
\end{figure}

The general setup of the fracture simulations performed here, which consists of an MD region  coupled to an FE region in the $x$- and $y$-direction, is specified in \Figref{fig:Silica_FEMD_SSP_modeI_Setup_SHIK}. For the mode I and mode II tests (prescribed surface traction component: $\bar{\TractionCurrent}_{x}$ or $\bar{\TractionCurrent}_{y}$), we apply the plane strain condition, as classically assumed in linear elastic fracture mechanics (LEFM). In particular, the  $z$-component of the  displacement of surfaces with normals in the $z$-direction (denoted as $z$-surfaces) is set to zero. 
In this direction, the MD domain is periodic. 
Whenever lateral strains are prevented in the present study, the corresponding displacements of the lateral surfaces of the MD simulation box and the FE domain are set to zero.
For the mode III tests (prescribed surface traction: $\bar{\TractionCurrent}_{z}$), the boundaries with normals in $z$-direction (denoted as $z$-boundaries) are non-periodic, free surfaces. First, the MD samples   are created and prepared according to the procedure described in Section~\ref{ssec:Molecular dynamics}. 
\acite{Rountree2020} derived a size of the fracture process zone in silica glass of about \SI{100}{\angstrom}, which we take into account by choosing the MD inclusion size accordingly: 
For most of the studies conducted here, we choose the dimensions of the MD systems to $\LengthInitial^{\mathrm{MD}}_{x} \times \LengthInitial^{\mathrm{MD}}_{y} \times \LengthInitial^{\mathrm{MD}}_{z} \approx \SI{100}{\angstrom} \times \SI{100}{\angstrom} \times \SI{50}{\angstrom}$, consisting of \num{11131} Si as well as \num{22262} O atoms before  the notch is introduced. 
The influence of the specimen size is examined in Supplementary Section~S4.5.  
Whenever larger dimensions of the particle domain are used, the numbers of atoms are scaled correspondingly. After synthesizing the MD samples, stochastic boundary conditions \cite{Rahimi2011}, i.e., the DPD terms and the anchor points, are introduced in the directions to be coupled to the continuum.

Upon insertion of notches by deleting atoms in an semi-elliptical region with a width of $\LengthInitial_{x}^{\mathrm{crack}} = \SI{10}{\angstrom}$ and, if not otherwise specified, a length of $\CrackLengthLEFM= \SampleWidthLEFM / 2 = \SI{115}{\angstrom}$, the MD domains of the system size mentioned above comprise approximately \num{15500} atoms in the inner MD domain, another \num{16500} atoms in the DPD region, and \num{5800} anchor points. Before applying surface tractions at the Neumann boundary to pull the FE-MD systems, we equilibrate them at zero load, i.e.,  $\bar{\TractionCurrent}_{i} = 0$, for \SI{160}{\pico\second} (2000 load steps), see Supplementary Section~S4.1.
In the present study, we neglect the influence of the FE-MD equilibration on the notch, with a change of the notch length due to the equilibration up to \SI{1.5(1.4)}{\percent}.  The initial $x$- and $y$-dimensions of the MD systems are set equal, i.e., $\LengthInitial^{\mathrm{MD}}_{x} = \LengthInitial^{\mathrm{MD}}_{y}$, while the ratio between the initial total system dimensions $2\SampleLengthLEFM$ and $\SampleWidthLEFM$ in $x$- and $y$-direction, respectively, is varied in some tests. In all studies presented here, $\LengthInitial^{\mathrm{MD}}_{z}$ matches the thickness of the total system in $z$-direction $\SampleThicknessLEFM$. The atomistic properties are evaluated in a cuboid region in front of the crack tip with variable dimensions $\LengthInitial^{\mathrm{obs}}_{x}$ and $\LengthInitial^{\mathrm{obs}}_{y}$ and a fixed dimension $\LengthInitial^{\mathrm{obs}}_{z} = \LengthInitial^{\mathrm{MD}}_{z}$, the so-called observation region.

Unless otherwise stated, the analyzed MD samples are prepared with a quenching rate of $\dot{\Temperature} = \SI{0.26}{\kelvin\per\pico\second}$. \acite{Zhang2020} state that this quenching rate is comparably small, i.e., MD simulations using the SHIK potential for samples that were cooled at lower $\dot{\Temperature}$ yield similar mechanical properties.
As far as the rate of applied surface traction $\dotbar{\TractionCurrent}_{i}$ is concerned, we use a value of \SI{20}{\giga\pascal\per\nano\second} in most  studies. The latter is small enough for the DPD thermostat to ensure appropriate control of the temperature in the MD region, as demonstrated in
	Supplementary Figure~S9b).
	Additionally, according to \acite{Zhang2020}, a strain rate of \SI{50}{\percent\per\nano\second} is small enough for the mechanical properties to be almost independent of the loading rate. A loading rate of $\dotbar{\TractionCurrent}_{i} = \SI{20}{\giga\pascal\per\nano\second}$ is thus in a medium range: 
Based on the plane strain stiffness  \cite{Gross2018}
\nopagebreak\begin{align}
\label{eq:EPlane}
\YoungsModulus' = \dfrac{\YoungsModulus}{1-\PoissonsRatio^{2}}
\end{align}
of the unnotched systems, which varies with the quenching rate $\TemperatureRate$, cf.\ 
Section~\ref{ssec:Molecular dynamics}, a surface traction rate of   $\dotbar{\TractionCurrent}_{i} = \SI{20}{\giga\pascal\per\nano\second}$ corresponds to a range of strain rates of about \SI{26}{\percent\per\nano\second} to  \SI{29}{\percent\per\nano\second}. For the atomistic specimen sizes investigated (not larger than \SI{200}{\angstrom}), this results in loading velocities of up to about \SI{6}{\metre\per\second}.

\subsubsection{Setups and stress intensity factors}
\label{sssec:Stress intensity factors}

The calculations of stress intensity factors  $\StressIntensFactor_{i}\,(i=\mathrm{I,II,III})$  carried out here are mainly based on \acite{Sih1973}, where we also provide the original references. The only exception is the rectangular panel subjected to far-field stresses leading to mode II conditions, which is not considered in \cite{Sih1973}.
With the considerations presented in this contribution, we also test the general applicability of the formulas for the critical stress intensity factors for samples of finite size on an atomic length scale.

\paragraph{Strip under tension.}

The first mode I setup considered is a  single edge cracked plate tension specimen (SECT), which we also refer to as ``strip'', see \Figref{fig:Silica_FEMD_SSP_modeI_Setup_SHIK}. Based on \cite{Bowie1965,Bowie1973}, \acite{Sih1973} states the stress intensity factor in this case (denoted as ``single edge crack in a rectangular panel'') as 
\nopagebreak\begin{align}
	\label{eq:KIStrip}   
	\StressIntensFactor_{\mathrm{I}} = \dfrac{\ForceSih}{\SampleWidthLEFM} \sqrt{ \pi\CrackLengthLEFM} \GeometryFactorSih 
\end{align}
with the force per unit thickness $\ForceSih = \bar{\TractionCurrent}_{x} \LoadLengthLEFM$ and 
the geometry factor $\GeometryFactorSih=\GeometryFactorSih(\SampleWidthLEFM/\SampleLengthLEFM, \CrackSampleRatio, \LoadLengthLEFM/\SampleWidthLEFM )$. Here,~$\LoadLengthLEFM$ denotes the length of the boundary on which the load is applied, where $\LoadLengthLEFM = \SampleWidthLEFM$ in the present test setups. For instance, with $\SampleWidthLEFM/\SampleLengthLEFM = 2.0$, $\CrackSampleRatio = \CrackLengthLEFM/\SampleWidthLEFM = 0.5$ and $\LoadLengthLEFM/\SampleWidthLEFM = 1.0$, the tabulated geometry factor is $\GeometryFactorSih = 3.01$.
For other sample or crack dimensions applied in this contribution that are 
not tabulated in \cite{Sih1973}, the corresponding geometry factors $\GeometryFactorSih$ are determined according to \cite{Tada2000}. For the geometries  tabulated in \cite{Sih1973}, similar values for $\GeometryFactorSih$ are obtained using the relationship from \cite{Tada2000}, see Supplementary Section~S4.10. Note that after crack propagation has started, the crack length would have to be adjusted when calculating the stress intensity factor, which we neglect here for the sake of simplicity.
For the particle domains of the dimensions in the $x$- and $y$-direction equal to \SI{100}{\angstrom}, the initial overall sample dimensions are $\SampleWidthLEFM = 2\SampleLengthLEFM  = \SI{230}{\angstrom}$. Here, the FE domain is discretized by \num{984} nodes.
We examine the influence of the total specimen dimensions in Supplementary Figure~S22.
For the above-mentioned sample dimensions, the standard surface traction rate of $\dotbar{\TractionCurrent}_{i} = \SI{20}{\giga\pascal\per\nano\second}$ selected here leads to a rate of the applied stress intensity factor of approximately $\dot{\StressIntensFactor}_{\mathrm{I}} = \SI{11.5}{\mega\pascal\sqrt{\metre}}\,\si{\per\nano\second}$.

We assume that at the atomic resolution, the setup of a  single edge crack in a rectangular panel adequately approximates the stress state near the crack tip of macroscopic samples. However, to assess whether subjecting MD samples to more complex boundary conditions also leads to quantitatively realistic results, we additionally perform 3PB and 4PB tests as mode I scenarios which are also used for macroscopic specimens.

\paragraph{Three- and four-point bending.}

\begin{figure}[ht!]
	\centering
	\includetikz{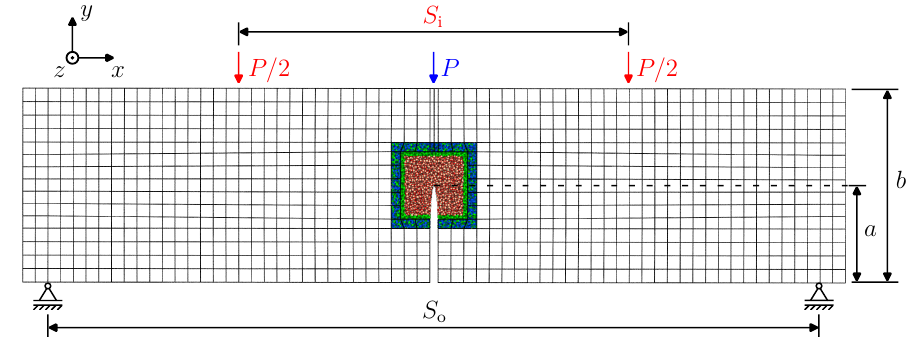}{1.0}
	\caption[Setup of the three-point and four-point  bending tests.]{Setup of the three-point (blue) and four-point (red) bending tests: Applied force $\ForceSih$ and  distances between the supports $\OuterSpan$ and the forces $\InnerSpan$.} 
	\label{fig:Silica_FEMD_SSP_3PB-4PB_Setup_SHIK}
\end{figure}

The general setup of the ``single-edge-cracked bend specimens'', i.e., the 3PB (blue) and 4PB tests (red), is illustrated in \Figref{fig:Silica_FEMD_SSP_3PB-4PB_Setup_SHIK}. The size of the MD domain, the notch length $\CrackLengthLEFM$, and the sample width $\SampleWidthLEFM$ are chosen to be identical to the values introduced 
above.

Based on the ASTM International standard ASTM STP 410 \cite{ASTMSTP410},
\acite{Sih1973}\footnote{Unlike experimental standards, \acite{Sih1973} does not state any validity restrictions on the formulas with regard to absolute specimen sizes or materials. A comparison of various references, including ASTM standards, is given in Supplementary Section~S4.10.} provides the stress intensity factor for 3PB and 4PB (i.e., pure bending) as
\nopagebreak\begin{align}
	\label{eq:KI_3/4PBSih1973}
	\StressIntensFactor_{\mathrm{I}} = \dfrac{6 \Moment_{z} \sqrt{\CrackLengthLEFM}}{\SampleWidthLEFM^{2} \SampleThicknessLEFM} \sum_{i=0}^{4} A_{i} \CrackSampleRatio^{i}
\end{align}
for $0 < \CrackSampleRatio = \CrackLengthLEFM/\SampleWidthLEFM \leq 0.6$. Initially, we choose the span between the supports to be $\OuterSpan = 4 \SampleWidthLEFM = \SI{920}{\angstrom}$. The corresponding FE meshes comprise  \num{4184} nodes for the 3PB and \num{4312} nodes for the 4PB tests. The maximum bending moment about the $z$-axis is $\Moment_{z} = \ForceSih \OuterSpan/4$ for the 3PB and $\Moment_{z} = \ForceSih [ \OuterSpan - \InnerSpan ]/4$ for the 4PB tests. In addition to the geometric specifications for the 3PB tests, the inner span $S_\mathrm{{i}} = 2 \SampleWidthLEFM$ applies to the 4PB tests.  The forces $\ForceSih$ are applied as surface tractions~$\bar{\TractionCurrent}_{y}$ over the entire sample thickness  $\SampleThicknessLEFM$, each distributed over a length of \SI{10}{\angstrom} in $x$-direction. To apply the same number of load increments for the 3PB and 4PB tests, we choose the increments in $\ForceSih$ for the 4PB tests to be twice as large as for the 3PB tests.   For the aspect ratio of outer span to sample width applied  in this contribution $\OuterSpan/\SampleWidthLEFM = 4$, the coefficients $A_{i}$ used in \Eqref{eq:KI_3/4PBSih1973} are given in \Tabref{tab:Coeff_3/4PB}. 

\begin{table}[ht!]
	\caption[Coefficients $A_{i}$ for calculating the stress intensity factor $\StressIntensFactor_{\mathrm{I}}$ for three-  and four-point bending.]{Coefficients $A_{i}$ for calculating the stress intensity factor $\StressIntensFactor_{\mathrm{I}}$ for three- (3PB) and four-point bending (4PB) for a ratio of outer span to sample width of $\OuterSpan/\SampleWidthLEFM = 4$ according to \cite{Sih1973}.}
	\label{tab:Coeff_3/4PB}
	\centering
	\begin{small}
		\begin{tabularx}{0.39\textwidth}{l r r r r r}
			\toprule
			& $A_{0}$ & $A_{1}$ & $A_{2}$ & $A_{3}$ & $A_{4}$ \\
			\midrule
			3PB & 1.93 & -3.07 & 14.53 & -25.11 & 25.80 \\
			4PB & 1.99 & -2.47 & 12.97 & -23.17 & 24.80 \\
			\bottomrule
		\end{tabularx}
	\end{small}
\end{table}

Since the MD domain is fixed in space during the FE calculations in the staggered scheme of the Capriccio method, the application of Dirichlet boundary conditions is not required to solve the corresponding equations \cite{Laubert2024}. Consequently, we can model the supports with Neumann boundary conditions, explicitly imposing the theoretical reaction forces  of $\ForceSih / 2$ caused by the supports. This is advantageous because the MD system  does not have to be moved over a large distance in space, 
which reduces the effect of the motion resistance of the bridging domain.

\paragraph{Mode II.}

To examine a rectangular panel (\Figref{fig:Silica_FEMD_SSP_modeI_Setup_SHIK}) subject to mode II conditions, we calculate the corresponding stress intensity factor $\StressIntensFactor_{\mathrm{II}}$ as
\nopagebreak\begin{align}
	\label{eq:KIITada1973}
	\StressIntensFactor_{\mathrm{II}} = \InPlaneShearStressLEFM \sqrt{\pi \CrackLengthLEFM} \GeometryFactorSih
\end{align}
based on \cite{Tada2000}, with the in-plane shear stress $\tau = \bar{\TractionCurrent}_{y}$ and the geometry factor
\nopagebreak\begin{align}
	\label{eq:FIITada1973}
	\GeometryFactorSih = \dfrac{1.122-0.561\CrackSampleRatio+0.085\CrackSampleRatio^{2}+0.180\CrackSampleRatio^{3}}{\sqrt{1-\CrackSampleRatio}}.
\end{align}
As in mode I, the plane strain state is enforced by preventing the $z$-surfaces from moving in the thickness direction. In addition, we do not allow any displacement of the $x$-surfaces in the $x$-direction.

\paragraph{Mode III.}

For the mode III tests, we employ a ``single edge crack in a rectangular beam subjected to longitudinal shear''. To this end, we use the setup introduced in \Figref{fig:Silica_FEMD_SSP_modeI_Setup_SHIK}, with out-of-plane shear stresses applied and without periodicity of the MD samples in $z$-direction. 
According to \cite{Sih1973} based on \cite{Westmann1967}, the stress intensity factor under mode III conditions can be derived from 
\nopagebreak\begin{align}
	\label{eq:KIII}
	\StressIntensFactor_{\mathrm{III}} = \LongitShearStressLEFM \sqrt{ \pi\CrackLengthLEFM} \GeometryFactorSih
\end{align}
with the longitudinal shear stress $\LongitShearStressLEFM = \bar{\TractionCurrent}_{z}$ and the tabulated value for the geometry factor $\GeometryFactorSih=\GeometryFactorSih( \CrackSampleRatio, \SampleLengthLEFM/\SampleWidthLEFM ) \approx 1.18$ for $\CrackSampleRatio = \CrackLengthLEFM/\SampleWidthLEFM = 0.5$ and $\SampleLengthLEFM/\SampleWidthLEFM = 0.5$. 
The simulations of mode III are performed by applying longitudinal shear to the $x$-surfaces of the FE domain while restricting their movement in $x$-direction. Hence, we keep the distance between the two loaded surfaces, i.e., $2\SampleLengthLEFM$, and their orientation constant during the load application  to avoid unrealistic deformations and rotations in space, referred to below as ``$x$-surface constraint''. 
Further details and an assessment of the influence of this constraint are provided in Supplementary Section~S4.9.

\subsection{Quantities for detecting the onset of crack propagation}
\label{ssec:Definition of the quantities of interest}

In the following, several measures 
for detecting the onset of crack propagation and thus the critical stress intensity factors $\StressIntensFactor_{i\mathrm{c}}\,(i=\mathrm{I,II,III})$ are introduced.

\subsubsection{Crack tip stress}
\label{sssec:Crack tip stress}

As emphasized by \acite{Rimsza2017}, the critical stress intensity factor is related to the initial events of bond breakage, i.e., the moment when the load exceeds the strength of the material. 
Consequently, we aim at determining $\StressIntensFactor_{\mathrm{c}}$ based on failure events near the crack tip. From a mechanical point of view, this should be observable by a peak value and a subsequent decrease in the stress recorded at the crack tip. To retrieve the stress at the atomically resolved crack tip, we evaluate the virial stress \cite{Thompson2009,Tadmor2011} 
\nopagebreak\begin{align}
\label{eq:TensileStress}
\mathbf{\CauchyStress}=-\dfrac{1}{\Volume} \left[ \sum\limits_{\AtomOne=1}^{\NumberAtoms}\sum\limits_{\substack{\AtomTwo=1\\\AtomTwo>\AtomOne}}^{\NumberAtoms} \ForceVector_{\AtomOne\AtomTwo} \otimes \left[\PositionVectorCur_{\AtomOne} - \PositionVectorCur_{\AtomTwo}\right]  + \sum\limits_{\AtomOne=1}^{\NumberAtoms} \Mass_{\AtomOne} \VelocityVector_{\AtomOne}\otimes\VelocityVector_{\AtomOne} \right]
\end{align}
with the number of considered atoms $\NumberAtoms$, the force between the atoms $\ForceVector_{\AtomOne\AtomTwo}$, their distance $\PositionVectorCur_{\AtomOne} - \PositionVectorCur_{\AtomTwo}$, the atomic mass $\Mass_{\AtomOne}$ and the velocity $\VelocityVector_{\AtomOne}$. 
Specifically, we consider the stress component $\CauchyStress_{xx}$ for mode I, while we evaluate the components $\CauchyStress_{xy}$ under mode II and $\CauchyStress_{xz}$ under mode III. As for the number of bonds, the kinetic energy, and the temperature (the evaluation of which is described in the next sections), the atoms located in the observation region in the initial configuration are used for the measurements of the quantities of interest, which corresponds to a Lagrangian perspective. For computing the virial stress according to  \Eqref{eq:TensileStress}, however, the constant volume
\nopagebreak\begin{align}
	\label{eq:VolObs}   
	\Volume = \Volume^{\mathrm{obs}} = \LengthInitial^{\mathrm{obs}}_{x} \LengthInitial^{\mathrm{obs}}_{y} \LengthInitial^{\mathrm{MD}}_{z} 
\end{align}
is used, even though the space occupied by the observed atoms may change during the mechanical loading.  This is a simplification that we consider justified until the crack starts to propagate.

\subsubsection{Number of bonds}

To verify that we are indeed tracking the first bond breakage events when finely resolving the virial stress at the crack tip, we measure the number of interactions between the observed crack tip atoms. As the SHIK potential \cite{Sundararaman2018,Sundararaman2019}  only contains ionic interactions, we refer to the number of pairwise interactions in the observation region at the crack tip as number of bonds $\NumberBonds$. While the samples are subjected to comparatively low external forces, we expect the number of bonds to fluctuate around a relatively constant value.  However, as soon as the crack propagates, 
a significant decrease in the bond count should be recognizable.

\subsubsection{Kinetic energy and temperature}
\label{sssec:Kinetic energy and temperature}

The temperature, which corresponds to the vibrational kinetic energy in MD simulations \cite{Tadmor2011}, is expected to increase during fracture due to the acceleration of atoms following bond breaking events at the crack tip \cite{Rice1969,Toussaint2016,VincentDospital2020}. 
Typical MD simulations cannot observe such a temperature increase in the fracture process zone because a thermostat is applied to the entire MD system. The Capriccio setup allows us to control the temperature only in the bridging region, i.e., in contrast to classical MD simulations, we do not blur temperature/kinetic energy changes in the crack tip region and can recognize temperature increases as an indicator of crack propagation.

Additionally, in Section~\ref{ssec:Temperature evolution}, we evaluate the temperature distribution in the MD samples, excluding the DPD region, by spatial averaging over an instantaneous temperature per atom \cite{Tadmor2011}
\nopagebreak\begin{align}
	\label{eq:TemperatureInstAtom} 
	\Temperature_{\mathrm{ins},\AtomOne} = \dfrac{\Mass_{\AtomOne} \VelocityVectorCurrent_{\AtomOne} \cdot \VelocityVectorCurrent_{\AtomOne}}{3 \BoltzmannConst}
\end{align} 
based on the atomic masses $\Mass_{\AtomOne}$ and velocities $\VelocityVectorCurrent_{\AtomOne}$ together with the Boltzmann constant $\BoltzmannConst$.

\subsubsection{Crack velocity}

To measure the crack velocity $\VelocityCurrent_{\mathrm{crack}}$ at the beginning of crack propagation at different external loads, we use a procedure resembling a creep test: after linearly increasing the external surface traction up to given $\StressIntensFactor_{\mathrm{I}}$, we keep the load constant for \SI{320}{\pico\second}, which corresponds to \num{4000} FE-MD iterations with 50 MD time steps each. The crack velocity is measured by dividing the length of the observation region in $y$-direction by the time that elapses from the moment the maximum virial stress $\CauchyStress_{xx}^{\mathrm{max}}$ is reached to the moment the virial stress drops to zero (i.e., when the crack has propagated through the observation region):
\nopagebreak\begin{align}
	\label{eq:Crackvelocity}
	\VelocityCurrent_{\mathrm{crack}} = \dfrac{\LengthInitial^{\mathrm{obs}}_{y}}{\Time(\CauchyStress_{xx}=0) - \Time(\CauchyStress_{xx}=\CauchyStress_{xx}^{\mathrm{max}})}
\end{align}
with $\Time(\CauchyStress_{xx}=0) \geq \Time(\CauchyStress_{xx}=\CauchyStress_{xx}^{\mathrm{max}})$. Hence, values of $\VelocityCurrent_{\mathrm{crack}} > 0$ are only to be expected when the crack propagates,
i.e., when the critical stress intensity factor is reached. Note that atoms may flow in and out of the evaluation region for this measurement, which corresponds to an Eulerian point of view.

\subsubsection{Crack opening displacement} 
\label{sssec:Crack opening displacement}

We measure the development of the crack opening displacement $\CrackOpeningDisp$ between the outermost FE nodes of the notch (highlighted in purple in \Figref{fig:Silica_FEMD_SSP_modeI_Setup_SHIK}), averaging over the displacements obtained for nodes with the same initial $x$-coordinate. This quantity can be directly compared with analytical predictions from LEFM. According to \cite{Tada2000}, the crack opening displacement for mode I fracture of the strip under tension can be determined as
\nopagebreak\begin{align}
	\label{eq:CODModeI}  \CrackOpeningDisp = \dfrac{4 \bar{\TractionCurrent}_{x} \CrackLengthLEFM}{\YoungsModulus'} V_{1} \quad \mathrm{for} \quad  \SampleLengthLEFM / \SampleWidthLEFM \geq 1.0
\end{align}
with the geometry factor
\nopagebreak\begin{align}
	\label{eq:CODModeIV} 
	V_{1} = \dfrac{1.46+3.42\left[ 1-\cos(\pi \CrackSampleRatio/2)\right]} {\left[\cos(\pi \CrackSampleRatio/2)\right]^{2}} 
\end{align}
and the stiffness under plane strain conditions $\YoungsModulus'$, see \Eqref{eq:EPlane}.
For $\CrackSampleRatio = \CrackLengthLEFM/\SampleWidthLEFM = 0.5$, $V_{1} \approx 4.92$ holds.
Due to the condition on the ratio of sample length to sample width  $\SampleLengthLEFM / \SampleWidthLEFM \geq 1.0$, the theoretical reference for the development of $\CrackOpeningDisp$ according to  \Eqref{eq:CODModeI} may not be fully comparable to the results obtained for   setups with  $\SampleLengthLEFM / \SampleWidthLEFM = 0.5$. However, as we note in Supplementary Sections~S4.5 and
S4.10, the geometry factors given by \acite{Sih1973} based on \acite{Bowie1973} and \acite{Bowie1965} are sufficiently similar to \cite{Tada2000} for the purpose of the present investigation. When it comes to the crack opening displacement $\CrackOpeningDisp$, 
crack growth is indicated by a change from the linear relationship between $\CrackOpeningDisp$ and the applied surface traction $\bar{\TractionCurrent}_{x}$ to a progressively nonlinear behavior, i.e., accelerated crack opening.

\section{Results}
\label{sec:Results}

\begin{figure}[ht!]
	\centering
	\includetikz{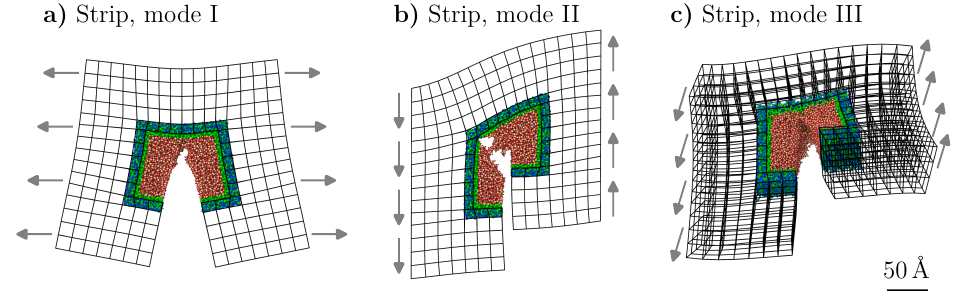}{1.0}
	\caption[Examples of deformed configurations of the strip simulations.]{Examples of deformed configurations of the strip simulations: a)~Mode I, b)~mode II, and c)~mode III conditions.}
	\label{fig:Silica_deformed_configurations_strip}
\end{figure}

We begin the presentation of the results obtained in the fracture simulations by examining the influence of the size of the observation region on the detection of crack propagation based on atomistic quantities (Section~\ref{ssec:Influence of the region of interest}). Subsequently, we evaluate the crack opening displacement (Section~\ref{ssec:Crack opening displacement}) and compare the predictions obtained from all the quantities considered (Section~\ref{ssec:Influence of the quantity of interest}).
Finally, we investigate the influence of the test setup (Section~\ref{ssec:Influence of the testing setup}).
Note that in this section we focus particularly on mode I. However, $\StressIntensFactor_{\mathrm{IIc}}$ and $\StressIntensFactor_{\mathrm{IIIc}}$ derived based on the procedure presented here are provided and discussed in Supplementary Section~S4 and Section~\ref{ssec:KIII/KI and KII/KI ratios}, respectively.
Examples of deformed configurations of the strip setups examined are shown in
\Figref{fig:Silica_deformed_configurations_strip}.

To determine a suitable analysis protocol, we focus on the influence of the size of the observation region specified in \Figref{fig:Silica_FEMD_SSP_modeI_Setup_SHIK} and the choice of the quantity of interest on the accuracy and precision of the measured critical stress intensity factor. To this end, we restrict this part of the study to the experimental setup shown in \Figref{fig:Silica_FEMD_SSP_modeI_Setup_SHIK}, i.e., a notched rectangular panel with an MD inclusion where the notch is continued to the center of the sample. 

\subsection{Influence of the size of the observation region}
\label{ssec:Influence of the region of interest}

\begin{figure}[ht!]
	\centering
	\hspace{-40pt}
	\includetikz{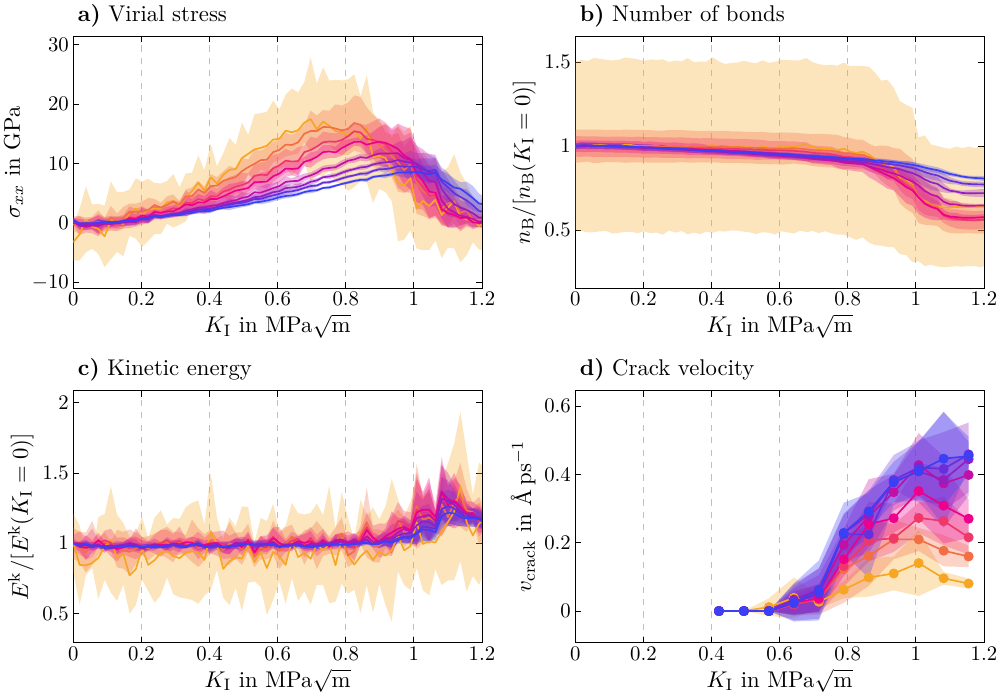}{0.9}
	\hspace*{-10pt}
	\includetikz{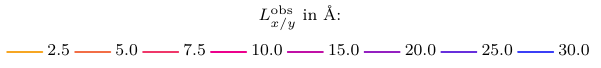}{0.5}
	\caption[Effect of the region of interest.]{Effect of the region of interest: a)~Virial stress $\CauchyStress_{xx}$, b)~number of bonds $\NumberBonds$, c)~kinetic energy $\KineticEnergy$, and d)~crack velocity $\VelocityCurrent_{\mathrm{crack}}$ over the stress intensity factor $\StressIntensFactor_{\mathrm{I}}$ for different sizes of the observation region in front of the crack tip $\LengthInitial^{\mathrm{obs}}_{x} = \LengthInitial^{\mathrm{obs}}_{y}$.
		Since the number of bonds and the kinetic energy are extensive quantities, we normalize them to the average value over five systems after equilibration at $\StressIntensFactor_{\mathrm{I}} = 0$.
	}
	\label{fig:Stress-xx-obs/Number-bonds/Kinetic-Energy/crackvelocity}
\end{figure}

In \Figref{fig:Stress-xx-obs/Number-bonds/Kinetic-Energy/crackvelocity}, we display four of the quantities of interest introduced in Section~\ref{ssec:Definition of the quantities of interest}, which are all evaluated in the observation region inside the particle domain: a)~the virial stress $\CauchyStress_{xx}$, b)~the number of bonds $\NumberBonds$, c)~the kinetic energy $\KineticEnergy$ and d)~the crack velocity $\VelocityCurrent_{\mathrm{crack}}$.

\subsubsection{Crack tip stress}

In \Figref{fig:Stress-xx-obs/Number-bonds/Kinetic-Energy/crackvelocity}a), we plot the virial stress $\CauchyStress_{xx}$ over the applied stress intensity factor $\StressIntensFactor_{\mathrm{I}}$, calculated from \Eqref{eq:KIStrip}, in evaluation regions of different sizes. A clear dependence on the size of the evaluation region is observed. In particular, the location of the maximum stress shifts to larger $\StressIntensFactor_{\mathrm{I}}$ values for larger evaluation regions, ranging from about \SI{0.73}{\mega\pascal\sqrt{\metre}} for $\LengthInitial_{x/y}^{\mathrm{obs}} = \SI{2.5}{\angstrom}$ up to \SI{1.01}{\mega\pascal\sqrt{\metre}} for $\LengthInitial_{x/y}^{\mathrm{obs}} = \SI{30}{\angstrom}$. Additionally, the value of the mean maximum stress decreases for larger regions ($\CauchyStress_{xx}^{\mathrm{max}} \approx \SI{17.5}{\giga\pascal}$ for $\LengthInitial_{x/y}^{\mathrm{obs}} = \SI{2.5}{\angstrom}$, while $\CauchyStress_{xx}^{\mathrm{max}} \approx \SI{8.6}{\giga\pascal}$ for $\LengthInitial_{x/y}^{\mathrm{obs}} = \SI{30.0}{\angstrom}$), indicating that the stress concentration becomes less detectable.  

\subsubsection{Number of bonds}

We evaluate the number of bonds $\NumberBonds$, i.e., the number of pairwise interactions in the observation region in \Figref{fig:Stress-xx-obs/Number-bonds/Kinetic-Energy/crackvelocity}b). On the one hand, a stronger relative decrease of $\NumberBonds$ for smaller evaluation regions can be discerned, to about \SI{80}{\percent} of the initial average value after equilibration of the systems for $\LengthInitial_{x/y}^{\mathrm{obs}} = \SI{30.0}{\angstrom}$ and to approximately \SI{56}{\percent} for $\LengthInitial_{x/y}^{\mathrm{obs}} = \SI{5.0}{\angstrom}$. This is reasonable as more intact bonds are tracked for larger regions during
crack propagation. In agreement with the virial stress, the value of $\StressIntensFactor_{\mathrm{I}}$ which indicates crack propagation, corresponding to a distinct change in slope in the case of $\NumberBonds$, shifts to larger values for larger observation regions. For $\LengthInitial_{x/y}^{\mathrm{obs}} = \SI{5.0}{\angstrom}$, we observe a decrease in the number of bonds at $\StressIntensFactor_{\mathrm{I}} \approx \SI{0.85}{\mega\pascal\sqrt{\metre}}$, which is in good agreement with the value for $\StressIntensFactor_{\mathrm{I}}$ leading to the maximum virial stress calculated in an observation region of the same size.

\subsubsection{Kinetic energy}

\Figref{fig:Stress-xx-obs/Number-bonds/Kinetic-Energy/crackvelocity}c)~shows the kinetic energy measured in the observation region during load application. Only when a significant number of bonds are broken,  the kinetic energy/temperature increases significantly, while the corresponding value of $\StressIntensFactor_{\mathrm{I}}$ appears to be similar for all sizes of observation regions. 
Therefore, the suitability of the kinetic energy to detect the onset of bond breaking, i.e., the beginning of crack propagation, is questionable. Larger observation regions blur this temperature increase more, since it is concentrated around the crack tip, cf.\ the temperature maps evaluated in Section~\ref{ssec:Temperature evolution}. For $\LengthInitial_{x/y}^{\mathrm{obs}} = \SI{2.5}{\angstrom}$, $\KineticEnergy$ increases  up to about \SI{40}{\percent} of the initial mean value, for $\LengthInitial_{x/y}^{\mathrm{obs}} = \SI{30.0}{\angstrom}$, it only increases to about \SI{20}{\percent}.

\subsubsection{Crack velocity}
\label{sssec:Crack velocity}

The results given in \Figref{fig:Stress-xx-obs/Number-bonds/Kinetic-Energy/crackvelocity}d)~for the crack velocity $\VelocityCurrent_{\mathrm{crack}}$ imply that the velocity increases with the size of the evaluation region. For $\LengthInitial_{x/y}^{\mathrm{obs}} = \SI{2.5}{\angstrom}$, the maximum mean velocity $\VelocityCurrent_{\mathrm{crack}}^{\mathrm{max}}$ is found to be about \SI{0.14}{\angstrom\per\pico\second}, while for $\LengthInitial_{x/y}^{\mathrm{obs}} = \SI{30.0}{\angstrom}$, $\VelocityCurrent_{\mathrm{crack}}^{\mathrm{max}} \approx \SI{0.45}{\angstrom\per\pico\second}$.
However,  after the first bond failure events, the crack propagates through the entire sample and thus through all observation regions. Therefore, the value of $\StressIntensFactor_{\mathrm{I}}$ at which the crack propagates for all samples (between 0.75 and \SI{0.8}{\mega\pascal\sqrt{\metre}}) is recorded equally for all sizes of observation regions. 
Since the load is kept constant during these measurements, the observed crack growth is unstable \cite{Anderson2017,Gross2018}. 

\subsection{Continuum-based evaluation: Crack opening displacement}
\label{ssec:Crack opening displacement}

\begin{figure}[ht!]
	\centering
	\includetikz{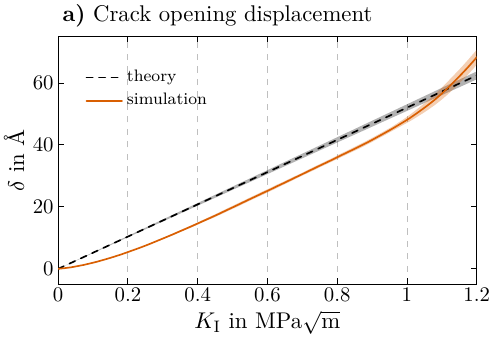}{0.48}
	\hspace{10pt}
	\includetikz{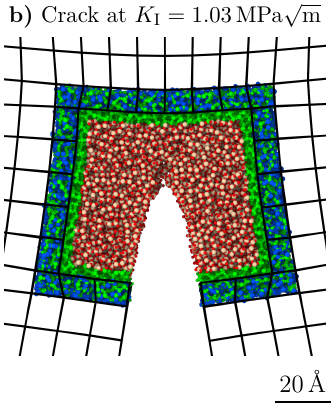}{0.4}
	\caption[Assessment of the overall deformation.]{Assessment of the overall deformation: a)~Resulting crack opening displacement~$\CrackOpeningDisp$  between the outermost finite element (FE) nodes of the notch over the stress intensity factor~$\StressIntensFactor_{\mathrm{I}}$  
		compared to the theoretical reference curve given by \Eqref{eq:CODModeI} and b)~snapshot of one of the samples at $\StressIntensFactor_{\mathrm{I}} = \SI{1.03}{\mega\pascal\sqrt{\metre}}$.
		For the present sample, the critical stress intensity factor, derived based on the virial stress as described in Section~\ref{ssec:Definition of the quantities of interest},
		is $\StressIntensFactor_{\mathrm{Ic}} = \SI{0.75}{\mega\pascal\sqrt{\metre}}$.
	}
	\label{fig:COD_red/Silica_FEMD_SSP_modeI_ts156000}
\end{figure}

\Figref{fig:COD_red/Silica_FEMD_SSP_modeI_ts156000}a) evaluates the evolution of the crack opening displacement  $\CrackOpeningDisp$, which is measured as introduced in Section~\ref{ssec:Fracture setups and associated stress intensity factors},
and compares it to the LEFM prediction given by \Eqref{eq:CODModeI}. In general, it is found that the magnitude of the measured crack opening displacement is in a similar range to the theoretical prediction of LEFM.
Initially, $\CrackOpeningDisp$ increases more slowly than the reference up to   $\StressIntensFactor_{\mathrm{I}} \approx \SI{0.3}{\mega\pascal\sqrt{\metre}}$. The subsequent evolution of $\CrackOpeningDisp$ shows a slope close to the target curve. Only when a significant number of bonds are broken, cf.\  \Figref{fig:COD_red/Silica_FEMD_SSP_modeI_ts156000}b),  $\CrackOpeningDisp$ becomes nonlinear and steeper than the LEFM reference.

\subsection{Detection of crack propagation: Comparison of the quantities considered}
\label{ssec:Influence of the quantity of interest}

\begin{figure}[ht!]
	\centering
	\includetikz{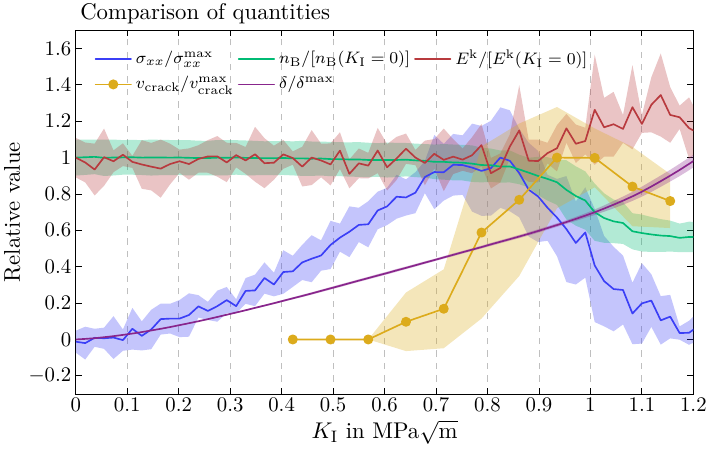}{1.0}
	\caption[Comparison of the evaluated quantities.]{Comparison of the evaluated quantities: Virial stress $\CauchyStress_{xx}$, number of bonds $\NumberBonds$, kinetic energy $\KineticEnergy$, crack velocity $\VelocityCurrent_{\mathrm{crack}}$, and crack opening displacement $\CrackOpeningDisp$ over the stress intensity factor $\StressIntensFactor_{\mathrm{I}}$, normalized either to the maximum of the mean curve in the considered range of $\StressIntensFactor_{\mathrm{I}}$ or to the value obtained at $\StressIntensFactor_{\mathrm{I}} = 0$, i.e., after equilibration of the samples.
	}
	\label{fig:Comparison_quantities}
\end{figure}

Based on the results for the atomistic quantities and the crack opening displacement $\CrackOpeningDisp$ given in \Figref{fig:Stress-xx-obs/Number-bonds/Kinetic-Energy/crackvelocity} and \Figref{fig:COD_red/Silica_FEMD_SSP_modeI_ts156000}a), possible estimates for the critical stress intensity factor $\StressIntensFactor_{\mathrm{Ic}}$ given in \Tabref{tab:KIc} can be derived according to the criteria defined in Section~\ref{ssec:Definition of the quantities of interest}. 
A direct comparison of the underlying curves is provided in
\Figref{fig:Comparison_quantities}.
It is evident that the virial stress $\CauchyStress_{xx}$, the number of bonds $\NumberBonds$, and the crack velocity $\VelocityCurrent_{\mathrm{crack}}$ suggest similar values for the critical stress intensity factor, i.e., about  $\StressIntensFactor_{\mathrm{Ic}}=\SI{0.8}{\mega\pascal\sqrt{\metre}}$ to \SI{0.9}{\mega\pascal\sqrt{\metre}}. In contrast, the kinetic energy $\KineticEnergy$  and the crack opening displacement $\CrackOpeningDisp$ only indicate crack propagation after the crack has propagated over a longer distance. In this case, a higher value for $\StressIntensFactor_{\mathrm{Ic}}$ between \SI{0.9}{\mega\pascal\sqrt{\metre}} and \SI{1.0}{\mega\pascal\sqrt{\metre}} would be determined.

\begin{table}[ht!]
	\caption[Comparison of the predictions obtained.]{Comparison of the predictions obtained: Values for the  critical stress intensity factor~$\StressIntensFactor_{\mathrm{Ic}}$ in \si{\mega\pascal\sqrt{\metre}} derived from the virial stress $\CauchyStress_{xx}$, the number of bonds $\NumberBonds$, the kinetic energy~$\KineticEnergy$, the crack velocity $\VelocityCurrent_{\mathrm{crack}}$, and the crack opening displacement $\CrackOpeningDisp$ evaluated in \Figref{fig:Comparison_quantities}.}
	\label{tab:KIc}
	\centering
	\begin{small}
		\begin{tabularx}{0.725\textwidth}{l c c c c c}
			\toprule
			Quantity of interest & $\CauchyStress_{xx}$ & $\NumberBonds$ & $\KineticEnergy$ & $\VelocityCurrent_{\mathrm{crack}}$ & $\CrackOpeningDisp$ \\
			\midrule
		 $\StressIntensFactor_{\mathrm{Ic}}$ in \si{\mega\pascal\sqrt{\metre}} & 0.80\,--\,0.90 & 0.85\,--\,0.90 & 0.90\,--\,1.00 & 0.75\,--\,0.80 & 0.90\,--\,1.00 \\
			\bottomrule
		\end{tabularx}
	\end{small}
\end{table}

In the following, we set the dimensions of the observation region  $\LengthInitial^{\mathrm{obs}}_{x} = \LengthInitial^{\mathrm{obs}}_{y}$ to \SI{5.0}{\angstrom}. 
Using this observation region, we identify the value of $\StressIntensFactor_{\mathrm{I}}$ that results in the maximum virial stress  as an indicator for the first bond breakage events and thus as the critical stress intensity factor~$\StressIntensFactor_{\mathrm{Ic}}$. We obtain it as the average of the independent evaluation of the individual replicas, which for the systems considered above yields $\StressIntensFactor_{\mathrm{Ic}} = \SI{0.80(0.08)}{\mega\pascal\sqrt{\metre}}$.

\subsection{Influence of the test setup}
\label{ssec:Influence of the testing setup}

In the following, we compare the results with those obtained in 3PB and 4PB tests as calculated from \Eqref{eq:KI_3/4PBSih1973}.

\subsubsection{Three- and four-point bending}

\begin{figure}[ht!]
	\centering
	\includetikz{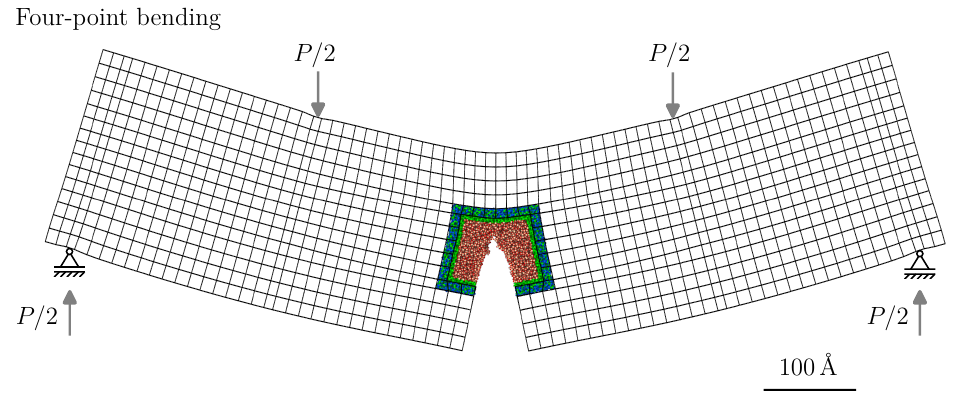}{1.0}
	\caption[Example of a deformed configuration of the four-point bending tests.]{Example of a deformed configuration of the four-point bending tests: The specimen shown here originates from a simulation in which Neumann boundary conditions were used to model the supports, loaded with a force $\ForceSih$.
	}
	\label{fig:Silica_deformed_configurations_bending}
\end{figure}

We contrast the results obtained for the above-discussed single edge crack in a rectangular panel with those observed when performing 3PB and 4PB tests as outlined in 
Section~\ref{ssec:Fracture setups and associated stress intensity factors}. The supports are modeled by Neumann boundary conditions, as displayed in \Figref{fig:Silica_deformed_configurations_bending}.
Note that we choose the same number of MD time steps per load step of $\MDTimeStepsPerIteration=50$ for the bending simulations as for the strip simulations. However, the selected load step sizes of $\Delta\bar{\Traction}_{x} = \SI{1.6}{\mega\pascal}$ lead to an approximately eight times smaller rate of the applied stress intensity factor $\dot{\StressIntensFactor}_{\mathrm{I}}$ according to \Eqref{eq:KI_3/4PBSih1973}. On the one hand, however, the application of larger load steps would lead to a decrease in calculation accuracy. On the other hand, conducting fewer MD time steps per load step leads to problems with temperature control (see Supplementary Figure~S9b). Since, based on the conclusions drawn from Supplementary Figure~S17a), a large effect of a change in the loading rate by this amount can be ruled out, we do not expect any significant impact on the results shown here.

\begin{figure}[ht!]
	\centering
	\includetikz{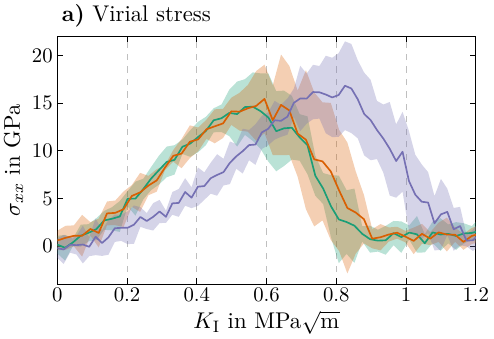}{0.48}
	\includetikz{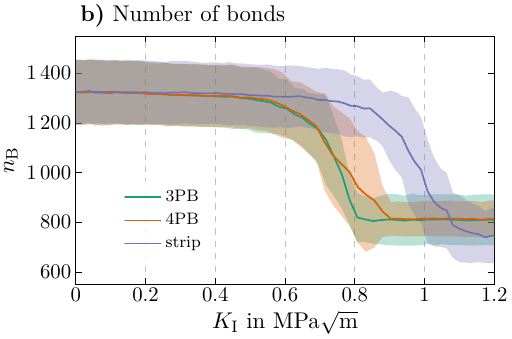}{0.48}
	\caption[Three- and four-point bending tests compared to tensile strip tests.]{Three- (3PB) and four-point bending (4PB) tests compared to tensile strip tests: a)~virial stress $\CauchyStress_{xx}$ and b)~number of bonds $\NumberBonds$ over the stress intensity factor $\StressIntensFactor_{\mathrm{I}}$.
        The supports acting in the 3PB and 4PB tests are modeled by applying Neumann boundary conditions.
	}
	\label{fig:Stress-xx-obs/Number-bonds_3PB-4PB_cracktip_n}
\end{figure}

In \Figref{fig:Stress-xx-obs/Number-bonds_3PB-4PB_cracktip_n},
the evolution of a)~the virial stress $\CauchyStress_{xx}$ and b)~the number of bonds $\NumberBonds$ near the crack tip during load application is displayed. 
We obtain lower values for the critical stress intensity factor of $\StressIntensFactor_{\mathrm{Ic}} = \SI{0.56(0.08)}{\mega\pascal\sqrt{\metre}}$ (3PB) and $\StressIntensFactor_{\mathrm{Ic}} = \SI{0.59(0.10)}{\mega\pascal\sqrt{\metre}}$ (4PB), which are about \SI{25}{\percent} smaller than for the strip under tension. 
Consequently, the motion resistance of the bridging domain, which may cause or at least exacerbate this discrepancy, seems to be less pronounced in the bending setup 
than in the strip under tension. 
The results that are obtained when modeling the supports using Dirichlet boundary conditions, where the MD domain does not remain in its initial position and the motion resistance becomes more pronounced, are reported in Supplementary Section~S4.7.

\subsubsection{Mode II, mode III, and assessment of further influencing factors}

Under mode II and mode III conditions, we obtain the critical stress intensity factors $\StressIntensFactor_{\mathrm{IIc}} = \SI{0.44(0.03)}{\mega\pascal\sqrt{\metre}}$, and $\StressIntensFactor_{\mathrm{IIIc}} = \SI{0.40(0.02)}{\mega\pascal\sqrt{\metre}}$ when we apply a loading rate $\dotbar{\TractionCurrent}_{y}$ or $\dotbar{\TractionCurrent}_{z}$, respectively, of  $\SI{20}{\giga\pascal\per\nano\second}$ to samples synthesized with a quenching rate of $\TemperatureRate = \SI{0.26}{\kelvin\per\pico\second}$. 
In the supplementary material, we provide the underlying evolutions of the virial stress in the observation region (Supplementary Sections~S4.2 and S4.8) together with an assessment of the influence of the loading rate and the quenching rate (Supplementary Section~S4.2). Moreover, we assess the impact of the load step size and the stiffness of the anchor point springs as well as the specimen dimensions on the results of the fracture simulations (Supplementary Sections~S4.4 and S4.5).

\section{Discussion}
\label{sec:Discussion}

Below, we discuss the results obtained in Section~\ref{sec:Results} and put them in a larger context. In particular, we first evaluate the size of the plastic zone at the beginning of crack propagation (Section~\ref{ssec:Size of the plastic zone}).
We then comment on the 
definition of an optimal analysis protocol (Section~\ref{ssec:Definition of an optimal analysis protocol}), the spatial temperature development in the atomistic systems  (Section~\ref{ssec:Temperature evolution}), 
and the detection of 
crack propagation based on the crack velocity measurements (Section~\ref{ssec:Crack velocity}). 
Subsequently, we calculate the ratios between the critical stress intensity factors under mode~I, mode~II, and mode~III, as these enable a comparison with experimental and theoretical values (Section~\ref{ssec:KIII/KI and KII/KI ratios}). In addition, the development of bond distances and angles under the three fracture modes is considered in the supplementary material (Supplementary Section~S4.3).

\subsection{Size of the plastic zone}
\label{ssec:Size of the plastic zone}

\begin{figure}[ht!]
	\centering
	\hspace{-50pt}
	\includetikz{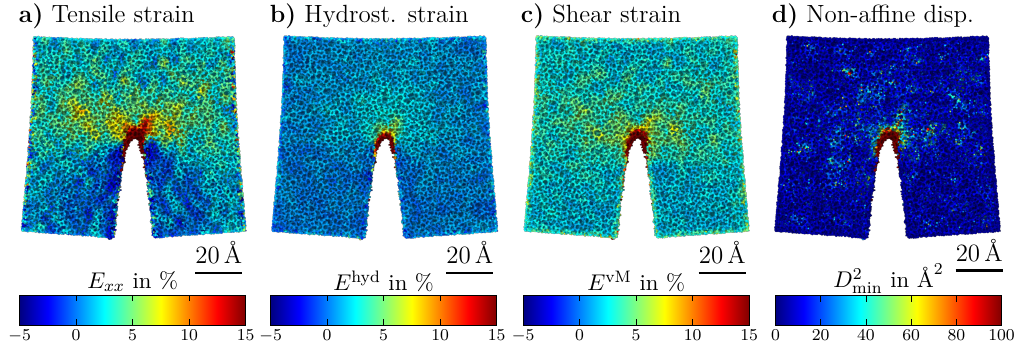}{0.9}
	\caption[Influence of the specimen size on the strain distribution.]{Influence of the specimen size on the strain distribution: Distribution of  a)~tensile strain $\GreeLagStra_{xx}$,  b)~hydrostatic strain $\VolStrain$, c)~von Mises local shear invariant $\ShearStrain$, and d)~the non-affine displacement measure $D^{2}_{\mathrm{min}}$  at an applied stress intensity factor of $\StressIntensFactor_{\mathrm{I}} = \SI{1.2}{\mega\pascal\sqrt{\metre}}$; dissipative particle dynamics/anchor point region excluded, cutoff of the strain evaluation $\SI{7}{\angstrom}$, size of the particle domain $\LengthInitial_{x/y}^{\mathrm{MD}} = \SI{200}{\angstrom}$.}
	\label{fig:Strain_contour_LxyMD200}
\end{figure}

According to LEFM, samples with finite dimensions must satisfy the condition that the crack length $\CrackLengthLEFM$ is sufficiently larger than the fracture process zone of radius~$r_{\mathrm{FPZ}}$, i.e.,
\nopagebreak\begin{align}
	\label{eq:SizesLEFM}
	r_{\mathrm{FPZ}} \ll \RadiusKDetField \ll \CrackLengthLEFM, 
\end{align}
where a common approximation for the radius of the $\StressIntensFactor$-determined field is $\RadiusKDetField \approx 0.15\CrackLengthLEFM$ \cite{Andric2018}. This corresponds to $\RadiusKDetField \approx \SI{69}{\angstrom}$ for the largest systems ($\SampleWidthLEFM = 2\SampleLengthLEFM = \SI{920}{\angstrom}$) investigated in Supplementary Figure~S22b). Furthermore, the applied far-field stresses must remain small enough not to cause material nonlinearities \cite{Andric2018}, which we verified in Supplementary Section~S2. 
To ensure that the plastic zone is significantly smaller than the overall dimensions of the sample, guide values for the specimen size as given in
\cite{Gross2018} can also be applied. For the crack length $\CrackLengthLEFM$, the ligament length $\SampleWidthLEFM-\CrackLengthLEFM$, and the sample thickness $\SampleThicknessLEFM$, the relation
\begin{align}
	\label{eq:PlasticZoneSample}
	\left\{ \CrackLengthLEFM, \SampleWidthLEFM-\CrackLengthLEFM, \SampleThicknessLEFM\right\} \geq 2.5 \left[\dfrac{\StressIntensFactor_{\mathrm{Ic}}}{\YieldStress}\right]^{2}
\end{align}
can be used. This describes a condition under which the crack tip is subjected to the plane strain state, resulting in a smaller plastic zone than in the plane stress state \cite{Gross2018}. 
By inserting the critical stress intensity factor~$\StressIntensFactor_{\mathrm{Ic}} $ of about  \SI{0.80}{\mega\pascal\sqrt{\metre}}, which was determined in Section~\ref{sec:Results}, and the yield stress~$\YieldStress$ of about \SI{10}{\giga\pascal} derived in Supplementary Section~S2, we obtain minimum sample dimensions of
\SI{160}{\angstrom}.
Since we explicitly prescribe the plane strain state by not allowing strains in the thickness direction, the condition for the sample thickness $\SampleThicknessLEFM$ is not applicable here. The minimum dimensions for the crack length $\CrackLengthLEFM$ and the ligament length $\SampleWidthLEFM-\CrackLengthLEFM$ given by \Eqref{eq:PlasticZoneSample} are smaller than the total dimensions of the larger samples investigated in Supplementary Figure~S22b), where no clear influence of the sample length was observed.
\Eqref{eq:PlasticZoneSample}  can also be utilized to obtain a relationship for the maximum size of the plastic zone at the critical stress intensity factor $\StressIntensFactor_{\mathrm{Ic}}$, i.e., \cite{Gross2018}
\begin{align}
	\label{eq:PlasticZoneCrit}
	r_{\mathrm{pc}} \lesssim 0.02 \left\{ \CrackLengthLEFM, \SampleWidthLEFM-\CrackLengthLEFM, \SampleThicknessLEFM\right\}, 
\end{align}
which leads to the requirement that $r_{\mathrm{pc}} \lesssim \SI{9.2}{\angstrom}$ 
for $\SampleWidthLEFM = 2\SampleLengthLEFM = \SI{920}{\angstrom}$.
\begin{figure}[ht!]
	\centering
	\hspace{-50pt}
	\includetikz{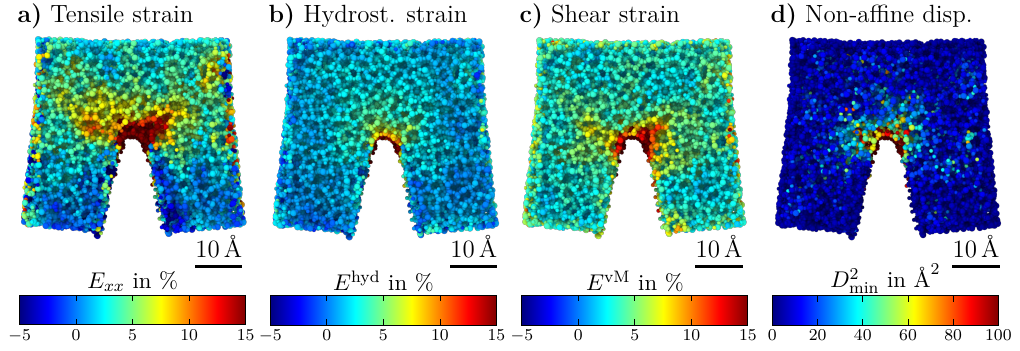}{0.9}
	\caption[Strain maps for the strip simulations under mode I conditions.]{Strain maps for the strip simulations under mode I conditions: Distribution of  a)~tensile strain $\GreeLagStra_{xx}$,  b)~hydrostatic strain $\VolStrain$, c)~von Mises local shear invariant $\ShearStrain$, and d)~the non-affine displacement measure $D^{2}_{\mathrm{min}}$  at an applied stress intensity factor of $\StressIntensFactor_{\mathrm{I}} = \SI{0.75}{\mega\pascal\sqrt{\metre}}$; dissipative particle dynamics/anchor point region excluded, cutoff of the strain evaluation $\SI{7}{\angstrom}$, size of the particle domain $\LengthInitial_{x/y}^{\mathrm{MD}} = \SI{100}{\angstrom}$.}
	\label{fig:Strain_contour}
\end{figure}
In the following, we evaluate the distributions of a)~the tensile strain $\GreeLagStra_{xx}$,  b)~the hydrostatic strain 
\nopagebreak\begin{align}
	\label{eq:GreenLagrangeStrainHyd}
	\VolStrain = \dfrac{1}{3} \tr (\GreeLagStraTen), 
\end{align}
and c)~the von Mises local shear invariant  \cite{Shimizu2007} 
\begin{align}
	\label{eq:MisesGreenLagrangeStrain}
	\ShearStrain =  \sqrt{\dfrac{1}{6} \left[\left[\GreeLagStra_{xx} - \GreeLagStra_{yy}\right]^{2} + \left[\GreeLagStra_{xx} - \GreeLagStra_{zz}\right]^{2} + \left[\GreeLagStra_{yy} - \GreeLagStra_{zz}\right]^{2}\right] +  \GreeLagStra^{2}_{xy} + \GreeLagStra^{2}_{xz} + \GreeLagStra^{2}_{yz}}
\end{align}
based on the Green-Lagrange strain tensor \cite{Holzapfel2000} as well as d)~the local deviation
from affine displacements  $\NonAffineDisp_{\mathrm{min}}^{2}$ according to \cite{Falk1998}. The contour plots shown in \Figref{fig:Strain_contour} and \Figref{fig:Strain_contour_LxyMD200}  were obtained in strip simulations under mode I conditions for sizes of the particle domain of $\LengthInitial_{x}^{\mathrm{MD}} = \LengthInitial_{y}^{\mathrm{MD}} = \SI{100}{\angstrom}$ and \SI{200}{\angstrom}, respectively. 
The samples with the larger particle domain  are described and evaluated in more detail in Supplementary Section~S4.5.
As Supplementary Figure~S5 suggests, especially the deviatoric contributions are responsible for plastic effects in the investigated material. Therefore, we consider in particular the von Mises local shear invariant $\ShearStrain$   and  $D^{2}_{\mathrm{min}}$  as suitable indicators for the size of the plastic zone. 
From the distributions obtained, a radius of the plastic zone of approximately up to  \SI{10}{\angstrom} can be inferred. This coincides well with the experimental value of \SI{6}{\angstrom} given in \cite{Wiederhorn1974}. 
Based on these considerations, the requirements specified in  \Eqref{eq:PlasticZoneSample} and \Eqref{eq:PlasticZoneCrit} for the sample dimensions in relation to the size of the plastic zone according to \cite{Gross2018} are met, although the sample dimensions are in the nanoscale. 
The size condition with respect to the  $\StressIntensFactor$-determined field given in \Eqref{eq:SizesLEFM} would be satisfied if we set $r_{\mathrm{FPZ}} \coloneq \RadiusPlasticZone$. Although estimating the size of the fracture process zone is beyond the scope of this study, considering the value $r_{\mathrm{FPZ}} = \SI{100}{\angstrom}$ obtained by \acite{Rountree2020} would lead to the conclusion that the samples tested are still too small and that crack lengths in the range of approximately \SI{1000}{\angstrom} are required. 
However, based on the results for different overall specimen sizes provided in Supplementary Figure~S22b), we do not expect a significant effect of the specimen dimensions on the results.

\subsection{Identifying an optimal analysis protocol}
\label{ssec:Definition of an optimal analysis protocol}

In the present study, the onset of crack propagation is identified by evaluating the stress intensity factor  $\StressIntensFactor_{i}\,(i=\mathrm{I,II,III})$ at the maximum virial stress of the individual FE-MD replicas. The resulting prediction for the critical stress intensity factor $\StressIntensFactor_{i\mathrm{c}}$ agrees well with the first bond failure events, see Section~\ref{ssec:Influence of the quantity of interest}.
At this level of loading, unstable crack propagation occurs in all samples, as indicated by the crack velocity measurements. 
In contrast,
an increase in temperature/kinetic energy as well as a change in the slope of the crack opening displacement only occur when a large number of bonds break and the samples fail on a comparatively large length scale.
From a mechanical point of view, 
it is reasonable to determine the critical stress intensity factor based on the moment at which the first bonds break \cite{Rimsza2017}, i.e., to resolve the quantities of interest at the crack tip as finely as possible. 
\acite{Rimsza2017} identified these first bond breakages based on the potential energy curve and obtained a critical stress intensity factor of $\StressIntensFactor_{\mathrm{Ic}} = \SI{0.76(0.16)}{\mega\pascal\sqrt{\metre}}$, which is in a similar range in terms of average and  variability as the values for the strip under tension determined in the present study. 
However, the exact choice of the region of interest influences not only the measured thermal fluctuations but also the detection of crack propagation, see \Figref{fig:Stress-xx-obs/Number-bonds/Kinetic-Energy/crackvelocity}. 
This implies that if the protocol and the quantity of interest for identifying the onset of bond breakage  in MD simulations are not correctly defined, the predictability of the  simulations is substantially reduced. 
Overall, the size of the observation region has a pronounced influence on the variance of the stress, the number of bonds and the kinetic energy, i.e., the variability decreases with increasing regions. In contrast, there is only a small size effect on the variability of the crack velocity.
While the maximum virial stress is a suitable indicator for the onset of crack propagation due to the first bond failures, a compromise must be found between a fine resolution of the sampling region
and the noise in the measurement due to the limited number of atoms evaluated. 

\subsection{Temperature evolution}
\label{ssec:Temperature evolution}

\begin{figure}[ht!]
	\centering
	\includetikz{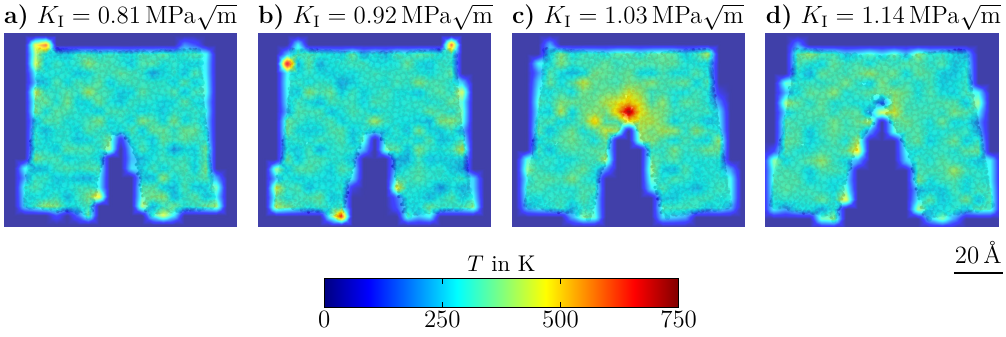}{1.0}
	\caption[Temperature maps for the strip under tension.]{Temperature maps for the strip under tension: Distribution of the spatially averaged per-atom temperature based on \Eqref{eq:TemperatureInstAtom}
		(dissipative particle dynamics/anchor point region excluded, $25 \times 25$ bins) for stress intensity factors a) $\StressIntensFactor_{\mathrm{I}} = \SI{0.81}{\mega\pascal\sqrt{\metre}}$, b)~\SI{0.92}{\mega\pascal\sqrt{\metre}}, c)~\SI{1.03}{\mega\pascal\sqrt{\metre}}, and d)~\SI{1.14}{\mega\pascal\sqrt{\metre}}.
		For the present sample, the critical stress intensity factor is $\StressIntensFactor_{\mathrm{Ic}} = \SI{0.75}{\mega\pascal\sqrt{\metre}}$.}
	\label{fig:Temperature_map}
\end{figure}

In \Figref{fig:Temperature_map}, temperature maps of one sample employed in the tensile strip tests are shown  at different stress intensity factors $\StressIntensFactor_{\mathrm{I}}$ beyond the sample's $\StressIntensFactor_{\mathrm{Ic}} = \SI{0.75}{\mega\pascal\sqrt{\metre}}$. At the crack tip, a temperature increase can be recognized. However, this only occurs at values of $\StressIntensFactor_{\mathrm{I}}$  considerably greater than $\StressIntensFactor_{\mathrm{Ic}}$, i.e., at about $\SI{1.0}{\mega\pascal\sqrt{\metre}}$, when a significant number of bonds are broken (see  \Figref{fig:Temperature_map}c)~for a visualization of an  atomistic structure obtained at this load). Obviously, the evolution of the kinetic energy and thus of the temperature lags behind the small number of first bond breakage events occurring at the onset of crack growth, which seems to be physically justified. Consequently, however, an increase in temperature/kinetic energy is not suitable for identifying $\StressIntensFactor_{\mathrm{Ic}}$ based on the first bond-breaking events. As mentioned in 
Section~\ref{ssec:Definition of the quantities of interest}, such an increase in the crack tip temperature can only be detected because we control the temperature only at the outer margin of the MD domain, not in the entire sample. At the moment depicted in \Figref{fig:Temperature_map}d), the temperature increase due to the bond breakage events in c) appears to have been eliminated by the temperature control.

\subsection{Crack velocity}
\label{ssec:Crack velocity}

The crack velocity measurements presented in 
Section~\ref{sssec:Crack velocity} allow conclusions to be drawn about the applied loading rates. The magnitude of the values determined for the crack speed $\VelocityCurrent_{\mathrm{crack}}$ indicates that the loading rate $\dotbar{\TractionCurrent}_{x}$ is small enough to achieve physically meaningful results. According to \cite{Achenbach1999}, the phase velocity of Rayleigh waves can be approximated by
\nopagebreak\begin{align}
\label{eq:RayleighVelocity}
\VelocityCurrent_{\mathrm{R}} = \dfrac{0.862 + 1.14 \PoissonsRatio}{1 + \PoissonsRatio} \VelocityCurrent_{\mathrm{T}} 
\end{align}
with the phase velocity of transversal waves $\VelocityCurrent_{\mathrm{T}} =  \sqrt{\YoungsModulus/\left[2\MassDensity 
 [1 + \PoissonsRatio\right] ]}$.
Using the elastic properties determined  at a quenching rate of \SI{0.26}{\kelvin\per\pico\second}, see
Supplementary Table~S2, we arrive at $\VelocityCurrent_{\mathrm{R}} \approx \SI{33.50}{\angstrom\per\pico\second}$. 
\acite{Chowdhury2019} have found a terminal crack velocity of $\VelocityCurrent_{\mathrm{crack}} = \SI{2.21(0.04)}{\kilo\metre\per\second} = \SI{22.1(0.4)}{\angstrom\per\pico\second}$ in pure MD simulations under mode~I conditions at a strain rate of $\SI{5e9}{\per\second} = \SI{500}{\percent\per\nano\second}$ using a ReaxFF parametrization by \acite{Fogarty2010}. These values for the strain rate and the crack velocity are significantly larger than  in the present measurements, where we conduct creep tests with a preceding strain rate of about \SI{30}{\percent\per\nano\second}  and obtain crack velocities of up to about \SI{0.45}{\angstrom\per\pico\second}. Note that the crack, once it has propagated through the large observation regions, is already close to the boundary of the MD region. Here, it must be taken into account that the crack is stopped by the FE region, which cannot capture the crack propagation. Hence, once the crack tip gets close to the FE-MD bridging domain, the simulations become unphysical.
Since the present study focuses only on the onset of crack propagation, detected in an observation region at a considerable distance from the FE region, we consider that the artifacts caused by this do not substantially affect the results.

\subsection{Ratios of critical stress intensity factors}
\label{ssec:KIII/KI and KII/KI ratios}

For samples prepared with a quenching rate of $\TemperatureRate = \SI{0.26}{\kelvin\per\pico\second}$ and subjected to a loading rate of $\dotbar{\TractionCurrent}_{i} = \SI{20}{\giga\pascal\per\nano\second}$, we obtain the critical stress intensity factors  $\StressIntensFactor_{\mathrm{Ic}} = \SI{0.80(0.08)}{\mega\pascal\sqrt{\metre}}$, $\StressIntensFactor_{\mathrm{IIc}} = \SI{0.44(0.03)}{\mega\pascal\sqrt{\metre}}$, and $\StressIntensFactor_{\mathrm{IIIc}} = \SI{0.40(0.02)}{\mega\pascal\sqrt{\metre}}$, see Section~\ref{sec:Results}.
The experimentally measured values for the critical stress intensity factor $\StressIntensFactor_{\mathrm{Ic}}$ of silica glass  typically range between 0.7 to \SI{0.8}{\mega\pascal\sqrt{\metre}}
\cite{Wiederhorn1969,Moody1995,Celarie2007,Rimsza2017,Rouxel2017}. 
For the critical stress intensity factors of soda-lime-silica glass under mode I and II conditions, \acite{Sglavo2003} obtained $\StressIntensFactor_{\mathrm{Ic}} = \SI{0.74(0.05)}{\mega\pascal\sqrt{\metre}}$ and $\StressIntensFactor_{\mathrm{IIc}} = \SI{0.64(0.10)}{\mega\pascal\sqrt{\metre}}$,  in symmetric and asymmetric 4PB tests, respectively.
However, as \citeauthor{Sglavo2003} mention, limited experimental data is available for comparison. 
As far as $\StressIntensFactor_{\mathrm{IIIc}}$ is concerned, experimental values for pure silica are even rarer. 
In this section, we comment on the ratios of $\StressIntensFactor_{\mathrm{IIc}} / \StressIntensFactor_{\mathrm{Ic}}$ and $\StressIntensFactor_{\mathrm{IIIc}} / \StressIntensFactor_{\mathrm{Ic}}$ obtained from the present simulations by comparing them with the predictions of the LEFM, allowing us to draw conclusions about the results for mode III without experimental data for~$\StressIntensFactor_{\mathrm{IIIc}}$.

To obtain theoretical target values for the ratios $\StressIntensFactor_{\mathrm{IIc}} / \StressIntensFactor_{\mathrm{Ic}}$ and $\StressIntensFactor_{\mathrm{IIIc}} / \StressIntensFactor_{\mathrm{Ic}}$ predicted by LEFM,
we use the $\StrainEnergyDensityFactor_{\mathrm{c}}$-theory \cite{Sih1973,Sih1973a,Sih1991}: According to this approach, the strain energy density factor $\StrainEnergyDensityFactor$ is defined as
\begin{equation}
\label{eq:StrainEnDensFac}
\begin{aligned}
\StrainEnergyDensityFactor & = \StrainEnergyDensityFactor_{1}  & + \, & \StrainEnergyDensityFactor_{12} &  + \, & \StrainEnergyDensityFactor_{2} & + \, & \StrainEnergyDensityFactor_{3} \\
 & = a_{11} \StressIntensFactorSih_{1}^{2} \mspace{-15mu} & + \, & 2a_{12} \StressIntensFactorSih_{1} \StressIntensFactorSih_{2} \mspace{-15mu}  & + \, & a_{22} \StressIntensFactorSih_{2}^{2} \mspace{-15mu} & + \, & a_{33} \StressIntensFactorSih_{3}^{2}.
\end{aligned}
\end{equation}
Here, $\StressIntensFactor_{i} = \StressIntensFactorSih_{j}\sqrt{\pi}\,(i=\mathrm{I,II,III})$, while $a_{jk}\,(j,k=\mathrm{1,2,3})$ are functions of the elastic properties and the polar angle in the $\StressIntensFactor$-dominance zone.
The $\StrainEnergyDensityFactor_{\mathrm{c}}$-theory assumes that the crack starts to propagate for all fracture modes at the same critical strain energy density,  $\StrainEnergyDensityFactor_{1\mathrm{c}}\overset{!}{=} \StrainEnergyDensityFactor_{2\mathrm{c}} \overset{!}{=} \StrainEnergyDensityFactor_{3\mathrm{c}}$ is required. This leads to
\nopagebreak\begin{align}
	\label{eq:RatioStressCrit}
	\dfrac{\StressIntensFactor_{\mathrm{IIc}}}{\StressIntensFactor_{\mathrm{Ic}}} = \sqrt{\dfrac{3\left[1-2\PoissonsRatio\right]}{2\left[1-\PoissonsRatio\right]-\PoissonsRatio^{2}}} \quad \mathrm{and} \quad
	\dfrac{\StressIntensFactor_{\mathrm{IIIc}}}{\StressIntensFactor_{\mathrm{Ic}}} = \sqrt{1 - 2 \PoissonsRatio}. 
\end{align}
This procedure results in target ratios of about  \num{1.07(0.01)} and \num{0.76(0.01)}, respectively, for the MD systems of silica synthesized with a quenching rate of $\TemperatureRate = \SI{0.26}{\kelvin\per\pico\second}$ (Poisson's ratio $\PoissonsRatio = \num{0.208(0.006)}$, see
Supplementary Table~S2).

Under mode II conditions, based on the component $\CauchyStress_{xy}$ evaluated in the observation region in samples prepared at a quenching rate of $\TemperatureRate = \SI{0.26}{\kelvin\per\pico\second}$ and subjected to a loading rate of $\dotbar{\TractionCurrent}_{y} = \SI{20}{\giga\pascal\per\nano\second}$, we obtain $\StressIntensFactor_{\mathrm{IIc}} = \SI{0.44(0.03)}{\mega\pascal\sqrt{\metre}}$. Thus, the Capriccio simulations predict a ratio of $\StressIntensFactor_{\mathrm{IIc}} / \StressIntensFactor_{\mathrm{Ic}} = 0.55(0.04)$, which is low compared to the theoretical target value of \num{1.07(0.01)} and experimental data:
The values for $\StressIntensFactor_{\mathrm{Ic}}$ and $\StressIntensFactor_{\mathrm{IIc}}$ by \acite{Sglavo2003} mentioned above result in a ratio of $\StressIntensFactor_{\mathrm{IIc}} / \StressIntensFactor_{\mathrm{Ic}} \approx 0.86$.
\acite{Shetty1987} obtained a ratio of $\StressIntensFactor_{\mathrm{IIc}} / \StressIntensFactor_{\mathrm{Ic}} \approx 1.23$  for soda-lime glass by means of chevron-notched disk specimens subjected to diametral compression. However, it is noticeable that the theoretical prediction either overestimates or underestimates the experimental values mentioned.

As far as mode III is concerned, for the bulk metallic glass (BMG) $\mathrm{Zr}_{61}\mathrm{Ti}_{2}\mathrm{Cu}_{25}\mathrm{Al}_{12}$ (ZT1) examined by \citeauthor{Song2016} in \cite{Song2016} by applying torsion to cylinders with circumferential notches, one obtains a theoretical ratio of about $\StressIntensFactor_{\mathrm{IIIc}} / \StressIntensFactor_{\mathrm{Ic}} \approx 0.52$, using a Poisson's ratio of $\PoissonsRatio = 0.367$ \cite{He2011}. This overestimates the experimental ratio of $\StressIntensFactor_{\mathrm{IIIc}} / \StressIntensFactor_{\mathrm{Ic}} \approx 0.39$ determined by \acite{Song2016} by about \SI{33}{\percent}. 
In the present study, the theoretical reference value of \num{0.76(0.01)} predicted by LEFM also overestimates the ratio of $\StressIntensFactor_{\mathrm{IIIc}}/\StressIntensFactor_{\mathrm{Ic}} = \num{0.51(0.06)}$ obtained in the FE-MD simulations, specifically by about \SI{51(18)}{\percent}. 

A detailed analysis of the available criteria for defining relationships between stress intensity factors determined for different crack opening modes is not intended here. Further approaches to determine these relations can be found in the literature. In this context, it is known that an influence of the experimental test method \cite{Sglavo2003} and the theoretical criterion \cite{Shetty1987} is observable.
The Capriccio coupling might additionally influence the results.
For instance, in the case of mode III, the  bridging domain is displaced in the out-of-plane direction, unlike in the mode I scenarios. Under such loading conditions, the bridging domain's resistance to motion
might have a greater influence and thus affect the values obtained for   $\StressIntensFactor_{\mathrm{IIIc}}$ more pronouncedly. 
While the Capriccio method provides a tool for evaluating the influence of different loading conditions or, as a future perspective, other media, on the structural properties of amorphous solids during failure, the peculiarities of the FE-MD coupling do reduce the quantitative accuracy of the method to some extent. However, they do not fundamentally undermine the method's capabilities.

\section{Conclusions}
\label{sec:Conclusions}

In this contribution, we perform fracture simulations of edge-cracked silica glass samples using the Capriccio method, which couples molecular dynamics (MD) with a continuum described by the finite element method (FEM). In particular, we conduct  experiments of mode I (single edge cracked plates under tension and bending specimens) as well as mode II and mode III (single edge 
cracked plates subjected to in-plane or out-of-plane shear). The critical stress intensity factors $\StressIntensFactor_{\mathrm{Ic}}$, $\StressIntensFactor_{\mathrm{IIc}}$, and $\StressIntensFactor_{\mathrm{IIIc}}$ are derived based on the onset of crack propagation. 
To detect the latter, we consider various indicators: In an evaluation region in front of the atomically resolved crack tip, we measure the virial stress, the number of pair interactions,  the kinetic energy, and the crack velocity. In the finite element (FE) domain, we evaluate the crack opening displacement. When finely resolving the crack tip region, the maximum virial stress indicates the onset of crack propagation in a particularly objective and meaningful way. Therefore, we take the external surface stress applied to the continuum at this instance in time to calculate the critical stress intensity factor. 

With this study, we demonstrate that it is generally possible to derive reasonable values for the critical stress intensity factor under all three fracture modes by using the Capriccio method. Consequently, the latter can be considered as a versatile and flexible tool to apply boundary conditions typically encountered in engineering applications to chemically specific atomistic domains. 
For the mode I simulations, our predictions for the critical stress intensity factor $\StressIntensFactor_{\mathrm{Ic}}$ agree well with experimental reference values. For modes II and III, for which only limited and uncertain experimental data are available, the values obtained for $\StressIntensFactor_{\mathrm{IIc}}$ and $\StressIntensFactor_{\mathrm{IIIc}}$ are within a plausible range. This becomes particularly apparent 
when comparing the ratios 
$\StressIntensFactor_{\mathrm{IIc}}/\StressIntensFactor_{\mathrm{Ic}}$ and $\StressIntensFactor_{\mathrm{IIIc}}/\StressIntensFactor_{\mathrm{Ic}}$ with estimates based on LEFM. 
Moreover, by evaluating the distributions of various strain measures, we show that the plastic zone is sufficiently small in relation to the overall sample dimensions, which is a crucial prerequisite for the applicability of LEFM \cite{Anderson2017,Gross2018}. 

With regard to future applications, the overall dimensions of the samples, i.e., of the FE domains, should be increased to macroscopic length scales to test and exploit the full potential of the Capriccio method. Additionally, the present computational framework can be used, for instance, to study the effect of environmental media, e.g., water or air, on the (fracture) mechanical properties of glassy materials  \cite{Wiederhorn2011,To2021a}.
However, the present investigation also emphasizes that further improvement of the Capriccio coupling is crucial to obtain quantitatively clearer results. In particular, problems arising from the motion resistance of the bridging domain, which results from the staggered FE-MD solution scheme of the Capriccio method and depends on the anchor point spring stiffness and the load step size, still need to be addressed \cite{Laubert2024,Laubert2024a}.
Importantly, the present study shows that the trajectory of the bridging domain significantly influences this behavior and that calibration of the Capriccio coupling based solely on tensile tests of unnotched specimens is insufficient.
Consequently, the tensile strip and bending tests lead to different results, which should be resolved by a more detailed investigation.
Further open questions are 
the influence of the crack shape and the notch generation, e.g., if introducing an atomically sharp crack instead of deleting atoms leads to deviations in the results. To be able to delete bonds in the notch region, MD models that explicitly consider bonded interactions would be required, which in turn would necessitate either reactive force fields such as ReaxFF \cite{Duin2001} or the definition of criteria for bond breaking.

\ifthenelse{\boolean{showauthors}}{

\clearpage\newpage

\section*{Acknowledgments}

We thank Fabrice Célarié, Jean-Pierre Guin and Theany To from the Institut de Physique de Rennes (IPR) for the fruitful discussions during Felix Weber's research stay. Likewise, we thank György Hantal, Bernd Meyer and Christian Wick from the Friedrich-Alexander-Universität Erlangen-Nürnberg (FAU) as well as Thomas Seelig from the Karlsruhe Institute of Technology.

\section*{Funding}

The authors gratefully acknowledge funding by various sources: 
The overall research was funded by the Deutsche Forschungsgemeinschaft (DFG, German Research Foundation) – 377472739/GRK 2423/2-2023.
Lukas Laubert and Sebastian Pfaller are funded by the DFG – 505866713 and the Agence nationale de la recherché (ANR, French Research Agency) – ANR-22-CE92-0049.
Sebastian Pfaller is furthermore funded by the DFG – 396414850 (Individual Research Grant ‘Identifikation von Interphaseneigenschaften in Nanokompositen’).
The international collaboration has been funded by an International Emerging Action (IEA) grant from Centre National de la Rercherche Scientifique (CNRS). 
In addition, scientific support and HPC resources have been provided by the Erlangen National High Performance Computing Center (NHR@FAU) of the Friedrich-Alexander-Universität Erlangen-Nürnberg (FAU) under the NHR project b136dc. NHR funding is provided by federal and Bavarian state authorities. NHR@FAU hardware is partially funded by the DFG project 440719683.

\section*{CRediT authorship contribution statement}

\textbf{Felix Weber:} Conceptualization, Data curation, Formal analysis, Investigation, Methodology, Software, Validation, Visualization, Writing~-~original draft, Writing~-~review~\&~editing. \textbf{Maxime Vassaux:} Conceptualization, Formal analysis, Funding acquisition, Investigation, Methodology, Project administration, Resources, Software, Supervision, Validation, Writing~-~original draft, Writing~-~review~\&~editing. \textbf{Lukas Laubert:} Formal analysis, Investigation, Methodology, Validation, Writing~-~review~\&~editing. \textbf{Sebastian Pfaller:} Conceptualization, Formal analysis, Funding acquisition, Investigation, Methodology, Project administration, Resources, Supervision, Validation, Writing~-~original draft, Writing~-~review~\&~editing.

}{

}

\section*{Simulation software and data}
In this contribution, the Capriccio code (version 2.0.1) \cite{Pfaller2024a} was employed, using LAMMPS (version 2 August 2023) \cite{Plimpton1995,Thompson2022} for the MD and Matlab (version R2022a) \cite{Matlab} for the FE simulations. The FE meshes were generated in Abaqus/CAE \cite{Abaqus} (version 2021.HF7). Ovito \cite{Stukowski2009} was applied for the visualization of the FE-MD samples. The data associated with this contribution are available via Zenodo \cite{Weber2025a}. 

\section*{Declaration on generative AI and AI-assisted technologies in the writing process}

During the preparation of this work the authors used DeepL in order to paraphrase drafted sentences, provide wording suggestions, and assist with punctuation and grammar. After using this tool, the authors reviewed and edited the content as needed and take full responsibility for the content of the published article.

\clearpage\newpage

\renewcommand{\headeright}{supplementary material}
\renewcommand{\undertitle}{supplementary material}
\section*{Supplementary material}

\setcounter{figure}{0}
\renewcommand\thefigure{S\arabic{figure}} 
\setcounter{table}{0}
\renewcommand\thetable{S\arabic{table}} 
\renewcommand\thesubsection{S\arabic{subsection}}

\setcounter{equation}{0}
\renewcommand\theequation{S\arabic{equation}}

In the following, we provide supporting information on the molecular dynamics (MD) interaction potentials utilized (Section~\ref{ssec:APP.Molecular dynamics potentials}), on the
characterization of the material
based on pure MD simulations (Section~\ref{ssec:APP.Elastic properties}), the setup of the Capriccio coupling (Section~\ref{ssec:APP.Capriccio parameters}), and additional data obtained from the fracture simulations under mode I, mode II, and mode III conditions (Section~\ref{ssec:APP.Fracture simulations}). 

\subsection{Molecular dynamics potentials}
\label{ssec:APP.Molecular dynamics potentials}

The SHIK potential \cite{Sundararaman2018,Sundararaman2019} is given as
\nopagebreak\begin{align}
	\label{eq:PotentialSHIK}    
	V (r_{\AtomOne\AtomTwo})=A_{\AtomOne\AtomTwo}\exp(-B_{\AtomOne\AtomTwo}r_{\AtomOne\AtomTwo})-\dfrac{C_{\AtomOne\AtomTwo}}{r_{\AtomOne\AtomTwo}^{6}}+\dfrac{D_{\AtomOne\AtomTwo}}{r_{\AtomOne\AtomTwo}^{24}}+V^{\mathrm{W}}(r_{\AtomOne\AtomTwo}). 
\end{align}
On the one hand, the short-range interactions are described by the Buckingham potential, which generalizes the Lennard-Jones potential \cite{Griebel2007}.
On the other hand, the Wolf truncation method \cite{Wolf1999} evaluates the long-range Coulomb interactions as
\nopagebreak\begin{align}
	\label{eq:PotentialWolf}    V^{\mathrm{W}}(r_{\AtomOne\AtomTwo})=\dfrac{q_{\AtomOne} q_{\AtomTwo}}{4\pi\varepsilon_{0}}\left[\dfrac{1}{r_{\AtomOne\AtomTwo}}-\dfrac{1}{r_{\mathrm{c}}^{\mathrm{W}}}+\dfrac{r_{\AtomOne\AtomTwo}-r_{\mathrm{c}}^{\mathrm{W}}}{\left[r_{\mathrm{c}}^{\mathrm{W}}\right]^{2}}\right]. 
\end{align}
The charges and masses of the silicon (Si) and oxygen (O) atoms are defined as $\Charge_{\mathrm{Si}} = \SI{1.7755}{\elementarycharge}$ and $\Charge_{\mathrm{O}} = \SI{-0.88775}{\elementarycharge}$ as well as $\Mass_{\mathrm{Si}} = \SI{28.086}{\gram\per\mole}$ and $\Mass_{\mathrm{O}} = \SI{16.0}{\gram\per\mole}$, respectively. The repulsive term $D_{\AtomOne\AtomTwo} / r_{\AtomOne\AtomTwo}^{24}$ prevents the potential from diverging when atoms come close to each other, especially at high temperatures. Furthermore, the potential formulation specified in \Eqref{eq:PotentialSHIK} is multiplied by the smoothing function 
\nopagebreak\begin{align}
	\label{eq:SmoothingSHIK}   
	G_{r_{\mathrm{c}}} (r_{\AtomOne\AtomTwo}) = \exp\left( - \dfrac{\gamma^{2}}{\left[r_{\AtomOne\AtomTwo} - r_{\mathrm{c}} \right]^{2}}\right) 
\end{align}
with the width of the smoothing function for both the short- and long-range interactions $\gamma = \SI{0.2}{\angstrom}$. The short-range cutoff is $r_{\mathrm{c}}^{\mathrm{Buck}} = \SI{8}{\angstrom}$, while the long-range cutoff is $r_{\mathrm{c}}^{\mathrm{W}} = \SI{10}{\angstrom}$. Moreover, the MD time step size is $\Delta\Time_{\mathrm{MD}} = \SI{1.6}{\femto\second}$. \Tabref{tab:ParametersSHIK} lists the specific values used for the parameters in \Eqref{eq:PotentialSHIK} for the Si--Si, Si--O, and O--O interactions \cite{Sundararaman2019}. 

\begin{table}[ht!]
	\caption[Parameters of the SHIK potential.]{Parameters of the SHIK potential  according to \cite{Sundararaman2019}.}
	\label{tab:ParametersSHIK}
	\centering
	\begin{small}
		\begin{tabularx}{0.55\textwidth}{l r r r r}
			\toprule
			Bond & $A_{\AtomOne \AtomTwo}$ in \si{\eV} & $B_{\AtomOne \AtomTwo}$ in \si{\per\angstrom} & $C_{\AtomOne \AtomTwo}$ in \si{\eV \angstrom^{6}} & $D_{\AtomOne \AtomTwo}$ in \si{\eV \angstrom^{24}} \\ \midrule
			Si--Si & \num{2798.0} & \num{4.4073} & \num{0.000} & \num{3423204} \\
			Si--O & \num{23108.0} & \num{5.0979} & \num{139.700} & \num{66} \\
			O--O & \num{1120.5} & \num{2.8927} & \num{26.132} & \num{16800} \\
			\bottomrule
		\end{tabularx}
	\end{small}
\end{table}

\subsection{Mechanical characterization}
\label{ssec:APP.Elastic properties}

\subsubsection{Time-proportional uniaxial tension}

\begin{figure}[ht!]
	\centering
	\includetikz{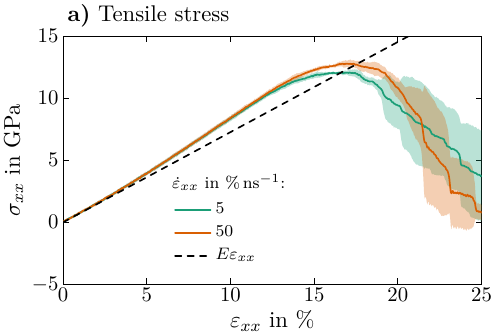}{0.48}
	\includetikz{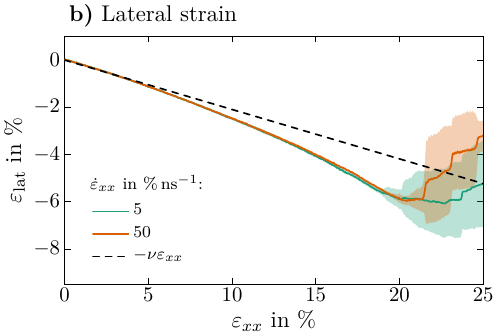}{0.48}
	\caption[Uniaxial tension under free lateral contraction.]{Uniaxial tension under free lateral contraction: a)~Virial stress $\CauchyStress_{xx}$ and b)~mean lateral strain $\SmallStrain_{\mathrm{lat}}$   over the applied tensile strain $\SmallStrain_{xx}$ obtained in molecular dynamics uniaxial tensile simulations at two different strain rates $\dot{\SmallStrain}_{xx}$ allowing for lateral contraction at zero external pressure.
		Young's modulus $\YoungsModulus$ and Poisson's ratio $\PoissonsRatio$ are derived by fitting a linear function to the data obtained at both strain rates applying the Matlab \cite{Matlab} function \texttt{polyfit} in the interval $0<\SmallStrain_{xx}\leq\SI{1.0}{\percent}$.
	}
	\label{fig:ut_free_lateral_rate}
\end{figure}

To ensure that the FE and MD domains share the same elastic properties, we determine Young's modulus  $\YoungsModulus$ as well as Poisson's ratio $\PoissonsRatio$ using uniaxial tensile tests with pure MD under periodic boundary conditions, taking into account lateral contraction at zero external pressure as \acite{Zhang2020} and averaging the results over five samples. 
Prepared at a quenching rate of  $\TemperatureRate = \SI{0.26}{\kelvin\per\pico\second}$, we test samples with dimensions of about $ \SI{80}{\angstrom} \times \SI{80}{\angstrom} \times \SI{80}{\angstrom}$ (\num{11398} Si and \num{22796} O atoms) and pull them at strain rates \SI{5}{\percent\per\nano\second} and \SI{50}{\percent\per\nano\second}. In \Figref{fig:ut_free_lateral_rate}, a)~the virial stress $\CauchyStress_{xx}$ and b)~the mean lateral strain
\nopagebreak\begin{align}
	\label{eq:MeanLatCont}   
    \SmallStrain_{\mathrm{lat}} = \dfrac{1}{2} \left[ \SmallStrain_{yy} + \SmallStrain_{zz} \right]
\end{align} 
are given over the applied strain $\SmallStrain_{xx}$, where $\SmallStrain_{ii} = [\LengthCurrent_{i}^{\mathrm{MD}} - \LengthInitial_{i}^{\mathrm{MD}} ] / \LengthInitial_{i}^{\mathrm{MD}}$ with the initial and current lengths of the MD simulation box $\LengthInitial_{i}^{\mathrm{MD}}$ and $\LengthCurrent_{i}^{\mathrm{MD}}$, respectively. The elastic constants are determined by fitting linear functions to the virial stress in the entire MD box $\CauchyStressTensor = - \PressureTensor$ \cite{Thompson2009,Tadmor2011} 
and to~$\SmallStrain_{\mathrm{lat}}$ obtained at the two strain rates in the strain interval $0<\SmallStrain_{xx}\leq\SI{1.0}{\percent}$, applying the Matlab \cite{Matlab} function \texttt{polyfit}. For the quenching rate $\TemperatureRate = \SI{0.26}{\kelvin\per\pico\second}$, we obtain $\PoissonsRatio = \num{0.209(0.003)}$ and $\YoungsModulus = \SI{72.8(1.4)}{\giga\pascal}$. 
Since we apply the LAMMPS \cite{Plimpton1995,Thompson2022} ``metal'' unit set, we use, e.g., $\YoungsModulus = \SI{0.454}{\eV\per\cubic\angstrom}$ as the Young's modulus of the FE domain for the systems obtained at this quenching rate, where
$\SI{1}{\eV\per\cubic\angstrom} = \SI{160.2}{\giga\pascal}$ \cite{Schmiermund2020}. 

\begin{figure}[ht!]
	\centering
	\includetikz{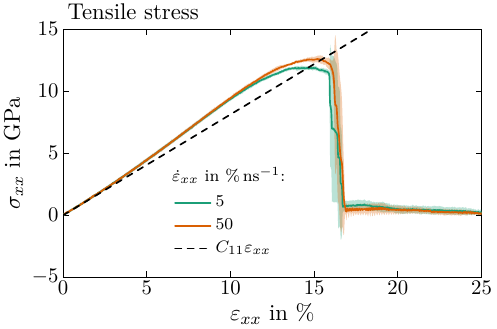}{0.48}
	\caption[Uniaxial tension under blocked lateral contraction.]{Uniaxial tension under blocked lateral contraction: Virial stress $\CauchyStress_{xx}$ over the applied tensile strain $\SmallStrain_{xx}$ obtained in molecular dynamics uniaxial tensile simulations at two different strain rates $\dot{\SmallStrain}_{xx}$ with impeded lateral contraction.
		The longitudinal modulus $\Stiffness_{11}$ is derived by fitting a linear function to the data obtained at both strain rates applying the Matlab \cite{Matlab} function \texttt{polyfit} in the interval $0<\SmallStrain_{xx}\leq\SI{1.0}{\percent}$.
	}
	\label{fig:ut_fix_lateral_rate}
\end{figure}

Tensile simulations without lateral contraction yield the longitudinal modulus $\Stiffness_{11} = \SI{81.4(1.5)}{\giga\pascal}$, see \Figref{fig:ut_fix_lateral_rate}, which is close to the 
analytical prediction based on Young's modulus $\YoungsModulus$ and Poisson's ratio $\PoissonsRatio$ according to linear elasticity as \cite{Tadmor2011}
\nopagebreak\begin{align}
	\label{eq:LongitudinalModulusEnu}
	\Stiffness_{11} = \dfrac{\YoungsModulus \left[1-\PoissonsRatio\right]}{\left[1+\PoissonsRatio\right] \left[1-2\PoissonsRatio\right]}
\end{align}
which results in $\Stiffness_{11} = \SI{81.8}{\giga\pascal}$.
The elastic constants determined are consistent with the values given by \acite{Zhang2020} in their study comparing the results obtained using the SHIK potential and other MD models for silica and sodium silicate glasses. There, a good agreement of various mechanical properties with experimental values was demonstrated when using the SHIK potential.

\begin{figure}[ht!]
	\centering
	\includetikz{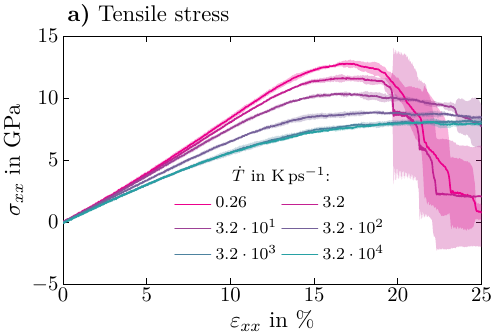}{0.48}
	\includetikz{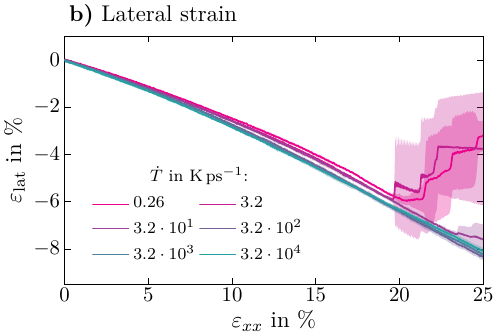}{0.48}
	\caption[Influence of the quenching rate.]{Influence of the quenching rate: a)~Virial stress $\CauchyStress_{xx}$ and b)~mean lateral strain $\SmallStrain_{\mathrm{lat}}$ over the applied tensile strain $\SmallStrain_{xx}$ obtained in molecular dynamics uniaxial tensile simulations  allowing for lateral contraction at zero external pressure for samples prepared at various quenching rates $\TemperatureRate$ at
		strain rate $\dot{\SmallStrain}_{xx} = \SI{25}{\percent\per\nano\second}$.
	}
	\label{fig:ut_free_lateral_cooling_rate}
\end{figure}

To retrieve the elastic properties of the systems prepared at quenching rates other than $\TemperatureRate = \SI{0.26}{\kelvin\per\pico\second}$, the same samples of size about $ \SI{100}{\angstrom} \times \SI{100}{\angstrom} \times \SI{50}{\angstrom}$ (\num{11131} Si and \num{22262} O atoms)  used in the FE-MD fracture simulations are examined under uniaxial tension at a strain rate of $\dot{\SmallStrain}_{xx} = \SI{25}{\percent\per\nano\second}$. At a quenching rate of   $\TemperatureRate = \SI{0.26}{\kelvin\per\pico\second}$ this would yield $\YoungsModulus = \SI{73.0(0.9)}{\giga\pascal}$ and $\PoissonsRatio = \num{0.208(0.006)}$, which does not differ significantly from the value obtained for the specimen size of approximately $\SI{80}{\angstrom} \times \SI{80}{\angstrom} \times \SI{80}{\angstrom}$. In \Figref{fig:ut_free_lateral_cooling_rate}, we evaluate a)~the virial stress $\CauchyStress_{xx}$ and b)~the mean lateral strain $\SmallStrain_{\mathrm{lat}}$ for different quenching rates $\TemperatureRate$.  

\Tabref{tab:ElasticConst_cooling_rate} summarizes the mass density $\MassDensity$ after synthesis of the glass samples as well as Young's modulus $\YoungsModulus$ and Poisson's ratio $\PoissonsRatio$ that we derive for the several $\TemperatureRate$.  The values for the mass density~$\MassDensity$ are obtained by applying the Matlab \cite{Matlab} function \texttt{polyfit} to the density signal recorded during the last \SI{320}{\pico\second} of the glass synthesis.
Here,  $\YoungsModulus$ reduces with increasing quenching rate  while $\MassDensity$ and $\PoissonsRatio$ increase. From the uniaxial tension simulations using pure MD,  it is evident that Hooke's law is sufficient to describe the system behavior in a stress range up to about \SI{3}{\giga\pascal}, which is well above the critical applied surface traction of $\bar{\TractionCurrent}_{x\mathrm{c}} \approx 1-\SI{2}{\giga\pascal}$ for the mode I strip setups studied here.
Note that increasing the quenching rate from \SI{3222.66}{\kelvin\per\pico\second} to \SI{32226.56}{\kelvin\per\pico\second} does not change the already highly ductile stress-strain behavior.

\begin{table}[ht!]
	\caption[Mass density $\MassDensity$, Young's modulus $\YoungsModulus$ and Poisson's ratio $\PoissonsRatio$ depending on the quenching rate $\TemperatureRate$.]{Mass density $\MassDensity$, Young's modulus $\YoungsModulus$ and Poisson's ratio $\PoissonsRatio$ depending on the quenching rate $\TemperatureRate$.}
	\label{tab:ElasticConst_cooling_rate}
	\centering
	\begin{small}
		\begin{tabularx}{0.52\textwidth}{r r r r}
			\toprule
		 $\TemperatureRate$ in \si{\kelvin\per\pico\second} & $\MassDensity$ in \si{\gram\per\cubic\centi\metre} & $\YoungsModulus$ in \si{\giga\pascal} & $\PoissonsRatio$\\
			\midrule
			\num{0.26} & $\num{2.224(0.009)}$ & $\num{73.0(0.9)}$ & $\num{0.208(0.006)}$ \\
			\num{3.22} & $\num{2.233(0.008)}$ & $\num{72.3(0.9)}$ & $\num{0.220(0.004)}$\\
			\num{32.23} & $\num{2.244(0.007)}$ & $\num{70.9(0.4)}$ & $\num{0.226(0.005)}$\\
			\num{322.27} & $\num{2.259(0.003)}$ & $\num{67.1(1.2)}$ & $\num{0.233(0.005)}$\\
			\num{3222.66} & $\num{2.268(0.006)}$ & $\num{64.7(1.1)}$ & $\num{0.233(0.003)}$\\
			\num{32226.56} & $\num{2.277(0.004)}$ & $\num{64.3(2.5)}$ & $\num{0.233(0.012)}$\\
			\bottomrule
		\end{tabularx}
	\end{small}
\end{table}

\subsubsection{Time-periodic tension-compression}

LEFM and thus the concept of the stress intensity factor $\StressIntensFactor$ is only applicable if both the plastic zone and the fracture process zone are small compared to the overall dimensions of the specimens \cite{Anderson2017,Gross2018}.  
To estimate possible plastic contributions and at which deformations they occur approximately, we perform time-periodic tension-compression tests on the pure MD samples with dimensions of $\SI{80}{\angstrom} \times \SI{80}{\angstrom} \times \SI{80}{\angstrom}$. The pressure in the lateral directions is kept constant at zero, i.e., lateral strains are permitted.
The tests are carried out at three strain amplitudes $\SmallStrain^{\mathrm{A}}_{xx}$, i.e.,   \SI{5}{\percent}, \SI{10}{\percent}, and \SI{15}{\percent}, with a maximum strain rate of $\dot{\SmallStrain}^{\mathrm{max}}_{xx} = \SI{25}{\percent\per\nano\second}$. The strain-time behavior in the loading direction is given by 
\nopagebreak\begin{align}
	\label{eq:CyclicLoading}   
	\SmallStrain_{xx}(\Time) = \SmallStrain^{\mathrm{A}}_{xx} \sin \Bigg(\dfrac{\dot{\SmallStrain}^{\mathrm{max}}_{xx}}{\SmallStrain^{\mathrm{A}}_{xx}}\Time \Bigg).
\end{align}
First, the samples are subjected to four loading cycles, followed by a relaxation phase at a strain of $\SmallStrain_{xx}=0$ with the same duration as a loading cycle.

\begin{figure}[ht!]
	\centering
	\hspace{-40pt}
	\includetikz{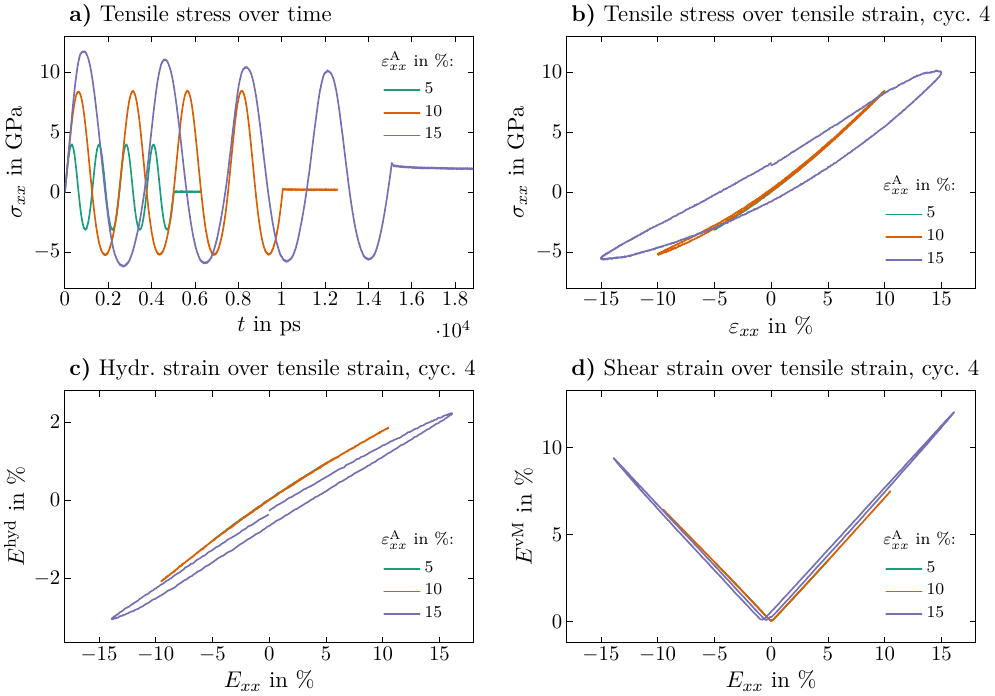}{0.9}
	\caption[Time-periodic tension-compression tests.]{Time-periodic tension-compression tests: Virial stress $\CauchyStress_{xx}$ over a)~time $\Time$ and b)~applied tensile strain $\SmallStrain_{xx}$ as well as c)~hydrostatic strain $\VolStrain$ and d)~von Mises strain $\ShearStrain$ over Green-Lagrange strain $\GreeLagStra_{xx}$ obtained in molecular dynamics simulations with uniaxial tension-compression loading at strain amplitudes $\SmallStrain^{\mathrm{A}}_{xx}$ of \SI{5}{\percent}, \SI{10}{\percent}, and \SI{15}{\percent} and a maximum strain rate of $\dot{\SmallStrain}^{\mathrm{max}}_{xx} = \SI{25}{\percent\per\nano\second}$. The samples are subjected to four cycles and a subsequent relaxation phase at a strain of $\SmallStrain_{xx} = 0$, allowing lateral strains at a pressure of zero.}
	\label{fig:Stress-xx_cycl_epsx}
\end{figure}

\Figref{fig:Stress-xx_cycl_epsx} shows the tensile stress $\CauchyStress_{xx}$ over a)~time $\Time$ and~b)~the applied strain $\SmallStrain_{xx}$. 
From \Figref{fig:Stress-xx_cycl_epsx}b), it can be noted that between strain amplitudes of $\SmallStrain^{\mathrm{A}}_{xx} = \SI{10}{\percent}$ and \SI{15}{\percent} a hysteresis develops in the stress-strain behavior, which indicates dissipative, and thus inelastic, effects.
In the fracture simulations conducted and discussed in the following sections, we evaluate the strains and thus the extent of local inelastic deformations in the vicinity of the crack tip using atomic strain measures calculated in Ovito \cite{Stukowski2009}. This evaluation includes, among others, the 
hydrostatic strain $\VolStrain$ (\Eqref{eq:GreenLagrangeStrainHyd}) and the von Mises local shear invariant $\ShearStrain$ (\Eqref{eq:MisesGreenLagrangeStrain}).
To assess the magnitude of the local strains at the crack tip, we compare them with the results obtained when calculating $\VolStrain$ and $\ShearStrain$ based on the total box strains occurring in the cyclic pure MD simulations. The corresponding strain measures as a function of the applied strain $\SmallStrain_{xx}$ are given in   \Figref{fig:Stress-xx_cycl_epsx}c) and~d). At the strains of the MD boxes at which inelasticity begins to occur according to \Figref{fig:Stress-xx_cycl_epsx}b), i.e., between strain amplitudes of $\SmallStrain^{\mathrm{A}}_{xx}=\SI{10}{\percent}$ and \SI{15}{\percent}, $\VolStrain$ measures about \SI{-2.1}{\percent} to \SI{-3.1}{\percent} in the compression phase and about  \SI{1.9}{\percent} to \SI{2.3}{\percent} in the tension phase, while $\ShearStrain$ is in a range of \SI{6.5}{\percent} to \SI{9.5}{\percent} under compression and approximately \SI{7.5}{\percent} to \SI{12.1}{\percent} under tension.

\begin{figure}[ht!]
	\centering
	\includetikz{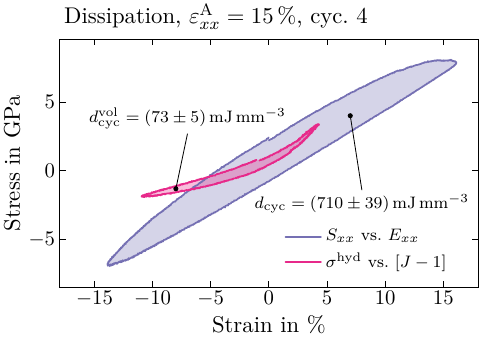}{0.48}
	\caption[Volumetric dissipation.]{Volumetric dissipation: Second Piola-Kirchhoff stress $\SecPioKirStress_{xx}$ and hydrostatic stress $\VolStress$ over Green-Lagrange strain $\GreeLagStra_{xx}$ or volume change $[\Jac-1]$, respectively, for time-periodic tests at a strain amplitude of $\SmallStrain^{\mathrm{A}}_{xx} = \SI{15}{\percent}$. The fourth loading cycle is evaluated here.}
	\label{fig:Dissip_cycl_strain}
\end{figure}

In a next step, we evaluate the importance of $\VolStrain$ and $\ShearStrain$, i.e., the role of volumetric and deviatoric deformations, as indicators of inelastic effects based on \cite{Pfaller2021}. The 
dissipated energy during a period $\PeriodDuration$ starting at time $\Time_{0}$ reads
\begin{align}
	\label{eq:DissipEnCyc}
	\DissipEn_{\mathrm{cyc}} = \int\displaylimits_{\Time_{0}}^{\Time_{0}+\PeriodDuration} \KirchhoffStressTensor : \RateDefTens \InfinitesInt \Time = \int\displaylimits_{\Time_{0}}^{\Time_{0}+\PeriodDuration} \SecPioKirStressTensor : \dot{\GreeLagStraTen} \InfinitesInt \Time
\end{align}
with the Kirchhoff stress tensor $\KirchhoffStressTensor$, the second Piola-Kirchhoff stress tensor $\SecPioKirStressTensor$, and the rate of deformation tensor $\RateDefTens$. The latter is obtained by the push-forward operation of the rate of the Green-Lagrange strain tensor $\dot{\GreeLagStraTen}$ as \cite{Holzapfel2000}
\begin{align}
	\label{eq:RateDefTens}
	\RateDefTens = \DefGradTensor^{-T} \cdot \dot{\GreeLagStraTen} \cdot \DefGradTensor^{-1}.
\end{align}
In addition, the volumetric portion of the dissipated energy is
\begin{align}
	\label{eq:DissipEnVolCyc}
	\DissipEn_{\mathrm{cyc}}^{\mathrm{vol}} = \int\displaylimits_{\Time_{0}}^{\Time_{0}+\PeriodDuration} \KirchhoffStressTensor^{\mathrm{vol}} : \RateDefTens \InfinitesInt \Time = \int\displaylimits_{\Time_{0}}^{\Time_{0}+\PeriodDuration} \CauchyStress^{\mathrm{hyd}} \dot{\Jac} \InfinitesInt \Time =  \oint\displaylimits_{\Time_{0}}^{\Time_{0}+\PeriodDuration}  \CauchyStress^{\mathrm{hyd}} \InfinitesInt \Jac,
\end{align}
using the Jacobian determinant $\Jac$ and the hydrostatic stress
\nopagebreak\begin{align}
	\label{eq:StressHyd}
	\CauchyStress^{\mathrm{hyd}} = \dfrac{1}{3} \tr (\CauchyStressTensor). 
\end{align}
Applying these principles, see \Figref{fig:Dissip_cycl_strain}, the ratio of volumetric dissipation in the fourth loading cycle with a strain amplitude of $\SmallStrain^{\mathrm{A}}_{xx} = \SI{15}{\percent}$ and averaged over five samples is $\DissipEn_{\mathrm{cyc}}^{\mathrm{vol}} / \DissipEn_{\mathrm{cyc}} = \SI{10.3(0.4)}{\percent}$. 
Consequently, the deviatoric contributions to the deformation dominate the inelastic effects. Without further investigation, we do not distinguish between plastic and viscous effects, so that we consider the onset of plasticity at an absolute strain level between \SI{10}{\percent} and \SI{15}{\percent}.

\subsubsection{Time-proportional simple shear}

\begin{figure}[ht!]
	\centering
	\includetikz{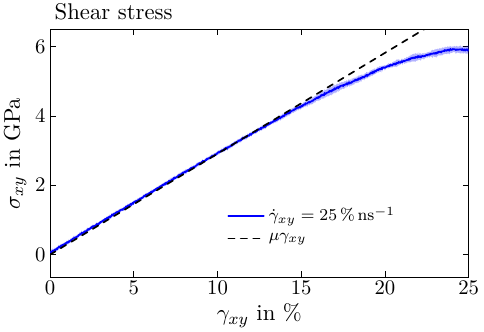}{0.48}
\caption[Simple shear.]{Simple shear: Virial stress $\CauchyStress_{xy}$ over engineering shear strain $\EngineeringShearStrain_{xy}$ obtained in molecular dynamics simple shear simulations  under impeded lateral contraction.
	The shear modulus $\ShearModulus$ is obtained by  applying the Matlab \cite{Matlab} function \texttt{polyfit} in the interval $0<\EngineeringShearStrain_{xy}\leq\SI{1.0}{\percent}$.
}
	\label{fig:sshear_fix_lateral}
\end{figure}

Besides, to assess the validity of the linear elastic constitutive law under shear conditions, we apply an engineering shear strain $\EngineeringShearStrain_{xy} \approx \Displacement_{x} / \LengthInitial_{y}^{\mathrm{MD}}$ (displacement $\Displacement_{x}$ applied between the two $y$-faces of the simulation box and constant length $\LengthInitial_{y}^{\mathrm{MD}}$ of the simulation box in $y$-direction) at a strain rate of $\dot{\EngineeringShearStrain}_{xy} = \SI{25}{\percent\per\nano\second}$ to five independent samples of size about $ \SI{100}{\angstrom} \times \SI{100}{\angstrom} \times \SI{50}{\angstrom}$  prepared under a quenching rate of $\TemperatureRate = \SI{3.2}{\kelvin\per\pico\second}$. Here, as in the study by \acite{Mantisi2012}, we keep the volume  constant, which results in a state of simple shear \cite{Holzapfel2000}. 
The shear stress response $\CauchyStress_{xy}$ in the systems  is evaluated in \Figref{fig:sshear_fix_lateral}, from which we obtain the shear modulus $\ShearModulus = \SI{29.11(0.15)}{\giga\pascal}$, which is in good agreement with the simulations of \acite{Zhang2020} using the SHIK potential. 
This value also corresponds well with the analytical prediction according to \cite{Tadmor2011}
\nopagebreak\begin{align}
	\label{eq:LameConst}		
	\ShearModulus =
	\dfrac{\YoungsModulus}{2\left[1+\PoissonsRatio\right]},
\end{align}
which leads to $\ShearModulus = \SI{29.63}{\giga\pascal}$.
Based on the results for the shear stress, we conclude that the linear elastic constitutive law is still valid up to $\bar{\TractionCurrent}_{z} \approx \SI{4}{\giga\pascal}$, which is well above the critical shear stress applied in the mode III simulations ($\bar{\TractionCurrent}_{z\mathrm{c}} \approx \SI{1.5}{\giga\pascal}$).

\subsection{Capriccio parameters and coupling studies}
\label{ssec:APP.Capriccio parameters}

\subsubsection{Overarching settings}
\label{sssec:Overarching settings}

This section provides additional information on the coupled FE-MD systems.
The continuum is discretized using hexahedral elements with eight nodes and two Gauss points per spatial direction.  In Section~\ref{sssec:APP.Quasi-1D simulations}, we examine the influence of the FE mesh resolution on the results obtained with the Capriccio coupling. 
Within the bridging domain, where the FE and MD regions overlap, the anchor points are  distributed uniformly by assigning them to the MD particles with a probability of 0.5 each.   For the setups investigated in the present study, the anchor point region has a width of \SI{10.0}{\angstrom} in the coupled directions. The width of the bridging domain is chosen so that its FE elements envelop at least all anchor points. In Section~\ref{sssec:APP.Quasi-1D simulations}, we demonstrate the influence of the bridging domain width. In this investigation, the new anchor point positions are determined by a moving least squares (MLS) approximation, where the four nearest anchor points are considered at each Gauss point. 
The DPD region covers the anchor point region and extends further into the MD domain, resulting in a   width of the DPD region of \SI{15.0}{\angstrom}. 
Due to missing atomic interaction partners towards the FE domain, we assume a loss of strain energy density of the MD region within the bridging domain. To compensate for this, we blend the FE and MD energies. Therefore, the FE energy within the bridging domain is gradually reduced from \SI{100}{\percent} to 0  towards the pure MD domain by applying a linear weighting function. However, the assumption made here that the stiffness of the MD domain decreases linearly to zero at its boundary should be further investigated in future studies. The anchor point spring stiffnesses remain unaffected by the energy weighting.
Further details on the treatment of the boundaries of the MD domain and their influence on the results of the fracture simulations can be found in 
Section~\ref{sssec:Treatment of the MD boundaries} and Section~\ref{sssec:APP.Treatment of the MD boundaries}, respectively.

\subsubsection{Treatment of the MD boundaries}
\label{sssec:Treatment of the MD boundaries}

\begin{figure}[ht!]
	\centering
    \hspace*{10pt}
	\includetikz{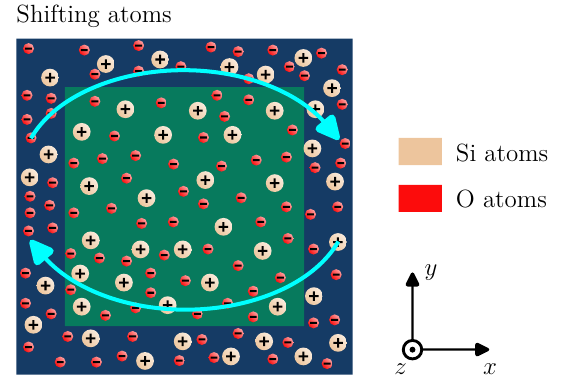}{0.6}
	\caption{Shifting of atoms to reduce the initial dipole moment.
	}
	\label{fig:Silica_dipole_correction}
\end{figure}

The MD systems, which consist of charged atoms, are cut at their boundaries to be coupled to the continuum. Additionally, in the fracture simulations introduced in Section~\ref{ssec:Fracture setups and associated stress intensity factors}, notches are created by removing atoms. Both procedures lead to artificial dipole moments and thus to an electric field that may influence the fracture behavior. The dipole moment of a set of discrete charges  $\Charge_{\AtomOne}$ with positions $\PositionVectorCur_{\AtomOne}$ is defined as \cite{Purcell2013}
\nopagebreak\begin{align}
	\label{eq:DipoleMoment} 
	\DipoleMomentVector = \sum\limits_{\AtomOne} \Charge_{\AtomOne} \PositionVectorCur_{\AtomOne}.
\end{align} 	
When preparing the samples, we take these artificial dipole moments into account in the following way: First, if applicable, we introduce a semi-elliptical notch by removing atoms on each of the two halves of the notch in the $x$-direction. To ensure a  stoichiometric ratio of deleted atoms, additional atoms necessary to be removed are chosen randomly along the crack surface. This also ensures that the systems remain charge neutral, i.e., their monopole is zero. Second, we shift some of the outermost atoms from one end of the MD region to the other one (cf.\ \Figref{fig:Silica_dipole_correction}) until the component of the dipole moment vector in the respective coupled direction is smaller than the absolute value of the O atom charge $\Charge_{\mathrm{O}}$ multiplied by the box length in that direction, i.e., $\left| \DipoleMoment_{i} \right| \overset{!}{\leq} \left|\Charge_{\mathrm{O}}\right| \LengthInitial_{i}^{\mathrm{MD}}$ in the directions  $i$ which are coupled to the continuum. 
Then, we perform an energy minimization using the Polak-Ribi\`{e}re version \cite{Polak1969} of the conjugate gradient algorithm to account for overlapping atoms. A convergence study in which we artificially vary the dipole moment is provided in Section~\ref{sssec:APP.Treatment of the MD boundaries}.

Furthermore, we use reflective walls \cite{Plimpton1995,Thompson2022} at the boundary of the MD region to account for individual atoms which would otherwise occasionally be pushed out of the bulk material and thus lead to higher computational costs. 
Based on the results presented in Section~\ref{sssec:APP.Quasi-1D simulations} and~\ref{sssec:APP.Treatment of the MD boundaries}, however, we show that the reflective walls do not significantly alter the structural properties at the boundary of the particle domain and the overall mechanical response of FE-MD systems.

\subsubsection{``Sandwich'' simulations}
\label{sssec:APP.Quasi-1D simulations}

In the following, the parameters essential for the Capriccio method are determined on the basis of a setup applying the FE-MD coupling in only one direction 
(``sandwich'' setup). 
Such a setup is advantageous for the investigation of the coupling mechanisms, since it excludes effects from the coupling in the lateral directions. However,
we also validate the obtained parameters in uniaxial tensile simulations applying the FE-MD coupling in two spatial directions while the MD domain is exposed to periodic boundary conditions in the third direction (``burrito'' setup, Section~\ref{sssec:APP.Quasi-2D simulations}) 
which are closer to the fracture tests. 

From recent studies using simplified realizations of the Capriccio method by \citeauthor{Laubert2024}~\cite{Laubert2024,Laubert2024a}, it is evident that the staggered solution scheme of the Capriccio method introduces a resistance with respect to moving the MD system, or more precisely the bridging domain, in space.
\citeauthor{Laubert2024} refer to this phenomenon  as ``motion resistance''.
It results in particular from the fact that the anchor point spring stiffness $\AnchorSpringStiffness$ not only penalizes a spatial mismatch between the MD particles and the anchor points as a penalty parameter,
but also, undesirably, the anchor point displacements \cite{Pfaller2021}.
It was found in \cite{Laubert2024,Laubert2024a} that this problem is a time-independent effect and can be mitigated, among others, by using smaller load steps or, albeit at the cost of domain adherence, a lower cumulative anchor point spring stiffness 
\nopagebreak\begin{align}
	\label{eq:kAPCumu}
	k^{\mathrm{AP}}_{\mathrm{cumu}} = n^{\mathrm{AP}} \AnchorSpringStiffness.
\end{align}
Further details on the cause of 
the motion resistance and possible remedies are provided in  \cite{Laubert2024,Laubert2024a}. There, a more detailed examination of the Capriccio coupling is also carried out, with an assessment of the strains that arise in the individual domains of ``sandwich'' systems.

\begin{figure}[ht!]
	\centering
    \hspace*{10pt}
	\includetikz{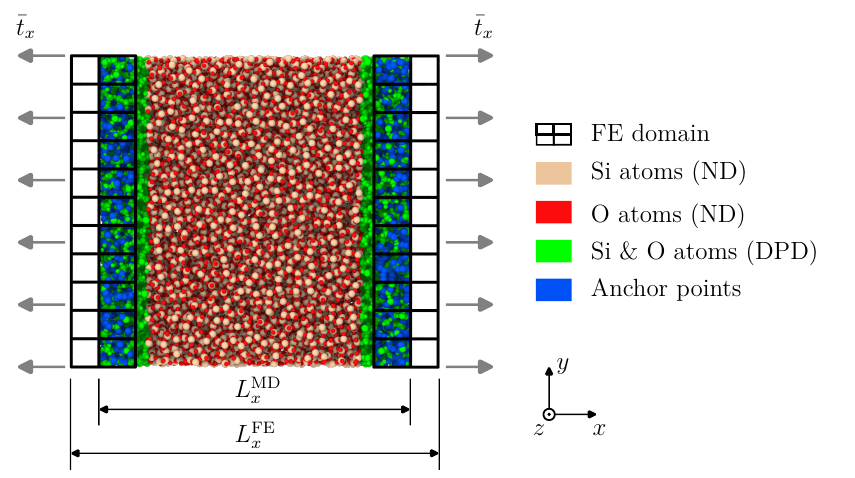}{1.0}
	\caption[Setup of the ``sandwich'' simulations.]{Setup of the ``sandwich'' simulations: A molecular dynamics (MD) domain periodic in $y$- and $z$-direction and consisting of silicon (Si) and oxygen (O) atoms described by Newtonian particle dynamics (ND) in its inner part and dissipative particle dynamics (DPD) in its outer part is concurrently coupled to a finite element (FE) domain by means of the Capriccio method. The transfer of displacements and forces between the FE and MD domain is ensured by the auxiliary anchor points. The initial system dimensions in $x$-direction are denoted as $\LengthInitial_{x}^{\mathrm{FE}}$ and $\LengthInitial_{x}^{\mathrm{MD}}$. An external tensile load is applied by imposing  surface tractions $\bar{\Traction}_{x}$.} 
	\label{fig:Silica_FEMD_SPP_Setup_SHIK}
\end{figure}

To obtain a reasonable set of parameters to couple the MD samples with FE domains using the Capriccio method, we perform a sensitivity study for the load step size $\Delta\bar{\TractionCurrent}_{x}$ and the anchor point  spring stiffness $\AnchorSpringStiffness$ in a ``sandwich'' setup. Therein, an MD region which is only periodic in $y$- and $z$-direction, see \Figref{fig:Silica_FEMD_SPP_Setup_SHIK}, is coupled to an FE region at each of its two free, non-periodic $x$-boundaries. The systems are deformed by subjecting them to Neumann boundary conditions, applying the surface traction $\bar{\TractionCurrent}_{x}$ stepwise. 
For instance, for the simulations of samples of size about $ \SI{100}{\angstrom} \times \SI{100}{\angstrom} \times \SI{50}{\angstrom}$ introduced in Section~\ref{ssec:Molecular dynamics},
approximately \num{3300} anchor points are inserted into each system. 
The overall system dimensions, i.e., the distances between the two loaded FE surfaces, are $\LengthInitial_{x}^{\mathrm{FE}} = \SI{120}{\angstrom}$ for the MD samples with dimensions $\LengthInitial_{x}^{\mathrm{MD}} \approx \LengthInitial_{y}^{\mathrm{MD}} \approx \SI{100}{\angstrom}$ and $\LengthInitial_{z}^{\mathrm{MD}} \approx \SI{50}{\angstrom}$, while $\LengthInitial_{x}^{\mathrm{FE}} = \SI{220}{\angstrom}$ applies for larger MD samples with dimensions of $\LengthInitial_{x}^{\mathrm{MD}} \approx \LengthInitial_{y}^{\mathrm{MD}} \approx \SI{200}{\angstrom}$ and $\LengthInitial_{z}^{\mathrm{MD}} \approx \SI{50}{\angstrom}$. 
In the ``sandwich'' simulations presented in this study, we do not allow for lateral contraction, partly because the application of the Berendsen barostat \cite{Berendsen1984} in the periodic directions leads to numerical instabilities. First, the systems are brought into equilibrium for a duration of \SI{160}{\pico\second} at $\bar{\TractionCurrent}_{x} = 0$, with FE and MD runs being carried out alternately, calculating the same number of MD time steps per FE load step as during the application of the mechanical loading. After this duration, the total energy in the particle domain $\TotalEnergy$ has converged, see \Figref{fig:Total-Energy}, where $\TotalEnergy$ is exemplarily evaluated during the equilibration of the fracture specimens.

\begin{figure}[ht!]
	\centering
	\includetikz{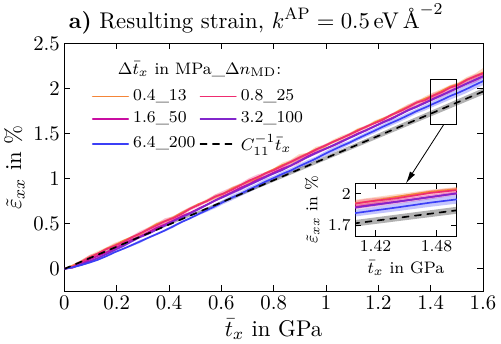}{0.48}
	\includetikz{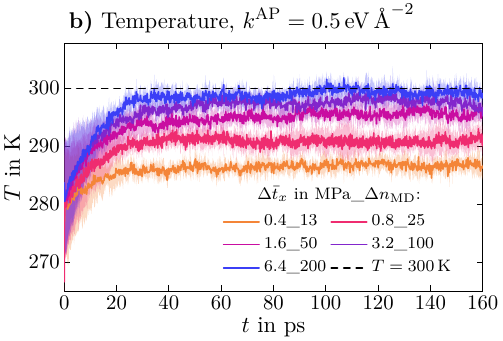}{0.48}
	\caption[Influence of the load step size  and the associated number of MD time steps per load step  in the ``sandwich'' simulations.]{Influence of the load step size $\Delta\bar{\TractionCurrent}_{x}$ and the associated number of MD time steps per load step $\MDTimeStepsPerIteration$ in the ``sandwich'' simulations: a)~Effective strain $\tilde{\SmallStrain}_{xx}$ over applied surface traction $\bar{\TractionCurrent}_{x}$
		and b)~temperature $\Temperature$ in the MD system during equilibration at zero load (prescribed temperature: $\Temperature = \SI{300}{\kelvin}$) for anchor point spring stiffness $\AnchorSpringStiffness=\SI{0.5}{\eV\per\square\angstrom}$.
		For the effective strain $\tilde{\SmallStrain}_{xx}$, the slope of the goal curve is given by the compliance $\Stiffness^{-1}_{11}$.
	}
	\label{fig:Strain-xx-rec_inv/Temperature_k0.5_ls}
\end{figure}

After the initial equilibration, the samples are subjected to uniaxial tension by imposing the surface traction $\bar{\TractionCurrent}_{x}$. The impact of the load step size $\Delta\bar{\TractionCurrent}_{x}$ is shown in \Figref{fig:Strain-xx-rec_inv/Temperature_k0.5_ls} for the anchor point spring stiffness $\AnchorSpringStiffness=\SI{0.5}{\eV\per\square\angstrom}$. Here, we consider the influence of $\Delta\bar{\TractionCurrent}_{x}$  on a)~the resulting effective strain between the two Neumann boundaries
\nopagebreak\begin{align}
	\label{eq:EpsilonRes}   \tilde{\SmallStrain}_{xx} = \dfrac{\LengthCurrent^{\mathrm{FE}}_{x} - \LengthInitial_{x}^{\mathrm{FE}}}{\LengthInitial_{x}^{\mathrm{FE}}},
\end{align}
where $\LengthCurrent^{\mathrm{FE}}_{x}$ is the current distance between the two loaded FE surfaces obtained from averaging over the nodal positions and $\LengthInitial_{x}^{\mathrm{FE}}$ is the total length of the FE-MD system after equilibration at zero load\footnote{Since the stress is the applied quantity, we plot $\tilde{\SmallStrain}_{xx}$ over $\bar{\TractionCurrent}_{x}$ in these diagrams.}, and b)~the temperature in the MD domain $\Temperature$ which develops during the equilibration. As lateral contraction is hindered in the ``sandwich'' studies, we compare the resulting integral strain $\tilde{\SmallStrain}_{xx}$ with the linear target curve $\tilde{\SmallStrain}^{\mathrm{target}}_{xx} = \Stiffness^{-1}_{11} \bar{\TractionCurrent}_{x}$ determined by the longitudinal modulus~$\Stiffness_{11}$. The response in $\tilde{\SmallStrain}_{xx}$, cf.\  \Figref{fig:Strain-xx-rec_inv/Temperature_k0.5_ls}a), indicates that convergence is reached at a load step size of $\Delta\bar{\TractionCurrent}_{x} = \SI{0.8}{\mega\pascal}$. However, there is a deviation from the target strain, which we will adjust below by modifying the anchor point spring stiffness $\AnchorSpringStiffness$. Note that in the present simulations, the number of MD time steps per load step $\MDTimeStepsPerIteration$ is scaled according to \Eqref{eq:LoadingRateFExx} such that the loading rate remains constant, which leads to an increasingly inadequate temperature control of the systems at smaller load step sizes, see \Figref{fig:Strain-xx-rec_inv/Temperature_k0.5_ls}b). This is because a certain minimum number of MD time steps must be calculated for each load step in order not to fall below the relaxation time of the thermostat, see  \cite{Zhao2021a}. However, taking into account that the target temperature of \SI{300}{\kelvin} is well below the glass transition temperature of about \SI{2000}{\kelvin} \cite{Sundararaman2018} and taking into account experimental studies on the temperature dependence of the Young's modulus of silica, as given, for example, in \cite{Marx1953,Spinner1960}, we consider the temperature deviations of at most  \SI{15}{\kelvin} determined in the present study to be negligible. However, to keep the temperature deviation as well as the numerical accuracy within reasonable limits, we choose a moderate value for the load step size of $\Delta\bar{\TractionCurrent}_{x} = \SI{1.6}{\mega\pascal}$ in the following, which leads to a number of MD time steps per load step of $ \MDTimeStepsPerIteration = 50$.

\begin{figure}[ht!]
	\centering
	\includetikz{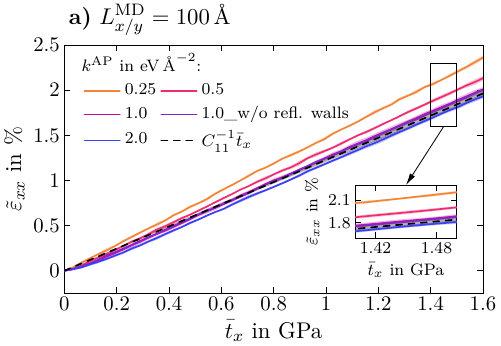}{0.48}
	\includetikz{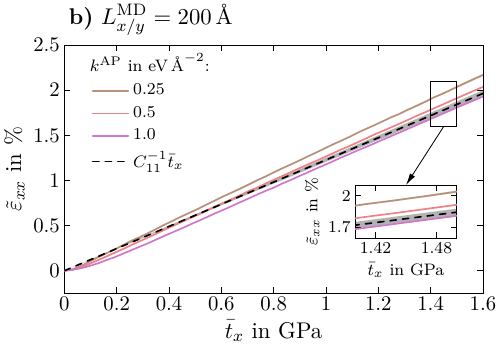}{0.48}
	\caption[Influence of the anchor point spring stiffness.]{Influence of the anchor point spring stiffness $\AnchorSpringStiffness$: Evaluation of the effective strain $\tilde{\SmallStrain}_{xx}$ over the applied surface traction $\bar{\TractionCurrent}_{x}$ in the ``sandwich'' simulations for load step size $\Delta\bar{\TractionCurrent}_{x} = \SI{1.6}{\mega\pascal}$  and atomistic specimen sizes a) $\LengthInitial^{\mathrm{MD}}_{x} \times \LengthInitial^{\mathrm{MD}}_{y} \times \LengthInitial^{\mathrm{MD}}_{z} \approx\SI{100}{\angstrom} \times \SI{100}{\angstrom} \times \SI{50}{\angstrom}$ and b)~ $\LengthInitial^{\mathrm{MD}}_{x} \times \LengthInitial^{\mathrm{MD}}_{y} \times \LengthInitial^{\mathrm{MD}}_{z} \approx\SI{200}{\angstrom} \times \SI{200}{\angstrom} \times \SI{50}{\angstrom}$. While the simulations generally use reflective walls in the coupled directions, a) provides a comparison with a series of simulations without reflective walls.
		The slope of the goal curve is given by the compliance $\Stiffness^{-1}_{11}$.
	}
	\label{fig:Strain-xx-rec_inv_k}
\end{figure}

In \Figref{fig:Strain-xx-rec_inv_k}, we analyze the influence of the anchor point spring stiffness  $\AnchorSpringStiffness$ for the previously defined load step size $\Delta\bar{\TractionCurrent}_{x} = \SI{1.6}{\mega\pascal}$ for atomistic specimen sizes of a)~$\LengthInitial^{\mathrm{MD}}_{x/y} \approx \SI{100}{\angstrom}$ and b)~$\LengthInitial^{\mathrm{MD}}_{x/y} \approx \SI{200}{\angstrom}$. Variation of $\AnchorSpringStiffness$ between \SI{0.25}{\eV\per\square\angstrom} and \SI{2.0}{\eV\per\square\angstrom} leads to the conclusion that $\AnchorSpringStiffness=\SI{1.0}{\eV\per\square\angstrom}$ is a reasonable value for the stiffness of the harmonic springs. 
Note that the effective strain between the loaded surfaces $\tilde{\SmallStrain}_{xx}$ provides only an estimate of the global system response.
For a more detailed assessment of the individual deformations in the MD, bridging and FE regions,~cf.~\cite{Laubert2024a}. 

\begin{figure}[ht!]
	\centering
	\hspace{-20pt}
	\includetikz{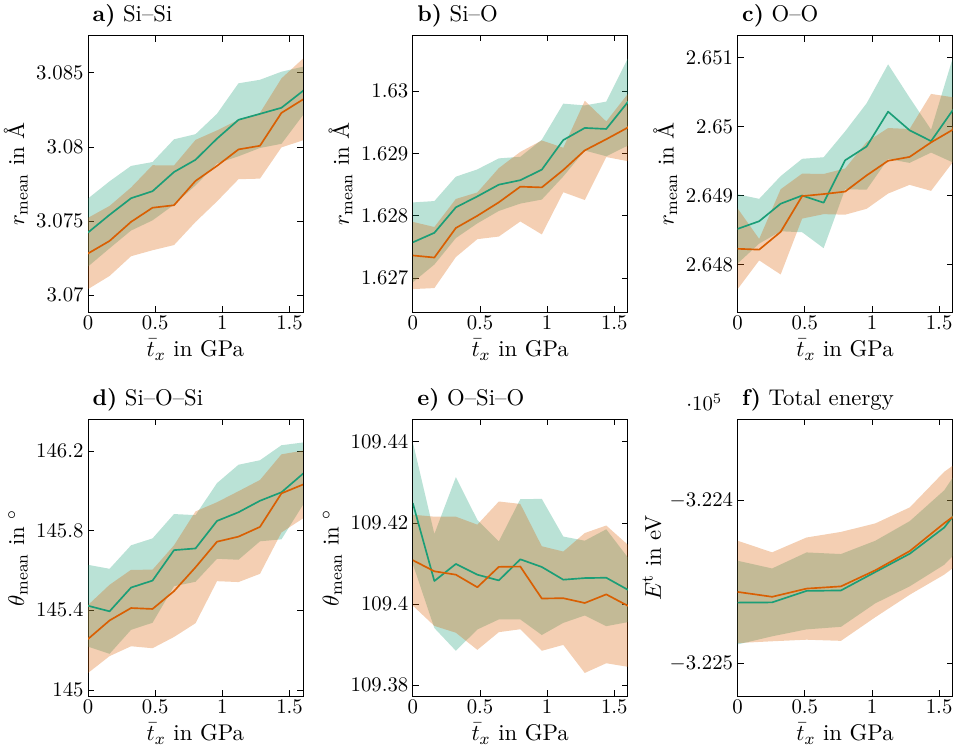}{0.9}
	\hspace*{30pt}
	\includetikz{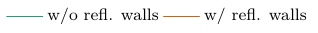}{0.5}
	\caption[Influence of the reflective walls concept on the structural properties.]{Influence of the reflective walls concept on the structural properties: Mean distance~$\InterparticleDistance_{\mathrm{mean}}$ obtained around the first peak of the radial distribution functions (RDFs) $\RDF(\InterparticleDistance)$ of the a)~Si--Si, b)~Si--O, and c)~O--O distances and mean  angle $\Angle_{\mathrm{mean}}$  obtained from the angle distribution functions (ADFs)  of the d)~Si--O--Si and e)~O--Si--O angles. Moreover, f)~shows the total energy of the particle domain $\TotalEnergy$. The interatomic distances and angles given in a) to e) are evaluated only in the dissipative particle dynamics region.
	}
	\label{fig:RDF/ADF_moment1_refl}
\end{figure}

In \Figref{fig:Strain-xx-rec_inv_k}a), we provide an additional comparison to a set of ``sandwich'' simulations that do not use the concept of reflective walls, which is generally applied in this contribution.
In the following, we further assess the influence of the reflective walls concept by analyzing the interatomic distances, interatomic angles as well as the total energy $\TotalEnergy$ of the particle domain in \Figref{fig:RDF/ADF_moment1_refl}. The radial distribution functions (RDFs) $\RDF(\InterparticleDistance)$ and the angle distribution functions (ADFs) are sampled in the DPD region of the ``sandwich'' systems. Based on these results, it can be concluded that the use of reflective walls does not significantly alter the structural properties at the boundary of the particle domain.
For the ADFs, the mean value  $\Angle_{\mathrm{mean}}$ of the entire distribution is evaluated, while for the RDFs, the mean distance around the first peak $\InterparticleDistance_{\mathrm{mean}}$ is determined in an interatomic distance range of up to  \SI{3.5}{\angstrom}, \SI{2.0}{\angstrom}, and \SI{3.0}{\angstrom}, respectively.

\begin{figure}[ht!]
	\centering
	\includetikz{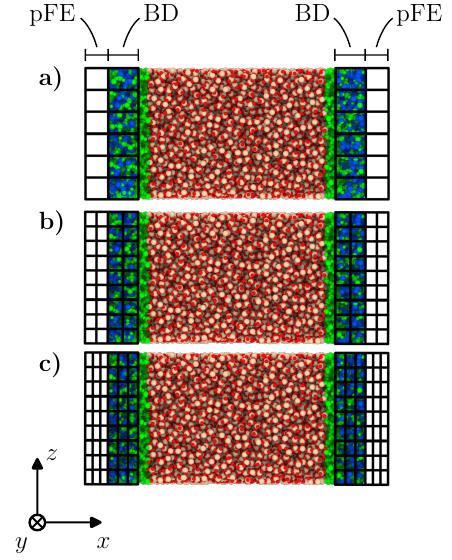}{0.48}
    \hspace{-20pt}
	\includetikz{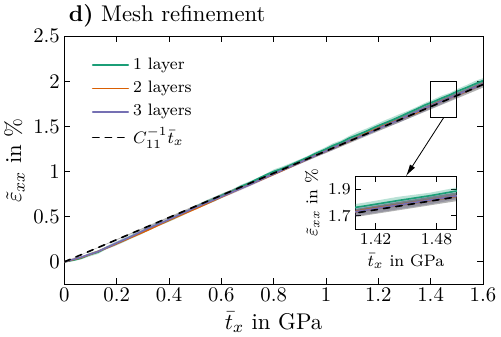}{0.48}	
	\caption[Mesh convergence study.]{Mesh convergence study: ``Sandwich'' setups with a)~one element layer each in the pure finite element (pFE) and bridging   domains  (BDs, \num{504} nodes in total), b)~two element layers (\num{1900} nodes), and c)~three element layers (\num{2660} nodes), with d)~the effective strain $\tilde{\SmallStrain}_{xx}$ over the applied surface traction $\bar{\TractionCurrent}_{x}$ during the uniaxial tensile simulations. The slope of the goal curve is given by the compliance $\Stiffness^{-1}_{11}$.
	}
	\label{fig:Strain-xx-rec_inv_FElayers}
\end{figure}

The ``sandwich'' systems with an initial overall length of $\LengthInitial_{x}^{\mathrm{FE}} = \SI{120}{\angstrom}$ are discretized by a total of \num{504} FE nodes, while \num{966} nodes are used for $\LengthInitial_{x}^{\mathrm{FE}} = \SI{220}{\angstrom}$.  The influence of the FE mesh resolution on the obtained results is shown in \Figref{fig:Strain-xx-rec_inv_FElayers}, where we refine the smaller system by using two or three layers of FE elements in both the pure FE regions and the bridging domains. We conclude that the overall deformation obtained with the Capriccio coupling is insensitive to the FE discretization.

\begin{figure}[ht!]
	\centering
	\includetikz{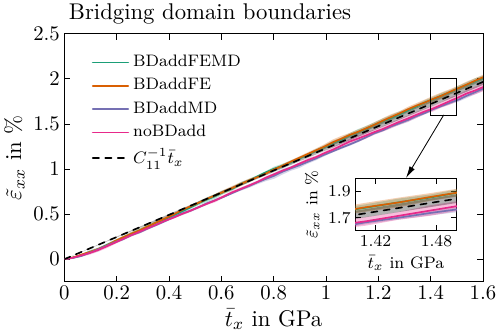}{0.48}
	
	\caption[Influence of the bridging domain boundaries.]{Influence of the bridging domain boundaries: Effective strain $\tilde{\SmallStrain}_{xx}$ over the applied surface traction $\bar{\TractionCurrent}_{x}$ during  the ``sandwich'' simulations if each  bridging domain  is enlarged by shifting its two respective boundaries adjacent to the finite element (FE) domain and the molecular dynamics (MD) domain by  \SI{1.0}{\angstrom} (into the FE and MD domain each, denoted as ``BDaddFEMD''), only one of the boundaries (into the FE domain, denoted as ``BDaddFE'' or into the MD domain, denoted as ``BDaddMD''), or if the boundaries coincide with those of the anchor point region (``noBDadd'').
		The slope of the goal curve is given by the compliance $\Stiffness^{-1}_{11}$.
	}
	\label{fig:Strain-xx-rec_inv_BDadd}
\end{figure}

In the simulations presented so far, the edges of the finite elements in the bridging domain  are set based on the positions of the outermost anchor points, but shifted to enlarge the bridging domain by \SI{1.0}{\angstrom} at both boundaries (denoted as ``BDaddFEMD''), i.e., to the pure FE and the pure MD region. The reason for this is to ensure that all anchor points are enclosed by the bridging domain. \Figref{fig:Strain-xx-rec_inv_BDadd} shows a comparison of this setting with the cases in which the bridging domain is only expanded by \SI{1.0}{\angstrom}  at the boundary to the FE domain (``BDaddFE''), the MD domain (``BDaddMD''), or at none of the two (``noBDadd''). It could be argued that the cases where the bridging domain boundary adjacent to the pure FE domain is not shifted (``BDaddMD'' and ``noBDadd'') agree better with the slope defined by the target compliance. Nevertheless, the cases where it is shifted (``BDaddFE'' and ``BDaddFEMD'') 
result in a deformation that is closer to the target overall. Since the case in which the bridging domain is only enlarged at its boundary with the FE region (``BDaddFE'') appears to produce a slightly better overall system deformation than the case in which both boundaries are shifted (``BDaddFEMD''), it is used for the following ``burrito'' and fracture simulations.

The Capriccio parameters defined for the following simulations in this contribution are summarized in \Tabref{tab:Parameters_Silica}.

\begin{table}[ht!]
	\caption{Parameters specified for the silica glass simulations.}
	\label{tab:Parameters_Silica}
	\centering
	\begin{small}
		\begin{tabularx}{0.59\textwidth}{l c c}
			\toprule
			Parameter & Symbol & Initial value \\
			\midrule
			Load step size & $\Delta \bar{\TractionCurrent}_{x}$ & \SI{1.6}{\mega\pascal} \\
			FE-MD iterations per load step & $\IterationsPerLoadStep$ & 1 \\
			Anchor point spring stiffness & $\AnchorSpringStiffness$ & \SI{1.0}{\eV\per\square\angstrom} \\
			Weighting function & $\WeightingFE/\bar{\WeightingFE}$ & $\bar{\WeightingFE}_{\mathrm{lin}}$ \\
			Remaining stiffness ratio & $\ModWeightingMD_{a}$ & 0 \\
			Number of quadrature points per direction & $\NumberGaussPoints$ & 2 \\
			DPD friction coefficient & $\DPDFricCoeff$ & \SI{0.1}{\eV \pico\second\per\square\angstrom}\\
			\bottomrule
		\end{tabularx}
	\end{small}
\end{table}

\subsubsection{``Burrito'' simulations}
\label{sssec:APP.Quasi-2D simulations}

\begin{figure}[ht!]
	\centering
    \hspace*{20pt}
	\includetikz{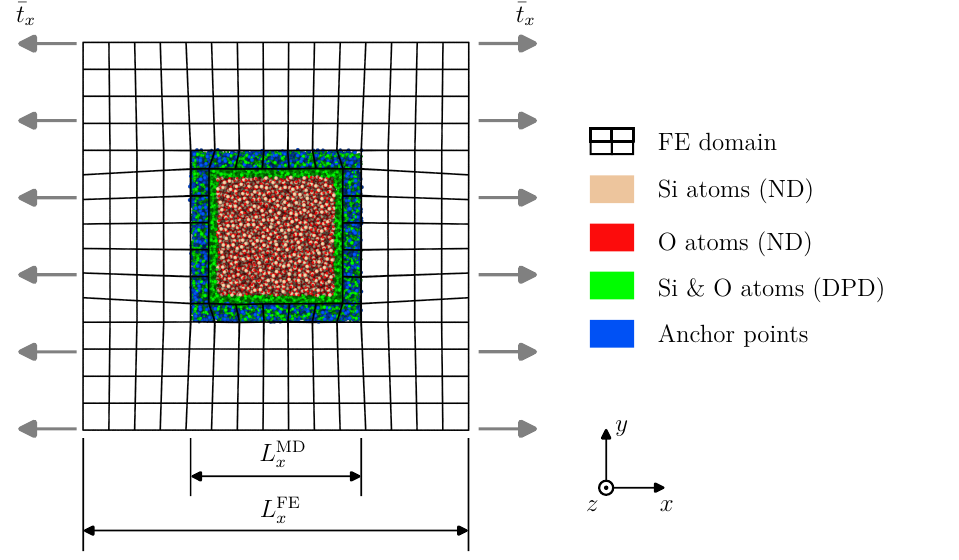}{1.0}
	\caption[Setup of the ``burrito'' simulations.]{Setup of the ``burrito'' simulations. 
	}
	\label{fig:Silica_FEMD_SSP_uncracked_Setup_SHIK}
\end{figure}

We validate the parameters selected on the basis of the ``sandwich'' simulations by means of so-called ``burrito'' simulations, where an MD system is coupled with an FE domain in both the $x$- and $y$-direction, see \Figref{fig:Silica_FEMD_SSP_uncracked_Setup_SHIK}. 

\begin{figure}[ht!]
	\centering
	\includetikz{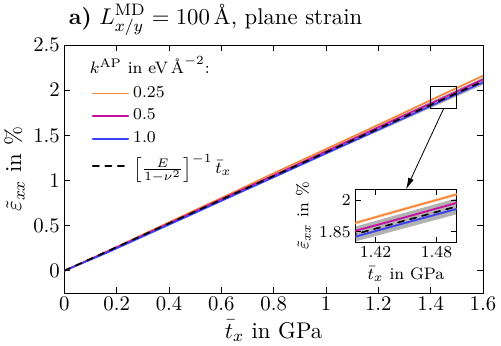}{0.48}
	\vspace{5pt}
	\includetikz{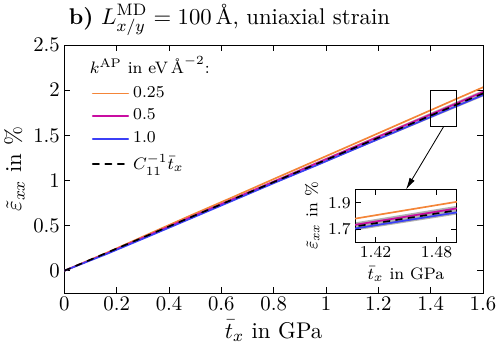}{0.48}
	\includetikz{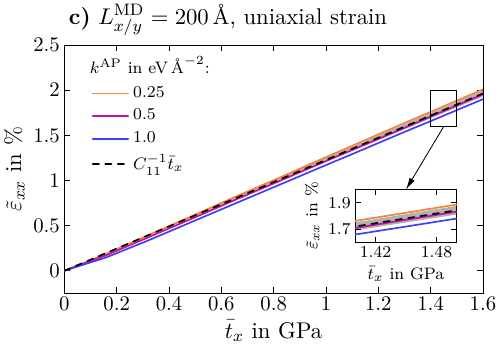}{0.48}
	\caption[Assessment of the Capriccio coupling in two spatial directions under free lateral contraction.]{Assessment of the Capriccio coupling in two spatial directions under free lateral contraction: Effective strain $\tilde{\SmallStrain}_{xx}$ over the applied surface traction $\bar{\TractionCurrent}_{x}$ during the ``burrito'' simulations for different anchor point spring stiffnesses $\AnchorSpringStiffness$  a)~in the plane strain case for an atomistic specimen size of $\LengthInitial^{\mathrm{MD}}_{x} \times \LengthInitial^{\mathrm{MD}}_{y} \times \LengthInitial^{\mathrm{MD}}_{z} \approx \SI{100}{\angstrom} \times \SI{100}{\angstrom} \times \SI{50}{\angstrom}$ and impeding the contraction of both lateral directions for b)~$\LengthInitial^{\mathrm{MD}}_{x/y} = \SI{100}{\angstrom}$ and c)~$\LengthInitial^{\mathrm{MD}}_{x/y} = \SI{200}{\angstrom}$.
		The slopes of the goal curves are given by the compliances $[\YoungsModulus/[1-\PoissonsRatio^{2}]]^{-1}$ and $\Stiffness^{-1}_{11}$, respectively.
	}
	\label{fig:Strain-xx-rec_inv_SSP}
\end{figure}

On the one hand, the plane strain state is prescribed, i.e.,  we only allow for lateral contraction  in the $y$-direction, see \Figref{fig:Strain-xx-rec_inv_SSP}a). Consequently, the slope of the target strain $\tilde{\SmallStrain}^{\mathrm{target}}_{xx}$ is defined by the compliance $ [\YoungsModulus/[1-\PoissonsRatio^{2}]]^{-1}$, cf.\ \Eqref{eq:EPlane},
which equals about $\SI{0.0131}{\per\giga\pascal}$ for the present samples. On the other hand, we perform tests that prevent deformations in both lateral directions 
for systems of two different sizes, cf.\ \Figref{fig:Strain-xx-rec_inv_SSP}b) and~c), where the corresponding compliance is given by $\Stiffness^{-1}_{11}$ as in the ``sandwich'' simulations. The initial system dimensions are  $\LengthInitial_{x}^{\mathrm{FE}} = \SampleWidthLEFM = \SI{230}{\angstrom}$ for  $\LengthInitial^{\mathrm{MD}}_{x} = \LengthInitial^{\mathrm{MD}}_{y} \approx \SI{100}{\angstrom}$ and $\LengthInitial_{x}^{\mathrm{FE}} = \SampleWidthLEFM = \SI{310}{\angstrom}$ for  $\LengthInitial^{\mathrm{MD}}_{x} = \LengthInitial^{\mathrm{MD}}_{y} \approx \SI{200}{\angstrom}$.
For the smaller samples, the FE mesh consists of \num{960} nodes, while \num{2530} nodes are used for the larger ones.
In general, the anchor point spring stiffness has less influence on the overall deformation compared to the ``sandwich'' systems. 
Moreover, the results for the smaller MD system with dimensions $\LengthInitial^{\mathrm{MD}}_{x} = \LengthInitial^{\mathrm{MD}}_{y} \approx \SI{100}{\angstrom}$ show that the parameters determined in ``sandwich'' simulations are also appropriate when the Capriccio coupling acts in two spatial directions\footnote{Since it only became apparent in the studies presented here and in \cite{Laubert2024,Laubert2024a} that the number of anchor points also influences the results (\Eqref{eq:kAPCumu}), we use $\AnchorSpringStiffness=\SI{1.0}{\eV\per\square\angstrom}$ for the following fracture simulations, even though the systems considered contain approximately twice as many anchor points as the ``sandwich'' systems.}. For the larger MD system with dimensions $\LengthInitial^{\mathrm{MD}}_{x} = \LengthInitial^{\mathrm{MD}}_{y} \approx \SI{200}{\angstrom}$, an only slightly better agreement of the strain for this larger system is suggested in the case of $\AnchorSpringStiffness = \SI{0.5}{\eV\per\square\angstrom}$, resulting from the larger number of anchor points and thus the larger cumulative anchor point spring stiffness, cf.\ \Eqref{eq:kAPCumu}.

\subsection{Fracture simulations}
\label{ssec:APP.Fracture simulations}

This section presents supplementary information on the fracture simulations using the Capriccio method. We first comment on the initial equilibration of the notched samples (Section~\ref{sssec:APP.Initial equilibration}) and 
the influence of the loading rate and the quenching rate under mode I and mode III conditions (Section~\ref{sssec:Influence of the loading and quenching rate}). Subsequently, we compare the evolution of the bond distances and angles under the different fracture modes (Section~\ref{sssec:Structural properties}) and 
consider the impact of the the load step size and the stiffness of the anchor point springs (Section~\ref{sssec:APP.Influence of load step size}) as well as the 
specimen dimensions (Section~\ref{sssec:Impact of the specimen dimensions}). 
After assessing the effect of the treatment of the MD boundaries with respect to the application of reflective walls and handling of dipole moments arising from the non-periodicity of the coupled atomistic samples (Section~\ref{sssec:APP.Treatment of the MD boundaries}), 
we illustrate the influence of using Dirichlet boundary conditions to model the supports in the bending tests (Section~\ref{sssec:Bending simulations: Supports modeled by Dirichlet boundary conditions}),
evaluate and the evolution of the crack tip stress and the number of bonds under mode II conditions (Section~\ref{sssec:Mode II}), 
and 
discuss the application of the compensative surface shift in the mode III simulations (Section~\ref{sssec:APP.Corrective displacement in mode III}).
Finally, we discuss challenges regarding the calculation of the stress intensity factors for various fracture mechanics test setups based on formulas available in the literature (Section~\ref{sssec:Calculation of stress intensity factors: issues encountered during the literature review}).

\subsubsection{Initial equilibration}
\label{sssec:APP.Initial equilibration}

\begin{figure}[ht!]
	\centering
	\includetikz{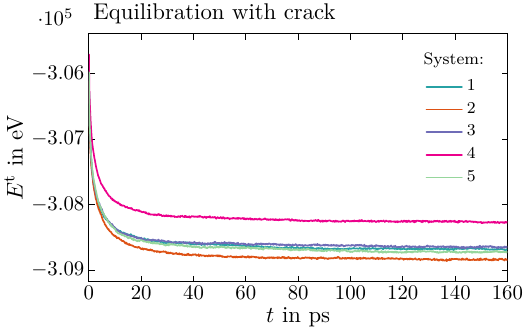}{0.48}
	\caption[Influence of the initial equilibration.]{Influence of the initial equilibration: Total energy $\TotalEnergy$ over time $\Time$ during equilibration of five pre-notched samples achieved by applying load steps under zero surface traction; strip under mode I conditions.
	}
	\label{fig:Total-Energy}
\end{figure}

As described in Section~\ref{sssec:Setup of the coupled particle-continuum systems}, we equilibrate the notched systems for a duration of \SI{160}{\pico\second}  by applying \num{2000} load steps with the respective surface traction $\bar{\TractionCurrent}_{i}=0$ and 50 MD time steps per load step. The evolution of the total energy $\TotalEnergy$ during the initial equilibration of the notched FE-MD systems for a duration of  $\SI{160}{\pico\second}$ is plotted for five individual samples in \Figref{fig:Total-Energy}. After this procedure, the total energy remains sufficiently constant.

\subsubsection{Influence of loading rate and quenching rate}
\label{sssec:Influence of the loading and quenching rate}

\paragraph{Influence of the loading rate, mode I.}

\begin{figure}[ht!]
	\centering
	\includetikz{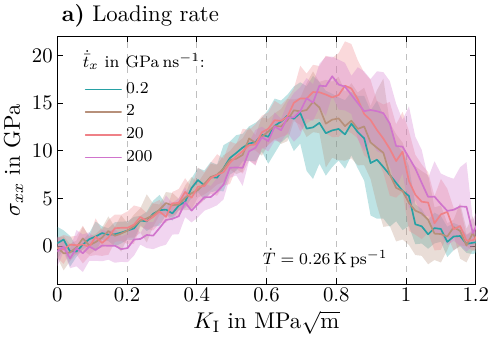}{0.48}
	\includetikz{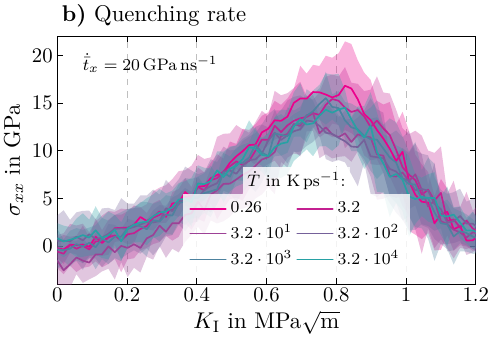}{0.48}
	\caption[Rate effects for the strip under mode I.]{Rate effects for the strip under mode I: Virial stress $\CauchyStress_{xx}$ over the stress intensity factor $\StressIntensFactor_{\mathrm{I}}$  for different a)~loading rates $\dotbar{\TractionCurrent}_{x}$ and b)~quenching rates $\dot{\Temperature}$.
	}
	\label{fig:Stress-xx-obs_rates}
\end{figure}

\Figref{fig:Stress-xx-obs_rates}a)~shows the influence of the loading rate $\dotbar{\TractionCurrent}_{x}$ on the crack tip stress $\CauchyStress_{xx}$ for loading rates 0.2, 2, 20, and \SI{200}{\giga\pascal\per\nano\second}.
We observe an increase in the critical stress intensity factor $\StressIntensFactor_{\mathrm{Ic}}$ with the loading rate, from \SI{0.73(0.11)}{\mega\pascal\sqrt{\metre}} for $\dotbar{\TractionCurrent}_{x}=\SI{0.2}{\giga\pascal\per\nano\second}$ to \SI{0.81(0.03)}{\mega\pascal\sqrt{\metre}} for $\dotbar{\TractionCurrent}_{x}=\SI{200}{\giga\pascal\per\nano\second}$. Thereby, the standard deviation over five replicas decreases considerably as the loading rate increases. 
In addition, the maximum stress, i.e., the strength, increases with increasing loading rate, from \SI{14.0(3.2)}{\giga\pascal} for $\dotbar{\TractionCurrent}_{x}=\SI{0.2}{\giga\pascal\per\nano\second}$ to \SI{17.8(2.1)}{\giga\pascal} for $\dotbar{\TractionCurrent}_{x}=\SI{200}{\giga\pascal\per\nano\second}$. This observation is consistent with the uniaxial tensile simulations using pure MD evaluated in \Figref{fig:ut_free_lateral_rate}a).
Note that different loading rates $\dotbar{\TractionCurrent}_{x}$ are achieved by adjusting the number of MD time steps computed to equilibrate the MD samples in each load step based on \Eqref{eq:LoadingRateFExx}. Consequently, for $\dotbar{\TractionCurrent}_{x} = \SI{200}{\giga\pascal\per\nano\second}$, the temperature in the MD region cannot be perfectly maintained at its target value of \SI{300}{\kelvin} because only five MD time steps are calculated per load step, recall
\Figref{fig:Strain-xx-rec_inv/Temperature_k0.5_ls}b).  The results suggest that $\dotbar{\TractionCurrent}_{x} = \SI{20}{\giga\pascal\per\nano\second}$ is a reasonable compromise between computational effort and a sufficiently low loading rate for the temperature to be controllable. 

\paragraph{Influence of the quenching rate, mode I.}

The effect of the quenching rate $\TemperatureRate$ on the stress in the observation region  $\CauchyStress_{xx}$ is shown in \Figref{fig:Stress-xx-obs_rates}b), where we vary $\TemperatureRate$ between 0.26 and \SI{3.2e4}{\kelvin\per\pico\second}. 
Even though there is no clear trend in terms of the toughness depending on the quenching rate, the systems appear to be slightly more brittle at higher quenching rates, with a maximum critical stress intensity factor of $\StressIntensFactor_{\mathrm{Ic}} = \SI{0.84(0.09)}{\mega\pascal\sqrt{\metre}}$ for $\TemperatureRate = \SI{3.2}{\kelvin\per\pico\second}$.
Moreover, the smaller quenching rates lead to the highest maximum stresses.
It can also be noticed that the variability of the results decreases with increasing $\TemperatureRate$. 

\paragraph{Influence of the loading rate, mode III.}

\begin{figure}[ht!]
	\centering
	\includetikz{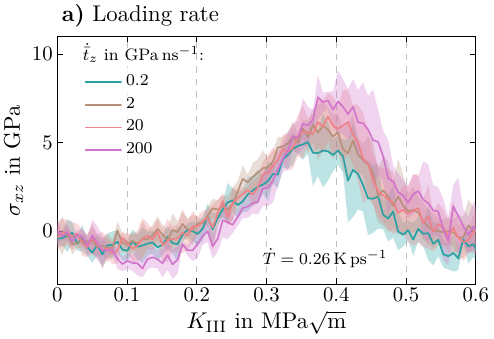}{0.48}
	\includetikz{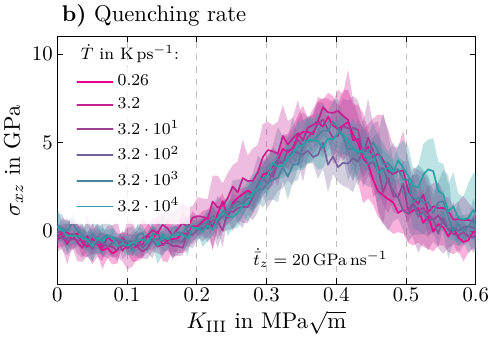}{0.48}
	\caption[Rate effects on the strip under mode III.]{Rate effects on the strip under mode III: Virial stress $\CauchyStress_{xz}$ over the stress intensity factor $\StressIntensFactor_{\mathrm{III}}$  for different a)~loading rates $\dotbar{\TractionCurrent}_{z}$ and b)~quenching rates $\dot{\Temperature}$.
	}
	\label{fig:Kinetic-Energy/Stress-xz-obs_rate}
\end{figure}

In \Figref{fig:Kinetic-Energy/Stress-xz-obs_rate}a), we determine the influence of the loading rate $\dotbar{\TractionCurrent}_{z}$ on the virial stress component~$\CauchyStress_{xz}$ in the observation region  for samples synthesized at a quenching rate of $\TemperatureRate = \SI{0.26}{\kelvin\per\pico\second}$. An increase in the critical stress intensity factor from $\StressIntensFactor_{\mathrm{IIIc}} = \SI{0.38(0.03)}{\mega\pascal\sqrt{\metre}}$ at a loading rate of $\dotbar{\TractionCurrent}_{z} = \SI{0.2}{\giga\pascal\per\nano\second}$ to $\StressIntensFactor_{\mathrm{IIIc}} = \SI{0.41(0.03)}{\mega\pascal\sqrt{\metre}}$ at  $\dotbar{\TractionCurrent}_{z} = \SI{200}{\giga\pascal\per\nano\second}$ is observable. However, a slight decrease between the loading rates $\dotbar{\TractionCurrent}_{z} = \SI{0.2}{\giga\pascal\per\nano\second}$ and $\dotbar{\TractionCurrent}_{z} = \SI{2}{\giga\pascal\per\nano\second}$ to $\StressIntensFactor_{\mathrm{IIIc}} = \SI{0.37(0.01)}{\mega\pascal\sqrt{\metre}}$ is evident. In contrast to the mode I simulations under different loading rates,
there is no consistent decrease in the standard deviation.
However, as with the mode I simulations, there is an increase in the maximum stress, i.e., from  \SI{5.0(0.6)}{\giga\pascal} for $\dotbar{\TractionCurrent}_{z}=\SI{0.2}{\giga\pascal\per\nano\second}$ to \SI{7.6(1.3)}{\giga\pascal} for $\dotbar{\TractionCurrent}_{z}=\SI{200}{\giga\pascal\per\nano\second}$.

\paragraph{Influence of the quenching rate, mode III.}

When evaluating the crack tip stress $\CauchyStress_{xz}$ for different quenching rates $\TemperatureRate$ at a loading rate of $\dotbar{\TractionCurrent}_{z} = \SI{20}{\giga\pascal\per\nano\second}$, see \Figref{fig:Kinetic-Energy/Stress-xz-obs_rate}b), we find no clear  influence of the quenching rate on the critical stress intensity factor, but a greater variability of the results for faster quenching, which in turn differs from the simulations of mode I at varying quenching rates. 
Similar to the mode I simulations, however, the highest maximum stresses are reached at the lower quenching rates.

\paragraph{Discussion.}

In this section, we address the effect of the quenching and loading rate for mode I and mode III observed in \Figref{fig:Stress-xx-obs_rates} and \Figref{fig:Kinetic-Energy/Stress-xz-obs_rate}, respectively. 
As far as the effect of the loading rate is concerned, the standard deviation of the crack tip stress $\CauchyStress_{xx}$ decreases with increasing loading rate under mode I conditions. The reason for this could be that there is less time for local rearrangements, which could lead to more pronounced intermediate stress drops (or stress avalanches) \cite{Bamer2019,Alshabab2023,Alshabab2023}, and thus more different stress responses among the samples. 

With respect to the quenching rate, we obtain the highest maximum stresses for the lowest quenching rates applied to produce the specimens. This is consistent with the observations of \acite{Ebrahem2018}, who attribute this increase in strength to an increase in medium ring sizes at the expense of small and large rings.
However, under both mode I and mode III conditions, we do not observe a distinct brittle-to-ductile transition when we increase the quenching rate. 
This contrasts with the uniaxial tensile simulations utilizing pure MD evaluated in \Figref{fig:ut_free_lateral_cooling_rate}, which, in accordance with \cite{Ebrahem2018} and \cite{Zhang2020}, consistently lead to more ductile behavior at higher quenching rates.
A transition from brittle to ductile was also observed by \acite{Richard2021} in their MD simulations on glass samples featuring central notches. 
Based on \cite{Alshabab2023,Alshabab2024}, we hypothesize that the edge-notch applied in the present study introduces a more pronounced singularity and thus stronger localization. 
This is where competition between fracture and rupture, i.e., strength-based damage, may come into play: For brittle fracture to occur, there must be a minimum crack length, otherwise the sample fails due to  rupture \cite{Sohail2023}.   The minimum notch length used in the present study, i.e., $\CrackLengthLEFM = \SI{115}{\angstrom}$, is considerably larger than the minimum crack length at which fracture occurs in silica, which reads \cite{Sohail2023}
\nopagebreak\begin{align}
	\label{eq:MinCrackLength}
	\CrackDimension_{\mathrm{min}} = \dfrac{\StressIntensFactor_{\mathrm{Ic}}^{2}}{\CauchyStress_{\mathrm{th}}^{2} \pi \GeometryFactorSih} \approx \SI{1.69}{\angstrom}, 
\end{align}
using the critical stress intensity factor $\StressIntensFactor_{\mathrm{Ic}} = \SI{0.80}{\mega\pascal\sqrt{\metre}}$, the  geometry factor $\GeometryFactorSih = 3.01$ and the theoretical strength $\CauchyStress_{\mathrm{th}} = \SI{20}{\giga\pascal}$ \cite{Wiederhorn1969,Silva2006,Brambilla2009}. 
However, the fact that the crack tip stresses in the present simulations do not decrease in a clearly brittle manner is characteristic of disordered materials \cite{Bamer2019}.
\subsubsection{Assessment of structural properties}
\label{sssec:Structural properties}

\begin{figure}[ht!]
	\centering
	\includetikz{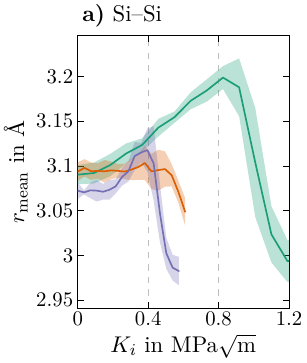}{0.3}
	\includetikz{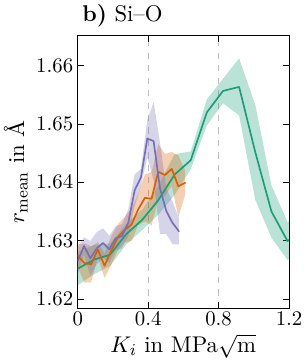}{0.3}
	\vspace{5pt}
	\includetikz{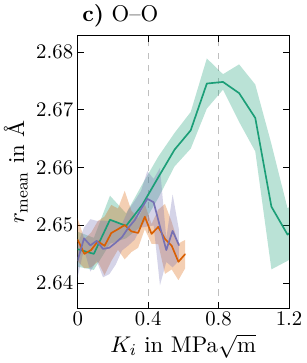}{0.3}
	\includetikz{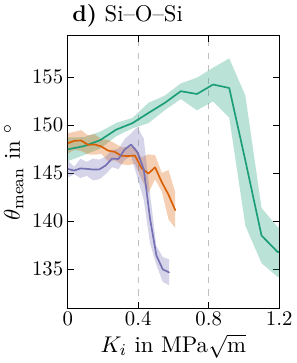}{0.3}
	\includetikz{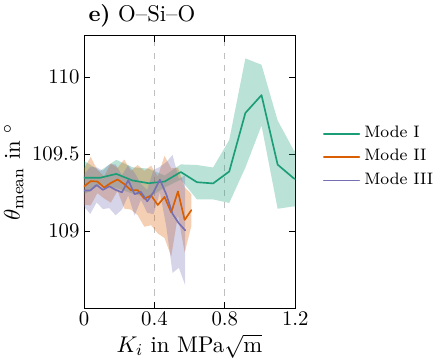}{0.3}
	\caption[Structural properties depending on the fracture mode.]{Structural properties depending on the fracture mode:		
	Mean distance  $\InterparticleDistance_{\mathrm{mean}}$ obtained around the first peak of the radial distribution functions (RDFs) $\RDF(\InterparticleDistance)$ of the a)~Si--Si, b)~Si--O, and c)~O--O distances and mean  angle $\Angle_{\mathrm{mean}}$  obtained from the entire angle distribution functions (ADFs)  of the d)~Si--O--Si and e)~O--Si--O angles over the  stress intensity factor $\StressIntensFactor_{i}\,(i=\mathrm{I,II,III})$.}
	\label{fig:RDF_ADF_moment1_modes}
\end{figure}

In the following, we analyze the atomic trajectories obtained from the mode I, II, and III fracture simulations in terms of structural properties. Such insights are only possible through the integration of chemical specificity and thus go beyond continuum-based approaches. Here, we consider the radial distribution functions (RDFs) of the Si--Si, Si--O, and O--O distances as well as the angle distribution functions (ADFs) of the Si--O--Si and the O--Si--O angles, using
an observation region of \SI{10}{\angstrom} and follow the atoms during mechanical loading. The results are displayed in \Figref{fig:RDF_ADF_moment1_modes}.
As with the ``sandwich'' simulations evaluated in \Figref{fig:RDF/ADF_moment1_refl}, for the RDFs we determine the mean distance around the first peak $\InterparticleDistance_{\mathrm{mean}}$ for interatomic distances of up to  \SI{3.5}{\angstrom}, \SI{2.0}{\angstrom}, and \SI{3.0}{\angstrom}, while for the ADFs we calculate the mean value $\Angle_{\mathrm{mean}}$ of the entire distribution.

The most pronounced widening of the structure under external load is observed for mode I. Here, all interatomic distances increase significantly, which also applies to the mean Si--O--Si angle. For mode I, the mean value of the Si--O--Si angles increases from \SI{147.5(1.2)}{\degree} to \SI{154.2(1.8)}{\degree}. The O--Si--O angles, in contrast, remain largely unaffected by the mode I loading conditions. However, while all other quantities begin to decrease upon reaching the critical stress intensity factor $\StressIntensFactor_{\mathrm{Ic}}$, the mean O--Si--O angles, which initially lie at \SI{109.3(0.1)}{\degree},  first increase slightly to \SI{109.9(0.2)}{\degree} before they return to their initial value.
The increase in bond length and its subsequent reduction during crack propagation found agrees with the results obtained by  \acite{Alshabab2023} in uniaxial tensile tests on samples with a central notch.
Performing mode I tests with a $\StressIntensFactor$-test setup in \cite{Rimsza2017}, also \citeauthor{Rimsza2017} observe an increase in Si--O distances and a corresponding increase in Si--O--Si and O--Si--O angles during loading of their samples. They attribute this behavior to elastic strains at the crack tip, followed by a decrease during fracture events. They point out that these observations are consistent with experimental studies and emphasize the need for further investigation of structural material descriptors of glasses.

Mode II is mostly invariant with respect to the Si--Si and O--O distances, while the Si--O distances increase by about half the value obtained for mode I, i.e., to about \SI{1.642(0.002)}{\angstrom}. At the start of crack propagation, however, there is a sharp reduction in the Si--Si distances, while the Si--O and O--O distances decrease only slightly. As far as the interatomic angles are concerned, no changes as rapid as those observed in the other two fracture modes are observed during crack propagation. The Si--O--Si angles decrease continuously during loading in mode II, while the O--Si--O angles decrease only slightly. 

In the simulations of mode III, a difference in the interatomic distances and angles is measured compared to the other two modes at  $\StressIntensFactor = 0$, i.e., after the equilibration of the FE-MD systems. 
This discrepancy can be attributed to the larger free surface area in the mode III arrangements \cite{Du2005}.
In mode III, all interatomic distances increase. 
In addition, we notice an increase in the Si--O--Si angles from an initial value of \SI{145.4(0.9)}{\degree} to \SI{148.0(1.2)}{\degree}, while the average O--Si--O angles remain practically constant before fracture.
At the beginning of crack propagation, a significant decrease in the Si--O--Si angles can be found, while the average O--Si--O angles show only a slight decrease.

Overall, the structure at the MD level opens up significantly more under mode I conditions, which may be the reason why the critical stress intensity factor $\StressIntensFactor_{\mathrm{Ic}}$ is greater than both $\StressIntensFactor_{\mathrm{IIc}}$ and~$\StressIntensFactor_{\mathrm{IIIc}}$. Nevertheless, further investigations beyond the scope of this work would be necessary, including a ring size analysis \cite{Ebrahem2018,Zhou2021}, to draw comprehensive conclusions from the atomic structures obtained in the present simulations.

\subsubsection{Influence of load step size and anchor point spring stiffness}
\label{sssec:APP.Influence of load step size}

\begin{figure}[ht!]
	\centering
	\includetikz{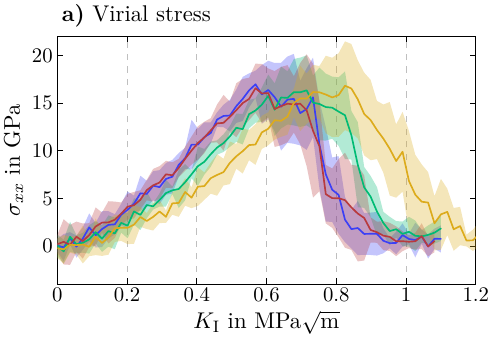}{0.48}
	\includetikz{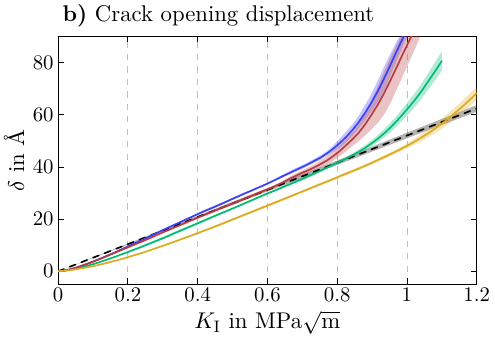}{0.48}
    \hspace*{-60pt}
	\includetikz{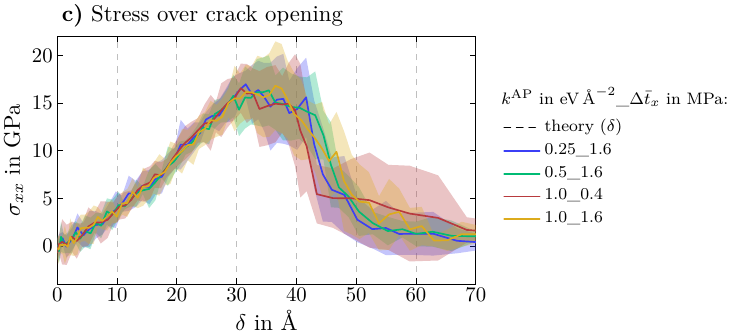}{0.48}
	
	\caption[Effect of the Capriccio setup on crack propagation.]{Effect of the Capriccio setup on crack propagation: Influence of the anchor point spring stiffness $\AnchorSpringStiffness$ and the load step size $\Delta\bar{\TractionCurrent}_{x}$ on a)~the virial stress $\CauchyStress_{xx}$ as well as~b) the crack opening displacement $\CrackOpeningDisp$ (theoretical reference curve given according to \Eqref{eq:CODModeI}) over the stress intensity factor $\StressIntensFactor_{\mathrm{I}}$ for a strip under mode I conditions. In c), the virial stress $\CauchyStress_{xx}$ is plotted over the crack opening displacement $\CrackOpeningDisp$.
	}
	\label{fig:Stress-xx-obs_ls_k_cracktip}
\end{figure}

In \Figref{fig:Stress-xx-obs_ls_k_cracktip}, we evaluate the effects of the load step size $\Delta\bar{\TractionCurrent}_{x}$ and the anchor point spring stiffness $\AnchorSpringStiffness$ for the case of a strip under mode I conditions.  For the load step size, we
test a smaller value of $\Delta\bar{\TractionCurrent}_{x} = \SI{0.4}{\mega\pascal}$ (otherwise \SI{1.6}{\mega\pascal}). 
From \Figref{fig:Stress-xx-obs_ls_k_cracktip}a), a large influence of the anchor point stiffness is evident, i.e., the samples break earlier at smaller values of  $\AnchorSpringStiffness$ because the coupling mechanism exerts less resistance to spatial motion inside the bridging domain, which implies that the MD region can move more seamlessly with the FE region. This is also apparent from the crack opening displacement $\CrackOpeningDisp$ evaluated in \Figref{fig:Stress-xx-obs_ls_k_cracktip}b). Moreover, values of the anchor point stiffness smaller than  $\AnchorSpringStiffness = \SI{1.0}{\eV\per\square\angstrom}$, which was obtained from the parameter identification procedure described in 
Section~\ref{ssec:Capriccio method},  yield a crack opening displacement that is closer to the LEFM target. As with the ``sandwich'' simulations, it becomes also clear here that a reduction in $\AnchorSpringStiffness$ has a  similar effect to a reduction in $\Delta\bar{\TractionCurrent}_{x}$ and thus a larger number of load steps, i.e., the results for $\AnchorSpringStiffness = \SI{0.25}{\eV\per\square\angstrom}$ are almost identical to those for $\Delta\bar{\TractionCurrent}_{x} = \SI{0.4}{\mega\pascal}$. 
However, when plotting the virial stress at the crack tip over the crack opening displacement, see \Figref{fig:Stress-xx-obs_ls_k_cracktip}c), the stress curves almost coincide, showing that the results are robust, especially when evaluating material properties depending on the current deformation state.  
It is also important that a direct proportionality between  $\CrackOpeningDisp$ and $\CauchyStress_{xx}$ is apparent.
Overall, the trends observed here are consistent with the in-depth studies on the influence of the load step size and the anchor point spring stiffness conducted in \cite{Laubert2024} and \cite{Laubert2024a}, respectively.

\subsubsection{Impact of the specimen dimensions}
\label{sssec:Impact of the specimen dimensions}

\begin{figure}[ht!]
	\centering
	\includetikz{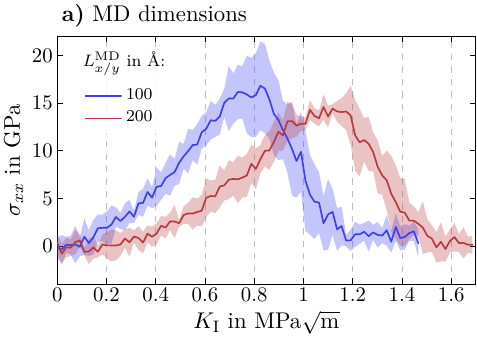}{0.48}
	\includetikz{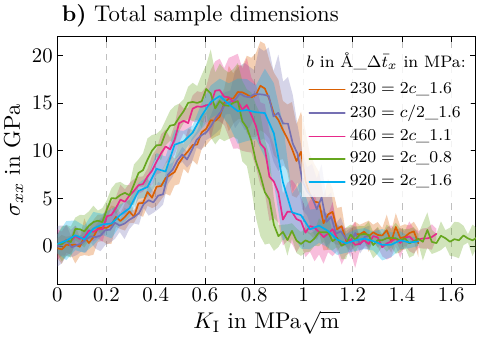}{0.48}
	\caption[Influence of the specimen size on crack propagation.]{
		Influence of the specimen size on crack propagation: Virial stress $\CauchyStress_{xx}$ over the stress intensity factor $\StressIntensFactor_{\mathrm{I}}$ depending on a) the dimensions of the molecular dynamics (MD) domain $\LengthInitial_{x}^{\mathrm{MD}}=\LengthInitial_{y}^{\mathrm{MD}}$ and b) the total sample dimensions $\SampleWidthLEFM$ and $\SampleLengthLEFM$  for a strip under mode I conditions.
	}
	\label{fig:Stress-xx-obs_LMD_LxFE_cracktip}
\end{figure}

In \Figref{fig:Stress-xx-obs_LMD_LxFE_cracktip}, we explore the influence of the sample dimensions introduced in \Figref{fig:Silica_FEMD_SSP_modeI_Setup_SHIK}.
Specifically, we investigate the influence of increasing the size of the MD domain to $\LengthInitial^{\mathrm{MD}}_{x} =  \LengthInitial^{\mathrm{MD}}_{y} \approx \SI{200}{\angstrom}$ (otherwise \SI{100}{\angstrom}) and the effect of different combinations of the sample width $\SampleWidthLEFM$ and the sample length  $2\SampleLengthLEFM$ (both otherwise \SI{230}{\angstrom}). 
For the in-plane dimensions of the MD domain of $\LengthInitial^{\mathrm{MD}}_{x} =  \LengthInitial^{\mathrm{MD}}_{y} \approx \SI{200}{\angstrom}$, the total sample dimensions are defined as  $\SampleWidthLEFM=2\SampleLengthLEFM = \SI{310}{\angstrom}$, where the FE domain comprises \num{2430} nodes. 
For the samples with dimensions $\SampleWidthLEFM=2\SampleLengthLEFM=\SI{460}{\angstrom}$, the FE domain is discretized by \num{4088} nodes, while \num{3928} nodes are used for $\SampleWidthLEFM=\SampleLengthLEFM/2=\SI{230}{\angstrom}$ and \num{15428} are employed for $\SampleWidthLEFM=2\SampleLengthLEFM=\SI{920}{\angstrom}$. Note that the geometry factor $\GeometryFactorSih$ is adapted for the case of $\SampleWidthLEFM=\SampleLengthLEFM/2=\SI{230}{\angstrom}$, i.e., $\SampleWidthLEFM / \SampleLengthLEFM = 0.5$,  according to \cite{Tada2000}
\nopagebreak\begin{align}
	\label{eq:YTada2000}
	\GeometryFactorSih =   \dfrac{0.752 + 2.02 \CrackSampleRatio + 0.37 \left[1 - \sin(\pi \CrackSampleRatio/2)\right]^{3}}{\cos(\pi \CrackSampleRatio/2)} \sqrt{\dfrac{2}{\pi \CrackSampleRatio} \tan(\pi \CrackSampleRatio/2)}.
\end{align}
However, \Eqref{eq:YTada2000}  yields similar geometry factors as the tabulated ones given in \cite{Sih1973} based on \cite{Bowie1965} and  \cite{Bowie1973}, which are valid for $\SampleLengthLEFM/\SampleWidthLEFM=0.5$, with a deviation of about $\SI{6}{\percent}$, i.e.,   $\GeometryFactorSih \approx 2.83$ for $\CrackSampleRatio = 0.5$. 
Since the ratio of notch length to sample width $\CrackSampleRatio=\CrackLengthLEFM/\SampleWidthLEFM$  is kept constant at 0.5 in these tests, the notch length increases with the sample width, which also leads to a change in the geometry factor $\GeometryFactorSih$. To increase the stress intensity factor $\StressIntensFactor_{\mathrm{I}}$ per load step equally in all simulations, i.e., with an increment of $\Delta \StressIntensFactor_{\mathrm{I}} = \SI{9.2e-4}{\mega\pascal\sqrt{\metre}}$, we reduce the surface traction step $\Delta \bar{\Traction}_{x}$ based on \Eqref{eq:KIStrip} accordingly. This results in $\Delta \bar{\Traction}_{x} = \SI{1.1}{\mega\pascal}$ for $\SampleWidthLEFM = \SI{460}{\angstrom}$ and $\Delta \bar{\Traction}_{x} = \SI{0.8}{\mega\pascal}$ for $\SampleWidthLEFM = \SI{920}{\angstrom}$.
When the size of the MD domain is changed in \Figref{fig:Stress-xx-obs_LMD_LxFE_cracktip}a), the influence of the cumulative anchor point spring stiffness $k^{\mathrm{AP}}_{\mathrm{cumu}}$, see \Eqref{eq:kAPCumu}, becomes apparent. Since the specimens contain approximately twice as many anchor points as the systems with $\LengthInitial^{\mathrm{MD}}_{x} =  \LengthInitial^{\mathrm{MD}}_{y} \approx \SI{100}{\angstrom}$, the greater motion resistance of the bridging domain leads to a delay in the mechanical loading of the MD domain. Consequently, only minor deformations are transferred to the crack tip up to a stress intensity factor of approximately $\StressIntensFactor_{\mathrm{I}} = \SI{0.4}{\mega\pascal\sqrt{\metre}}$. This ultimately causes the samples to fail at higher stress intensity factors $\StressIntensFactor_{\mathrm{I}}$.
Nevertheless, the size of the MD samples has no substantial influence on the maximum stress values at the crack tip, confirming that the selected MD samples are sufficiently large.

\Figref{fig:Stress-xx-obs_LMD_LxFE_cracktip}b) shows at first glance that samples with larger overall dimensions also tend to fracture earlier. However, the influence of the increment in the surface traction $\Delta\bar{\TractionCurrent}_{x}$ must be considered here: When using the same surface traction step of $\Delta\bar{\TractionCurrent}_{x}=\SI{1.6}{\mega\pascal}$, the cases $\SampleWidthLEFM=2\SampleLengthLEFM=\SI{230}{\angstrom}$, $\SampleWidthLEFM=\SampleLengthLEFM/2=\SI{230}{\angstrom}$, and $\SampleWidthLEFM=2\SampleLengthLEFM=\SI{920}{\angstrom}$ yield a similar stress history in the observation region. 

\subsubsection{Influence of the MD boundaries}
\label{sssec:APP.Treatment of the MD boundaries}

\begin{figure}[ht!]
	\centering
	\includetikz{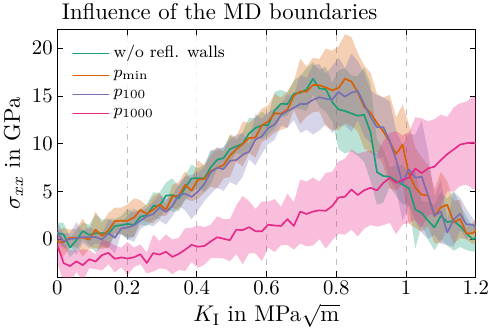}{0.48}
	\caption[Treatment of the particle domain boundaries.]{Treatment of the particle domain boundaries: Influence of the reflective walls concept and the dipole $\DipoleMomentVector$  on the virial stress $\CauchyStress_{xx}$  over the stress intensity factor $\StressIntensFactor_{\mathrm{I}}$; strip under mode I conditions.
		For the dipole moment, cases of the dipole minimized to fulfill $\left| \DipoleMoment_{i} \right| \leq  \left|\Charge_{\mathrm{O}}\right|\LengthInitial^{\mathrm{MD}}_{i}$ (``$\DipoleMoment_{\mathrm{min}}$'', default for the simulations conducted in this study) and artificially increased to values of $\left| \DipoleMoment_{i} \right| \approx 100 \left|\Charge_{\mathrm{O}}\right|\LengthInitial^{\mathrm{MD}}_{i}$ (``$\DipoleMoment_{100}$'') or $\left| \DipoleMoment_{i} \right| \approx 1000 \left|\Charge_{\mathrm{O}}\right|\LengthInitial^{\mathrm{MD}}_{i}$ (``$\DipoleMoment_{1000}$'') are compared.
	}
	\label{fig:Stress-xx-obs_p/refl_cracktip}
\end{figure}

In the following, we assess the influence of the reflective walls used at the non-periodic boundaries of the MD domain coupled to the continuum and the dipole moments arising due to the non-periodicity in \Figref{fig:Stress-xx-obs_p/refl_cracktip}. For the fracture simulations performed without reflective walls, we obtain $\StressIntensFactor_{\mathrm{Ic}} = \SI{0.78(0.07)}{\mega\pascal\sqrt{\metre}}$, which is not significantly different from the value of $\StressIntensFactor_{\mathrm{Ic}} = \SI{0.80(0.08)}{\mega\pascal\sqrt{\metre}}$ obtained with the reflective walls. 
The original components of the dipole of the default system  (atomistic specimen size $\SI{100}{\angstrom} \times \SI{100}{\angstrom} \times \SI{50}{\angstrom}$) are $\left| \DipoleMoment_{x} \right| = \SI{867.22(299.16)}{\elementarycharge\angstrom}$ and $\left| \DipoleMoment_{y} \right| = \SI{758.63(350.80)}{\elementarycharge\angstrom}$. We obtain a critical stress intensity factor of $\StressIntensFactor_{\mathrm{Ic}} = \SI{0.80(0.08)}{\mega\pascal\sqrt{\metre}}$ for a dipole moment that has been minimized by shifting atoms
to satisfy $\left| \DipoleMoment_{i} \right| \leq  \left|\Charge_{\mathrm{O}}\right|\LengthInitial^{\mathrm{MD}}_{i}$. To further investigate the influence of the dipole moment, we artificially increase it, again by shifting atoms. In this case, we obtain  $\StressIntensFactor_{\mathrm{Ic}} = \SI{0.78(0.09)}{\mega\pascal\sqrt{\metre}}$ for an artificially increased dipole moment in the $x$- and $y$-direction of $\left| \DipoleMoment_{i} \right| \approx 100 \left|\Charge_{\mathrm{O}}\right|\LengthInitial^{\mathrm{MD}}_{i}$. Only if an excessively large dipole moment is enforced ($\left| \DipoleMoment_{i} \right| \approx 1000 \left|\Charge_{\mathrm{O}}\right|\LengthInitial^{\mathrm{MD}}_{i}$), the systems become unstable. We conclude that the emerging dipole moments do not significantly affect the overall fracture properties when the non-periodic MD domains are derived from well-equilibrated amorphous systems, in which the dipole moment remains in a reasonable range even without additional modification of the atomistic structure.
However, we still minimize the initial dipole moment in the non-periodic directions of all systems studied here. Especially when larger MD systems are to be investigated, the problem of artificial dipole moments arising from the cutting of the samples becomes more important, since the dipole moment scales linearly with the system size, cf.\ \Eqref{eq:DipoleMoment}.

\subsubsection{Bending simulations: Supports modeled by Dirichlet boundary conditions}
\label{sssec:Bending simulations: Supports modeled by Dirichlet boundary conditions}

\begin{figure}[ht!]
	\centering
	\includetikz{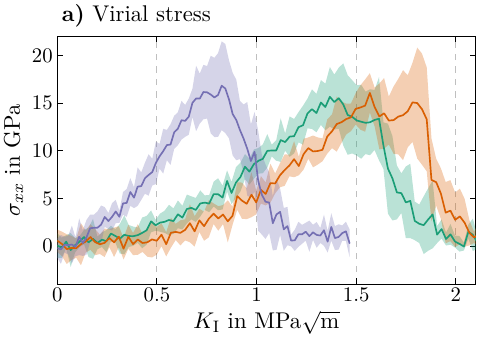}{0.48}
	\includetikz{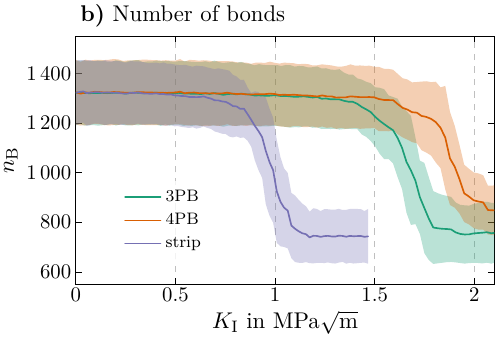}{0.48}
	\caption[Three- and four-point bending tests compared to tensile strip tests.]{Three- (3PB) and four-point bending (4PB) tests compared to tensile strip tests: a)~virial stress $\CauchyStress_{xx}$ and b)~number of bonds $\NumberBonds$ over the stress intensity factor $\StressIntensFactor_{\mathrm{I}}$.
        The supports acting in the 3PB and 4PB tests are modeled by applying Dirichlet boundary conditions.
	}
	\label{fig:Stress-xx-obs/Number-bonds_3PB-4PB_cracktip_d}
\end{figure}

For the classical case in which the supports applied in the three- (3PB) and four-point bending (4PB) tests are modeled with Dirichlet boundary conditions, \Figref{fig:Stress-xx-obs/Number-bonds_3PB-4PB_cracktip_d} shows the virial stress component $\CauchyStress_{xx}$ and the number of bonds $\NumberBonds$ evaluated in the observation region in front of the crack tip.  
Here, the displacement in the $y$-direction is set to zero for the nodes that initially share the $x$-positions of the supports.
We obtain high values for the critical stress intensity factor of  $\StressIntensFactor_{\mathrm{Ic}} = \SI{1.37(0.17)}{\mega\pascal\sqrt{\metre}}$ for 3PB and $\StressIntensFactor_{\mathrm{Ic}} = \SI{1.65(0.16)}{\mega\pascal\sqrt{\metre}}$ for 4PB. Due to the motion resistance of the bridging domain \cite{Laubert2024,Laubert2024a}, see Section~\ref{sssec:APP.Quasi-1D simulations},
the samples fail only at much higher applied loads than expected.
If the supports are modeled by fixing the $y$-displacement of the support points, the MD domain must be displaced in space over a large distance, while the motion resistance of the bridging domain \cite{Laubert2024,Laubert2024a} acts as a counterforce. An evaluation of the overall reaction force that develops at the supports shows that the coupling carries a significant share of the applied load: The reaction force is only about \SI{36.8(0.2)}{\percent} of the applied force at $\StressIntensFactor_{\mathrm{I}} \approx \SI{1.35}{\mega\pascal\sqrt{\metre}}$ for 3PB and \SI{64.7(0.3)}{\percent}  at $\StressIntensFactor_{\mathrm{I}} \approx \SI{1.66}{\mega\pascal\sqrt{\metre}}$  for 4PB. 
Taking into account the results obtained in \cite{Laubert2024,Laubert2024a}, modeling the supports using Dirichlet boundary conditions with the current state of the Capriccio method would actually require a much smaller anchor point spring stiffness, which would result in a worse domain adherence, or a much larger number of load steps, which would result in higher computational costs.

\subsubsection{Mode II}
\label{sssec:Mode II}

\begin{figure}[ht!]
	\centering	
	\includetikz{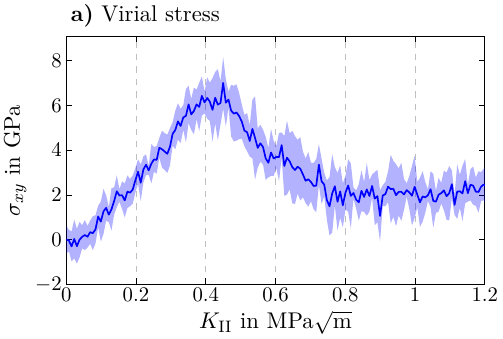}{0.48}
	\includetikz{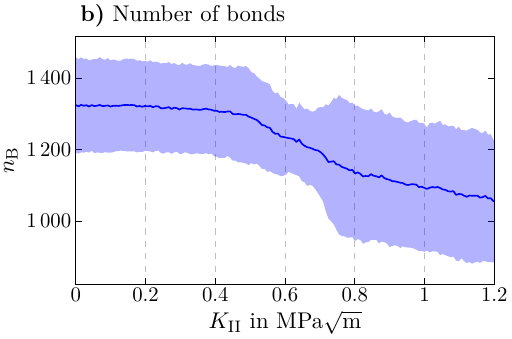}{0.48}
	\caption[Crack initiation under mode II conditions.]{Crack initiation under mode II conditions: a)~Virial stress $\CauchyStress_{xy}$ and b)~number of bonds $\NumberBonds$ over the stress intensity factor $\StressIntensFactor_{\mathrm{II}}$.}
	\label{fig:Stress-xy-obs/Number-bonds}
\end{figure}

To test the behavior of the rectangular panel samples under mode II conditions, we subject specimens that were prepared at a quenching rate of $\TemperatureRate = \SI{0.26}{\kelvin\per\pico\second}$ to a loading rate of $\dotbar{\TractionCurrent}_{y} = \SI{20}{\giga\pascal\per\nano\second}$. In \Figref{fig:Stress-xy-obs/Number-bonds}, we evaluate the shear stress component $\CauchyStress_{xy}$ and the number of bonds $\NumberBonds$ in the observation region. Based on the maximum virial stress, we obtain a critical stress intensity factor of $\StressIntensFactor_{\mathrm{IIc}} = \SI{0.44(0.03)}{\mega\pascal\sqrt{\metre}}$.

\subsubsection{Influence of the test setup: Lateral movement constraint under mode III}
\label{sssec:APP.Corrective displacement in mode III}

To assess the influence of the constraint in the $x$-direction, we apply a concept resembling a barostat to the FE domain. First, we measure the  evolution of the reaction stress averaged over the loaded positive $x$-surface  $\CauchyStress_{xx}$ during the simulations that impose the ``$x$-surface constraint'', keeping the distance $2\SampleLengthLEFM$ constant.  Then, in a new series of simulations studying the effect of the constraint, we prescribe a
new distance between the two surfaces that is necessary to eliminate  $\CauchyStress_{xx}$ based on linear elasticity,
cf.\ Section~\ref{sssec:Constitutive law}.
Specifically, we vary the distance between the surfaces by 
\nopagebreak\begin{align}
		\Delta \SampleLengthLEFM(\CauchyStress_{xx} ) &= 2\bar{\SampleLengthLEFM}(\CauchyStress_{xx} ) - 2\SampleLengthLEFM,
	\end{align}
	whereby with
	\nopagebreak\begin{align}
		\dfrac{2\SampleLengthLEFM - 2\bar{\SampleLengthLEFM}}{2\bar{\SampleLengthLEFM}} &\overset{!}{=} \dfrac{1 + \PoissonsRatio}{\YoungsModulus} \left[\CauchyStress_{xx} - \dfrac{\PoissonsRatio}{1 + \PoissonsRatio} \tr(\CauchyStressTensor)\right] 
	\end{align}
	it follows that
	\nopagebreak\begin{align}
		\label{eq:FEBarostatDistance}
		\Delta \SampleLengthLEFM(\CauchyStress_{xx} ) & =
		2\SampleLengthLEFM \left[  \dfrac{1+\PoissonsRatio}{\YoungsModulus} \left[ \CauchyStress_{xx} - \dfrac{\PoissonsRatio}{1+\PoissonsRatio} \tr(\CauchyStressTensor) \right] + 1 \right]^{-1} - 2\SampleLengthLEFM.
	\end{align}
Furthermore, since the $z$-surfaces of the FE-MD samples must be free surfaces under mode III conditions, we increase the sample thickness  compared to the mode I tests, i.e., to $\SampleThicknessLEFM \approx \SI{100}{\angstrom}$, to assess the influence of surface effects.

\begin{figure}[ht!]
	\centering
	\includetikz{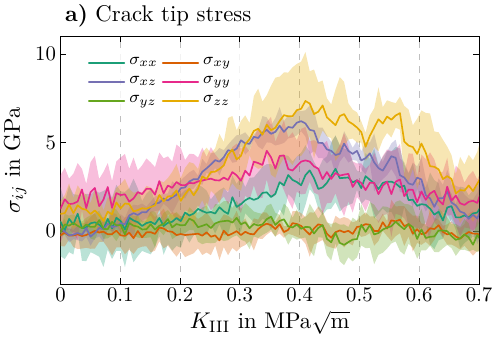}{0.48}
	\includetikz{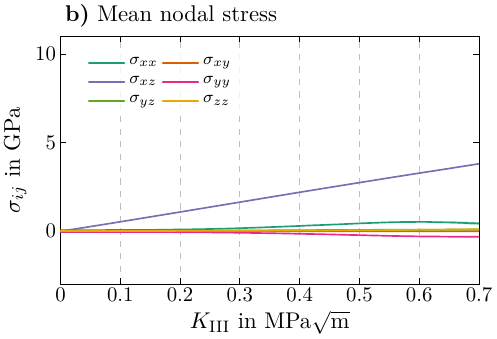}{0.48}
	\caption[Influence of the ``$x$-surface constraint'' on the mode III simulations.]{Influence of the ``$x$-surface constraint'' on the mode III simulations: a)~Virial stress components at the crack tip $\CauchyStress_{ij}$
		and b)~mean nodal stress components $\CauchyStress_{ij}$ obtained from averaging over the loaded positive $x$-face of the finite element domain over the stress intensity factor $\StressIntensFactor_{\mathrm{III}}$.
	}
	\label{fig:Stress-obs/stress}
\end{figure}

We here investigate the influence of the distance constraint between the two loaded surfaces, which was introduced in 
\Eqref{eq:FEBarostatDistance}. 
For this investigation, we double the thickness of the particle domain, resulting in an atomistic specimen size of $\SI{100}{\angstrom} \times \SI{100}{\angstrom} \times \SI{100}{\angstrom}$, with these systems being generated at a quenching rate of $\TemperatureRate = \SI{3.2}{\kelvin\per\pico\second}$.
In 
\Figref{fig:Stress-obs/stress}, we display a)~all the virial stress components that develop at the crack tip as well as b)~the nodal stress components averaged over the loaded positive $x$-surface of the FE domain during the simulations in which we apply the ``$x$-surface constraint''. It can be observed that the reaction stress  $\CauchyStress_{xx}$ resulting from the ``$x$-surface constraint'' is  small compared to the applied stress $\CauchyStress_{xz}$.   

\begin{figure}[ht!]
	\centering
	\includetikz{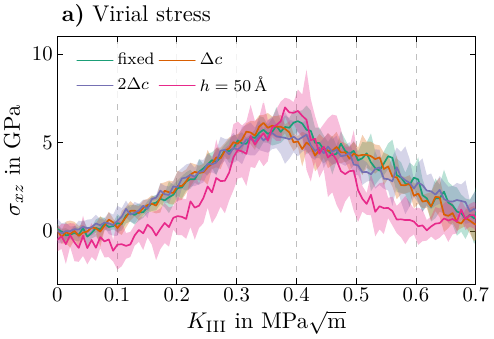}{0.48}
	\includetikz{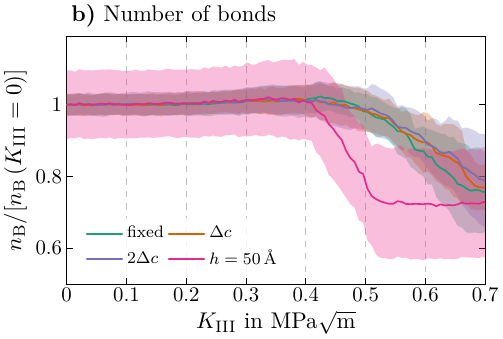}{0.48}
	\caption[Evaluation of different geometric specifications for the mode III simulations.]{Evaluation of different geometric specifications for the mode III simulations: a)~Virial stress $\CauchyStress_{xz}$ and b)~normalized number of bonds $\NumberBonds$
		over the stress intensity factor $\StressIntensFactor_{\mathrm{III}}$ for inhibiting the lateral movement of the loaded faces (``fixed'') and for applying a compensative surface shift $\Delta \SampleLengthLEFM$ according to \Eqref{eq:FEBarostatDistance} or twice its magnitude ($2 \Delta \SampleLengthLEFM$) between the two loaded faces. For these tests, the atomistic specimen size is $\SI{100}{\angstrom} \times \SI{100}{\angstrom} \times \SI{100}{\angstrom}$. The results are compared to those obtained using flatter samples  with a thickness of $\SampleThicknessLEFM = \SI{50}{\angstrom}$, where the lateral movement of the loaded  faces is inhibited.
	}
	\label{fig:Stress-xz-obs/Number-bonds_modeIII}
\end{figure}

\Figref{fig:Stress-xz-obs/Number-bonds_modeIII} provides the results for a)~the virial stress $\CauchyStress_{xz}$ and b)~the number of bonds $\NumberBonds$ at the crack tip. Since the number of bonds is an extensive quantity, we normalize it to the average value over five systems after equilibration at $\StressIntensFactor_{\mathrm{III}} = 0$. The results indicate that the ``$x$-surface constraint'' according to \Eqref{eq:FEBarostatDistance} does not significantly affect the location of the maximum stress and thus the value for the critical stress intensity factor, where the ``$x$-surface constraint'' yields $\StressIntensFactor_{\mathrm{IIIc}} = \SI{0.40(0.02)}{\mega\pascal\sqrt{\metre}}$ and the compensation by means of $\Delta \SampleLengthLEFM$ leads to $\StressIntensFactor_{\mathrm{IIIc}} = \SI{0.36(0.03)}{\mega\pascal\sqrt{\metre}}$. Applying double the value of $\Delta \SampleLengthLEFM$ (denoted as ``$2 \Delta \SampleLengthLEFM$'') results in $\StressIntensFactor_{\mathrm{IIIc}} = \SI{0.39(0.03)}{\mega\pascal\sqrt{\metre}}$.  Here, we also compare the curves for the different levels of compensative surface shifts with the results obtained for flatter specimens with a thickness of $\SampleThicknessLEFM = \SI{50}{\angstrom}$, where we apply the ``$x$-surface constraint'', i.e., $\Delta \SampleLengthLEFM = 0$. Since the sample thickness has no significant influence on the position of the maximum virial stress, i.e., we also obtain $\StressIntensFactor_{\mathrm{IIIc}} = \SI{0.40(0.02)}{\mega\pascal\sqrt{\metre}}$ for the flatter samples, we will limit the following considerations to the flatter samples to reduce the computational effort.  Furthermore, we continue to apply the ``$x$-surface constraint'' in the following, as it does not cause any significant perturbation and implies simpler boundary conditions.

\subsubsection{Calculation of stress intensity factors: Issues encountered during the literature review}
\label{sssec:Calculation of stress intensity factors: issues encountered during the literature review}

To ensure a mechanically sound comparability between the stress intensity factors for the different loading scenarios, we aim to refer to Sih's ``Handbook of stress intensity factors'' \cite{Sih1973} as a unified basis. Nevertheless, it is important to consider further literature since not all load cases can be found in a single handbook. However, a direct comparison is not always straightforward due to different sources and derivations, deviating definitions of stress intensity factors, and, partly, misprint identified in the references. Hence, we comment on these peculiarities in this section, focusing on modes I and III.

\paragraph{Basic remarks.}

Aside from Sih's handbook \cite{Sih1973}, which serves as the main source here, we also refer to the handbooks by \acite{Tada2000} and \citeauthor{Murakami1987}~\cite{Murakami1987a} and ASTM International standards \cite{ASTMSTP410,ASTMC1421-10} as well as the underlying research articles \cite{Sih1965a,Westmann1967,Gross1964,Brown1966,Bowie1965,Bowie1973,Srawley1976,Srawley1976a,Nisitani1985,Wang1992}. 
For mode I fracture, we consider (i)~the single edge cracked plate tension specimen (SECT), (ii)~the 3PB, and (iii)~the 4PB setups. For mode III, we consider a single edge crack in a rectangular beam under longitudinal shear.

\begin{figure}[ht!]
	\centering
	\hspace{-40pt}
	\includetikz{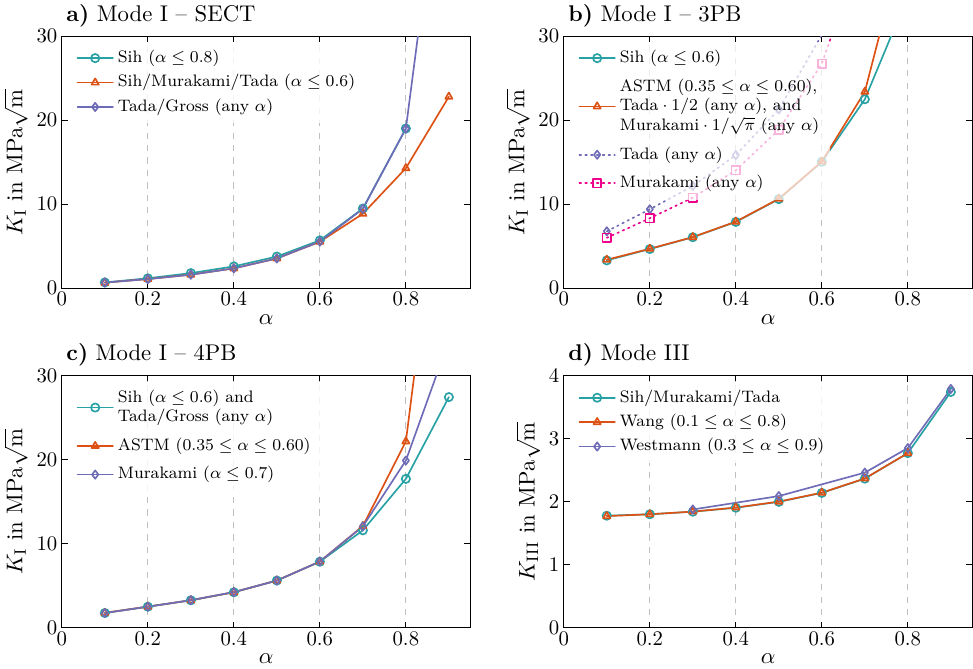}{0.9}
	\caption[Comparison of different references for calculating the stress intensity factor.]{Comparison of different references for calculating the stress intensity factor $\StressIntensFactor_{i}$ ($i=\mathrm{I,III}$): Evaluation of $\StressIntensFactor_{i}$ over the ratio of crack length over specimen width $\CrackSampleRatio = \CrackLengthLEFM / \SampleWidthLEFM$ for the a)~single edge cracked plate tension specimen (SECT), b)~three-point bending (3PB) setup, c)~four-point bending (4PB) setup, and d)~the mode III setup  provided in the references by  \acite{Sih1973}, \acite{Tada2000}, \acite{Gross2018}, \citeauthor{Murakami1987}~\cite{Murakami1987,Murakami1987a}, \acite{Westmann1967}, \acite{Wang1992} as well as the ASTM International standard ASTM C1421 \cite{ASTMC1421-10}.}
	\label{fig:comparison_references}
\end{figure}

To qualitatively compare the results for the stress intensity factors $\StressIntensFactor_{\mathrm{I}}$ and $\StressIntensFactor_{\mathrm{III}}$ obtained from the different references, we evaluate the given relations as a function of  $\CrackSampleRatio=a/b$ in \Figref{fig:comparison_references} for the following arbitrary values:  

\begin{itemize}
	\item Mode I: specimen width $\SampleWidthLEFM=\SI{1}{\metre}$,	 force $\ForceSih=\SI{1e6}{\newton}$, 	outer span $\OuterSpan = \SI{4}{\metre}$, inner span  $\InnerSpan = \SI{2}{\metre}$, and sample thickness $\SampleThicknessLEFM=\SI{1}{\metre}$  as well as
	\item Mode III: crack length $\CrackLengthLEFM=\SI{1}{\metre}$ and longitudinal shear stress $\LongitShearStressLEFM=\SI{1}{\mega\pascal}$. 
\end{itemize}
In the following, we provide a detailed comparison of the literature sources we consider in this contribution, i.e., for the SECT setup,
the 3PB 
and 4PB 
setups as well as the mode III setup.
The denotation of the system dimensions is introduced in  \Figref{fig:Silica_FEMD_SSP_modeI_Setup_SHIK} (SECT and mode III) and \Figref{fig:Silica_FEMD_SSP_3PB-4PB_Setup_SHIK} (3PB and 4PB).

\paragraph{Mode I -- SECT.}

For the SECT setup, we need to consider different formulas with varying accuracy depending on the ratio $\CrackLengthLEFM/\SampleWidthLEFM$ (crack length over specimen width, also referred to as $\CrackSampleRatio$) as well as $\SampleLengthLEFM/\SampleWidthLEFM$ (half specimen length over specimen width). As stated by \acite{Tada2000}, the effect of $\SampleLengthLEFM/\SampleWidthLEFM$ is ``practically negligible'' if $\SampleLengthLEFM/\SampleWidthLEFM \geq 1.0$. In the literature, we identify two groups of formulas, each of which renders almost identical results for the same range of $\CrackLengthLEFM/\SampleWidthLEFM$. In particular, group~A comprises formulas for $\CrackLengthLEFM/\SampleWidthLEFM \leq 0.6$, which are those of \acite{Sih1973} based on ASTM STP 410 \cite{ASTMSTP410} (labeled as ``single-edge-cracked plate under tension'' and valid for $\CrackLengthLEFM/\SampleWidthLEFM \leq 0.6$), by \acite{Tada2000} based on \acite{Gross1964} and \acite{Brown1966} (accuracy $\SI{0.5}{\percent}$ for $\CrackLengthLEFM/\SampleWidthLEFM \leq 0.6$) and by \acite{Murakami1987} based on \acite{Brown1966} (accuracy $\pm\SI{0.5}{\percent}$ for $\CrackLengthLEFM/\SampleWidthLEFM \leq 0.6$). It should be noted that the formulas of \acite{Tada2000} and \acite{Murakami1987} are identical. In contrast, group~B subsumes formulations that are valid for a wider range of $\CrackLengthLEFM/\SampleWidthLEFM$. These comprise those by \acite{Sih1973} based on \acite{Bowie1965} and \acite{Bowie1973} (labeled as ``single edge crack in a rectangular panel'' and given in tabulated form for $\CrackLengthLEFM/\SampleWidthLEFM = 0.1, 0.2, ..., 0.8$ without any statement about accuracy) at a ratio $\SampleLengthLEFM/\SampleWidthLEFM=0.5$ and by \acite{Tada2000} based on \cite{Tada1973} (accuracy ``better than \SI{0.5}{\percent} for any $\CrackLengthLEFM/\SampleWidthLEFM$''), which is identical to that given by \acite{Gross2018}. In fact, both groups yield almost the same results for ratios up to $\CrackLengthLEFM/\SampleWidthLEFM = 0.6$, see \Figref{fig:comparison_references}a).

\paragraph{Mode I -- 3PB.}

For the 3PB setup, we consider a ratio of $\OuterSpan / \SampleWidthLEFM = 4$ (span of supports over specimen width) at a ratio of  $\CrackLengthLEFM/\SampleWidthLEFM = 0.5$ (crack length over specimen width). We compare the formulas given by ASTM C1421 \cite{ASTMC1421-10} based on \acite{Srawley1976} (maximum error of \SI{2}{\percent} and $0.35 \leq \CrackLengthLEFM/\SampleWidthLEFM \leq 0.60$), by \acite{Sih1973} based on ASTM STP 410 \cite{ASTMSTP410} (labeled as ``single-edge-cracked bend specimen'' and valid for $ 0 < \CrackLengthLEFM/\SampleWidthLEFM \leq 0.6$), \acite{Tada2000} based on \acite{Srawley1976} (accuracy $\SI{0.5}{\percent}$ for any $\CrackLengthLEFM/\SampleWidthLEFM$) and by  \acite{Murakami1987} based on \acite{Srawley1976} (accuracy $\pm\SI{0.5}{\percent}$ for any $\CrackLengthLEFM/\SampleWidthLEFM$). In the case of \cite{Murakami1987}, it should be remarked that the given formula leads to results that deviate from the others by a factor of $\sqrt{2}$. For \cite{Tada2000},  the formula deviates by a factor of $2$, which erroneously appears   in the geometry factor $\GeometryFactorSih(\CrackLengthLEFM/\SampleWidthLEFM)$. Most likely, both issues are due to misprints in the references. In summary, as given in \Figref{fig:comparison_references}b), all formulas render almost the same results, where all formulations, except for that by \acite{Tada2000}, are identical if the misprints mentioned above are taken into account.

\paragraph{Mode I -- 4PB.}

For the 4PB setup, we again focus on a ratio of $\OuterSpan / \SampleWidthLEFM = 4$ (outer span over specimen width) and a ratio of $\CrackLengthLEFM/\SampleWidthLEFM = 0.5$ (crack length over specimen width). In addition, we choose the inner span  as $\InnerSpan=0.5 \OuterSpan$ according to the procedure of ASTM. Here, we consider the formulas given by ASTM C1421 \cite{ASTMC1421-10} based on \acite{Srawley1976a} (maximum error of \SI{2}{\percent} and $0.35 \leq \CrackLengthLEFM/\SampleWidthLEFM \leq 0.60$), by \acite{Sih1973} based on ASTM STP 410 \cite{ASTMSTP410} (labeled as ``single-edge-cracked bend specimen'' and valid for $ 0 < \CrackLengthLEFM/\SampleWidthLEFM \leq 0.6$), \acite{Tada2000} based on \cite{Tada1973} (accuracy ``better than $\SI{0.5}{\percent}$ for any $\CrackLengthLEFM/\SampleWidthLEFM$'') and by \acite{Murakami1987} based on \cite{Nisitani1985} (accuracy $\pm\SI{1.0}{\percent}$ for $\CrackLengthLEFM/\SampleWidthLEFM \leq 0.7$). Up to $\CrackLengthLEFM/\SampleWidthLEFM = 0.6$, all sources render almost identical results (\Figref{fig:comparison_references}c). For $\CrackLengthLEFM/\SampleWidthLEFM \geq 0.7$, the deviations become more severe. It should be noted that the formula provided by \acite{Tada2000} is identical to that given by \acite{Gross2018}.

\paragraph{Mode III.}

In the case of mode III, we compare the closed formulas provided by \acite{Sih1973} (referred to as ``strip with edge crack subjected to uniform longitudinal shear stresses''), \acite{Murakami1987a}, both based on \cite{Sih1965a} (labeled as an exact formula), and \acite{Tada2000} with the tabulated results given by \acite{Sih1973} (referred to as ``single edge crack in a rectangular beam subjected to longitudinal shear, flexure and torsion''), \acite{Westmann1967}, and \acite{Wang1992}. In contrast to the closed formulas of \cite{Sih1973}, \cite{Murakami1987a}, and \cite{Tada2000}, which do not specify the distance between the applied load and the crack, \acite{Westmann1967} and \acite{Wang1992} introduce the ratio $\SampleLengthLEFM/\SampleWidthLEFM$ (half specimen length over specimen width), which is similar to the SECT setup, and provide the stress intensity factors in tabulated form for varying $\SampleLengthLEFM/\SampleWidthLEFM$ and $\CrackLengthLEFM/\SampleWidthLEFM$. As displayed in \Figref{fig:comparison_references}d), all sources render almost identical results.

\clearpage\newpage
\bibliography{references.bib}

\end{document}